\documentclass[%
reprint, 
superscriptaddress,
preprintnumbers,
nofootinbib,
amsmath,
amssymb,
amsfonts,
aps,
prd, twocolumn,
floatfix
]{revtex4-2}

\usepackage{graphicx}
\usepackage{dcolumn}
\usepackage{bm}
\usepackage{hyperref} 
\usepackage{xcolor}
\usepackage[section]{placeins}
\usepackage{multirow}

\usepackage{graphicx}
\usepackage{hyperref}
\usepackage{latexsym}
\usepackage{bbm}
\usepackage{ulem}
\usepackage{pdfsync}
\usepackage{epsfig}
\usepackage{epstopdf}
\usepackage{subfigure}
\usepackage{color}
\usepackage{comment}
\usepackage{slashed}
\usepackage{braket}
\usepackage{setspace}

\newcommand{\dndM}{\frac{dn}{dM}(M,z)}

\hypersetup{linktocpage=true } 

\newcommand{\mbfk}{\mathbf{k}}
\newcommand{\mbfq}{\mathbf{q}}

\begin{document}


\preprint{\hbox{UTWI-13-2024}}
\title{Unlocking New Physics with Joint Power Spectrum and Voxel Intensity Distribution Forecasts in Line-Intensity Mapping}

\author{Vivian I. Sabla}
\email{vivian.sabla@austin.utexas.edu}
\affiliation{Texas Center for Cosmology and Astroparticle Physics, Weinberg Institute for Theoretical Physics, Department of Physics, The University of Texas at Austin, Austin, TX 78712, USA} 
\affiliation{Department of Physics and Astronomy, Dartmouth College, \\ HB 6127 Wilder Laboratory, Hanover, NH 03755 USA}
\author{Jos\'e Luis Bernal}
\affiliation{Instituto de Física de Cantabria (IFCA), CSIC-Univ. de Cantabria, Avda. los Castros s/n, E-39005 Santander, Spain}
\author{Gabriela Sato-Polito}
\affiliation{School of Natural Sciences, Institute for Advanced Study, Princeton, NJ 08540, United States}
\affiliation{William H. Miller III Department of Physics and Astronomy, Johns Hopkins University, 3400 North Charles Street, Baltimore, MD 21218, United States} 
\author{Marc Kamionkowski}
\affiliation{William H. Miller III Department of Physics and Astronomy, Johns Hopkins University, 3400 North Charles Street, Baltimore, MD 21218, United States}

\date{\today}

\begin{abstract}
The power spectrum and voxel intensity distribution (VID) are two
summary statistics that can be applied to condense the information encoded in line-intensity maps. The information contained in
both summary statistics is highly complementary, and their combination allows for a major increase in precision of parameter estimation from line-intensity mapping (LIM) surveys. Until recently, combination of these statistics required simulation-based estimations of their covariance. In this work we leverage an analytical model of covariance between these observables to run a joint Fisher forecast focusing on the CO(1-0) rotational line targeted by the COMAP survey and a wider, shallower hypothetical iteration. We consider a generalized phenomenological non-CDM model, models with axion dark matter, and local primordial non-Gaussianity, to highlight where a combined analysis of the power spectrum and VID can be most useful. Our results demonstrate improvements in sensitivity to beyond-$\Lambda$CDM physics over analyses using either the power spectrum or VID on their own, by factors ranging from 2 to 50, showcasing the potential of joint analyses in unlocking new insights into fundamental physics with LIM surveys.
\end{abstract}
 
\maketitle

\section{Introduction}

Line-intensity mapping (LIM) has emerged as a promising technique to quickly map large three-dimensional cosmological volumes all the way to cosmic dawn. This is achieved by measuring the integrated emission along the line of sight from bright spectral lines originating from all galaxies, including individually unresolved sources as well as from the diffuse intergalactic medium \cite{Kovetz:2017agg,Kovetz:2019uss,Bernal:2022jap}. 
Similarly to traditional large scale structure surveys, line-intensity fluctuations act as a biased tracer of the underlying matter distribution, but are also sensitive to astrophysical phenomena that trigger line emission, making LIM a probe of the formation and evolution of galaxies, and the properties of the intergalactic medium. As individual galaxy detection is not required, LIM can employ low-aperature telescopes with relatively low angular resolution to quickly map the sky with inexpensive surveys at high redshift, being optimal in the high-noise or high-confusion regimes~\cite{Cheng:2018hox, Schaan:2021hhy}. First developed as a technique to probe the 21-cm hyperfine transition in neutral hydrogen \cite{Chang:2007xk,Loeb:2008hg,Visbal:2008rg}, LIM has blossomed in recent years, with many experiments currently underway \cite{vanHaarlem:2013dsa,Bandura:2014gwa,DeBoer:2016tnn,MeerKLASS:2017vgf,Keating:2020wlx,Keating:2016pka,Cleary:2021dsp,CONCERTO:2020ahk,Gebhardt:2021vfo} and under construction \cite{,CCAT-Prime:2021lly,Sun:2020mco,Switzer:2021jeg,Vieira:2020tim,Dore:2014cca,Koopmans:2015sua,Newburgh:2016mwi}, each targeting atomic and molecular spectral lines from radio to optical wavelengths. 

As the experimental landscape opens, accurate theoretical modeling of LIM observables, covariances, and contaminants is of the utmost importance in order to derive accurate astrophysical and cosmological constraints. A number of techniques have been proposed for extracting cosmological information from line-intensity maps. The primary statistic used is the LIM power-spectrum, which captures the Gaussian information in the line-intensity maps. 
However, observational complications aside, the challenge for using LIM to constrain cosmology and astrophysics comes from their inherent interdependence --- uncertainties in the astrophysics
driving the line-emission serve as effective nuisance parameters in deriving constraints on cosmological parameters, and vice versa. For instance, to linear order the power spectrum depends only on the first two moments of the line-luminosity function, however it does so in a way that is degenerate with cosmological parameters, limiting its constraining power \cite{Padmanabhan:2018llf,Bernal:2019jdo,Schaan:2021hhy}. 

Alternative summary statistics, such as the voxel intensity distribution (VID), have been proposed to break these degeneracies while also accessing the non-Gaussian information in LIM observations. The VID is an estimator of the one-point probability distribution function (PDF) of the intensity measured within a voxel, which depends on convolutions of the line-luminosity function~\cite{Breysse:2016szq}. 
Through its sensitivity to the whole line-luminosity function, the VID is particularly useful for constraining the astrophysics of line emission, and very complementary to the power spectrum.  A joint analysis using both the power-spectrum and VID has the potential to break parameter degeneracies and significantly improve the inference of the line-luminosity function \cite{COMAP:2018kem,COMAP:2021lae,Breysse:2022alx}. Previous joint analyses empirically estimate the covariance from simulations \cite{COMAP:2018kem}, however Ref.~\cite{Sato-Polito:2022fkd} derived an analytic expression for the covariance which depends on the integrated bispectrum of one power of the emitter overdensity and two of the intensity fluctuations. 

In this work we aim to explore and highlight the gains of adding the VID to the LIM power spectrum in a joint analysis to constrain models beyond $\Lambda$CDM, taking advantage of the analytic covariance between both summary statistics. The dependence of the LIM power spectrum on new physics is twofold. At large scales, it inherits changes in the shape of the matter power spectrum or additional, scale-dependent, contributions to the halo bias. On the other hand, changes in the halo mass function, connected to the matter power spectrum, affect the line-luminosity function and therefore the amplitude of the line-intensity fluctuations and their power spectrum. This second effect also affects the VID due to the changes in the luminosity function.\footnote{Note however that the connection between an intensity bin in the VID and halo mass is not trivial, since a given intensity bin in the VID receives contribution from a wide range of halo masses (see Ref.~\cite{Sato-Polito:2022wiq})} The sensitivity of the VID to cosmologically-sourced changes in the halo mass function have been explored, including the sensitivity to the amplitude of the primordial power spectrum at small scales~\cite{Libanore:2022ntl}, and the boost of the clustering at small scales created by primordial magnetic fields~\cite{Adi:2023qdf}.\footnote{Both the power spectrum and the VID are also sensitive to contributions to the signal from exotic radiative decays~\cite{Bernal:2020lkd,Bernal:2021ylz}, although the physical process involved in the effects on the measurements are very different from those considered in this work.} 

Given the sensitivity and complementarity of the power spectrum and the VID, we focus on cosmological models that modify the halo mass function in its light and massive end. Other modifications affecting only the background evolution can be probed using baryon-acoustic oscillation analyses of the LIM power spectrum (see e.g., Ref.~\cite{Bernal:2019gfq}). First, we focus on non-cold dark matter (nCDM) models. There are many nCDM candidates, each with their own phenomenology (see e.g., Ref.~\cite{DelPopolo:2021bom} for a review), but a general feature is a suppression of the matter power spectrum at small scales, which reduces the abundance of light halos. This feature has motivated phenomenological parametrizations at specific redshifts (see e.g,. Ref.~\cite{Murgia:2017lwo}): in particular, we employ a two-parameter cut off in the matter power spectrum. In addition, we consider a fully physical case of ultralight axion dark matter \cite{Marsh:2015xka, Kim:2015yna, Chun:2014xva, Kawasaki:2014sqa, Hwang:2009js, Visinelli:2009kt, Kamionkowski:1997zb, Preskill:1982cy,Bauer:2020zsj}. While previous studies have investigated the constraining power of line-intensity mapping regarding 
non-CDM models (see e.g., Refs.~\cite{Carucci:2015bra, Creque-Sarbinowski:2018ebl, Bauer:2020zsj, Bernal:2020lkd,Murakami:2024jyi}), this is the first study considering the combination of the power spectrum and VID. 
Second, we consider the sensitivity to local primordial non-Gaussianity, which is a signature to discriminate between different inflationary models \cite{Byrnes:2010em}. The level of local primordial non-Gaussianity in strongly constrained by
\textit{Planck} observations, but its potential signatures in the large-scale structure ---affecting the halo bias at large scales and the abundance of the most massive halos--- can be used to obtain complementary, independent, and eventually stronger constraints. 

The methods and conclusions we present in this study are general to any spectral line except for 21 cm before reionization. Nonetheless, we focus on the CO(1-0) rotational emission line being measured by the COMAP survey at redshifts $2.4 < z < 3.4$ \cite{Cleary:2021dsp}, as an example of near-term observational capabilities. CO is second most abundant molecule in galaxies after molecular hydrogen making it a good tracer of the distribution of matter, as well as star formation and stellar mass \cite{Cleary:2021dsp,Bernal:2022jap,Li:2015gqa}. Similarly, the potential of the combination of the VID and the power spectrum extends to about any flavor of new physics, and we restrict our study here to the examples listed above.

This paper is organized as follows. In Sec.~\ref{sec: observables}, we review  the formalism needed to compute the LIM power spectrum and VID, as well as the covariance between the two observables. Next, we describe the beyond-$\Lambda$CDM cosmological scenarios considered, and discuss their impact on the LIM observables in Sec. \ref{sec: cosmologies}. In Sec.~\ref{sec: forecasting}, we present the survey specifications and astrophysical model of CO emission we use to derive constraints, give details on our Fisher forecast and present our results for each cosmological model considered. We end in Sec.~\ref{sec: discussion} with a discussion of our results. Further details on the noise and survey specifications are provided in the appendix.

\section{LIM Observables}
\label{sec: observables}
In this Section we review the modeling of the power spectrum and VID, as well as their covariance. The brightness temperature of a given radiation source at a redshift $z$ with a rest-frame frequency $\nu$ is related to the local luminosity density $\rho_L(z)$ via 
\begin{equation}   
    T(z) = \frac{c^3(1+z)^2}{8\pi k_B \nu^3 H(z)} \rho_L(z) \equiv X_{LT}\rho_L(z),
\end{equation}
where $c$ is the speed of light, $k_B$ is the Boltzmann constant, $H(z)$ is the Hubble parameter at the target redshift, and we define $X_{LT}$ in our second equality to simplify subsequent expressions.

The average luminosity density can be computed using the halo mass function $dn/dM$, assuming a relationship between the luminosity of the spectral line and the mass $M$ of the host halo:
\begin{equation}
    \braket{\rho_L}(z) = \int dM L(M,z)\dndM. 
\end{equation}
Using these expressions, we can derive the one- and two-point summary statistics of line-intensity maps, described below. 

\subsection{Power Spectrum}
\label{sec: Pk theory}
The power spectrum $P(k)$ is the Fourier transform of the two-point correlation function of perturbations of the temperature $\delta T \equiv T-\braket{T}$. $\delta T$ is a biased tracer of the underlying matter density perturbations, contributing a clustering component to the overall LIM power spectrum; due to the discrete distribution of line emitters, there is an additional scale-independent shot-noise contribution, resulting in $P(k,\mu,z) = P_\text{clust}(k,\mu,z) + P_\text{shot}(z)$, where $k$ is the magnitude of the Fourier mode, and $\mu$ is the cosine of the angle between the wave mode vector $\boldsymbol{k}$ and its line-of-sight component $k_\parallel$.

To first order, line-intensity and matter-density perturbations are related by a linear bias. 
We use the one- and two-halo terms of the halo model to model the power spectrum~\cite{Cooray:2002dia, Asgari:2023mej} as
\begin{multline}
    P_{\rm clust} = \left[X_{\rm LT}\int{\rm d}M \frac{{\rm d}n}{{\rm d}M}L(b_h+f\mu^2)\mathcal{U}(k)\right]^2P_{\rm m}(k) \\
    + X_{\rm LT}^2\int{\rm d}M \frac{{\rm d}n}{{\rm d}M}L^2\mathcal{U}(k) 
    \ ,
    \label{eqn: pred Pk}
\end{multline}
where $\mathcal{U}(k)$ is the Fourier transform of the NFW spherical density profile of a halo of mass $M$, which we calculate assuming a halo-mass concentration- relation given by Ref.~\cite{Diemer:2018vmz}. The first and second terms correspond to the two-halo and one-halo terms, respectively. In the equation above,  
$b_h$ denotes the halo bias, $f(z)$ is the linear growth rate, $P_m$ is the linear power spectrum of cold dark matter and baryons, and all quantities in the integral except for $f$ and $\mu$ depend on the halo mass. We neglect the redshift-space distortions at small scales, known as the fingers of God, for simplicity, especially regarding the analytic covariance with the VID. Assuming Poissonian shot noise, the second contribution to the power spectrum is\footnote{Stochastic contributions to the power spectrum and halo exclusion introduces scale-dependence in the shot noise, which also becomes non Poissonian. See e.g., Refs.~\cite{MoradinezhadDizgah:2021dei, Obuljen:2022cjo} for further details in the context of LIM.}
\begin{equation}
    \label{eq: Pshot}
    P_{\rm shot} = X_{\rm LT}^2\int{\rm d}M \frac{{\rm d}n}{{\rm d}M}L^2\,.
\end{equation}


Lastly, in our analysis we consider the monopole of the power spectrum, which can be computed as $P_0(k) = \frac{1}{2} \int d\mu P(k,\mu)$. 

\subsection{Voxel Intensity Distribution}
\label{sec: VID theory}
We follow the VID modeling from Ref.~\cite{Breysse:2022alx}, including the implementation of extended profiles from Ref.~\cite{Bernal:2023ovz}. We refer the interested reader to those references for further detail. We employ the following convention for the Fourier transforms for the VID computations. Consider $\tau$ as the Fourier conjugate of the brightness temperature $T$. The direct and inverse Fourier transforms of a function $f$ and its Fourier counterpart $\breve{f}$ are given by 
\begin{equation}
\begin{split}
    \breve{f}(\tau) & = \int {\rm d}T \ f(T)e^{-iT\tau}\,, \\
    f(T) & = \int \frac{{\rm d}\tau}{2\pi}\ \breve{f}(\tau)e^{i{\tau T}}\,. 
\end{split}
\end{equation}

Assuming that the emission line follows a Dirac delta function, each emitter contributes to the observed line-intensity map with a support given by the experimental resolution. We consider this profile to be a Gaussian determined by the frequency channel width $\delta\nu$ and the full-width half maximum $\theta_{\rm FWHM}$ of the telescope beam in the direction along and transverse to the line of sight, respectively; integrating to a volume $V_{\rm prof}$. 

If we denote the probability distribution function of the observed intensity from a single halo of mass $M$ with $\mathcal{P}_1^{(M)}(T)$ ---with units of inverse brightness temperature---, the characteristic function of the temperature distribution ---i.e., the Fourier transform of the probability distribution function for the temperature--- neglecting clustering is given by
\begin{equation}
    \breve{\mathcal{P}}^{(u)}(\tau)=\exp\left\lbrace\int{\rm d}M\frac{{\rm d}n}{{\rm d}M}V_{\rm prof}\left(\breve{\mathcal{P}}_1^{(M)}(\tau)-1\right)\right\rbrace\,,
\end{equation}
where $\tau$ is the Fourier conjugate of the temperature and we use $\breve{\mathcal{P}}$ to denote a characteristic function. 
Clustering can be taken into account promoting the halo mass function in the expression above to be density dependent, and taking the ensemble average over realizations. The ensemble average of an exponential results in the exponential of a series of powers of the cumulant of the matter distribution. Truncating that series at the second order, the characteristic function accounting for clustering is given by
\begin{equation}
    \frac{\breve{\mathcal{P}}(\tau)}{\breve{\mathcal{P}}^{(u)}(\tau)}=\exp\left\lbrace\left[\int{\rm d}M\frac{{\rm d}n}{{\rm d}M}V_{\rm prof}\left(\breve{\mathcal{P}}_1^{(M)}(\tau)-1\right)b_h\right]^2\frac{\sigma^2}{2}\right\rbrace\,,
\end{equation}
where $\sigma^2$ is the matter variance smoothed over the scales of a voxel in redshift space.  

Finally, there will be a thermal noise contribution to the total observed temperature per voxel which we model as a Gaussian distribution with standard deviation given by the effective instrumental noise per voxel, making the final total temperature PDF 
\begin{equation}
\mathcal{P}_\text{tot}(T) = (\mathcal{P}_\text{noise} \ast  \mathcal{P})(T).
\end{equation}
Before adding in the noise, we remove the mean temperature from the PDF to highlight only the fluctuations in the VID; this accounts, at zero-th order, for information loss in foreground subtraction.

In practice, $\mathcal{P}_\text{tot}$ is not directly measurable from intensity maps. However it can be estimated from the number of voxels $\mathcal{B}_i$ for which the measured intensity is within a given temperature bin $\Delta T_i$: 
\begin{equation}
    \mathcal{B}_i = \int_{\Delta T_i} dT \mathcal{P}_\text{tot}(T)\,.
\end{equation}

\subsection{Covariance}
\label{sec: Pk+VID covariance theory}
We assume a diagonal covariance for the power spectrum monopole, given by 
\begin{equation}
\sigma_0^2 = \frac{1}{2 N_{\rm modes}}\int \text{d}\mu
\tilde P_\text{tot}(k,\mu)^2\,,
\end{equation}
where the tilde denotes the observed power spectrum accounting for the experimental resolution and survey window (see discussion in Appendix \ref{sec: noise/survey specs}), and 
\begin{equation}
N_\text{modes}(k,\mu) = \frac{k^2\Delta k 
}{4\pi^2}V_\text{field}
\end{equation}
is the number of modes per bin $\Delta k$ in $k$ in the observed volume.

We assume a diagonal covariance for the VID due to the temperature binning of the observed signal.\footnote{We ignore the subdominant supersample covariance discussed in Ref.~\cite{Sato-Polito:2022fkd}, as well as the physical covariance between different bins; the interested reader can find a simulation-based estimation of this latter contribution in Refs.~\cite{COMAP:2018kem, Bernal:2023ovz}.} 
Assuming that the temperature in a voxel is a Poisson sampling of the line-intensity PDF, the diagonal variance of the VID is given by $\mathcal{B}_i(1-\mathcal{B}_i)/N_{\rm vox}$ where $N_{\rm vox}$ is the number of voxels in the line-intensity map.\footnote{We define the voxel size to be determined by $\theta_{\rm FWHM}$ and $\delta\nu/0.4247$, as is the optimal value to minimize the correlation between different intensity bins while maximizing the statistical information~\cite{Vernstrom:2013vva, Bernal:2023ovz}. Therefore, $N_{\rm vox} \equiv \Omega_\text{field}\Delta\nu/(\theta_\text{FWHM}^2\delta\nu/0.4248)$, with $\Omega_{\rm field}$ being the solid angle on the sky covered by the survey, and $\Delta\nu$ the frequency band of the experiment.} 
Ref.~\cite{Bernal:2023ovz} checked that this approximation is very accurate for the diagonal of the covariance of the VID.  

The covariance between the power spectrum monopole and the VID was derived in Ref.~\cite{Sato-Polito:2022fkd}, and can be understood as the response of the measured power spectrum to the mean temperature perturbation. This is because the VID is the distribution of the temperature fluctuation $\delta T$, which depends on the perturbation $\delta_h$ in the number density of emitters  on scales of the volume probed, and their line-luminosity PDF, while 
the power spectrum depends on two powers of $\delta T$. Here we derive the covariance for the improved modeling of the VID presented in Refs.~\cite{Breysse:2022alx, Bernal:2023ovz}.

To simplify the computation of the covariance, like the derivation in Ref.~\cite{Sato-Polito:2022fkd}, we promote $\breve{\mathcal{P}}^{(u)}$ to depend on the local halo perturbation transforming ${\rm d}n/{\rm d}M\rightarrow {\rm d}n/{\rm d}M(1+\delta_h)$ and expand the characteristic function to linear order in $\delta_h$.
Therefore, the resulting covariance is proportional to the integrated bispectrum $ \braket{\delta_h\delta T \delta T}$ with an additional factor accounting for the luminosity function:
\begin{widetext}
\begin{multline}
    \sigma_{B_i,P_0} = 
    \frac{1}{V_\text{field}^2} \Upsilon_i \int \frac{d^2\Omega_{\hat\mbfk}}{4\pi}\int \frac{d^3\mbfq_1}{(2\pi)^3} \int \frac{d^3\mbfq_2}{(2\pi)^3} \int \frac{d^3\mbfq_3}{(2\pi)^3}
    W_\text{vol}(\mbfq_1) W_\text{vol}(\mbfq_2) W_\text{vol}(\mbfq_3)\times \\ 
    W_{\rm res}(-\mbfq_1) W_{\rm res}(\mbfk-\mbfq_2) W_{\rm res}(-\mbfk-\mbfq_3) \braket{\delta_h(-\mbfq_1)\delta T(\mbfk-\mbfq_2)\delta T(-\mbfk-\mbfq_3)}, 
\end{multline}
\end{widetext}
where
\begin{equation}
\Upsilon_i = \int_{\Delta_T}{\rm d}T\int\frac{{\rm d}\tau}{2\pi}\breve{\mathcal{P}}_{\rm noise}\breve{\mathcal{P}}^{(u)}\log\left(\breve{\mathcal{P}}^{(u)}\right)e^{i T \tau}\,,
\end{equation}
and $W_{\rm vol}$ and $W_{\rm res}$ are the window functions corresponding to the volume probed and the experimental resolution, respectively, in Fourier space (see discussion in Appendix \ref{sec: noise/survey specs}).\footnote{Note that we do not include the one-halo term in our calculation of the bispectrum. We have checked that this has little impact on our results and does not affect our conclusions.} 

We refer the reader to Ref.~\cite{Sato-Polito:2022fkd} for an equivalent step-by-step derivation for an older VID modeling, and the expression for the integrated bispectrum used. 

\section{Beyond-$\Lambda$CDM Cosmologies}
\label{sec: cosmologies}
In this Section we describe the deviations from the standard $\Lambda$CDM model that we consider. Throughout this work, we assume the standard cosmological parameters to be given by the \textit{Planck} 2018 TT,TE,EE+lowE+lensing+BAO results with $\{ \omega_b, \omega_c, n_s, \log(10^{10} A_s), h \} = \{0.02242,0.11933,0.9665, 0.6766\}$ 
and consider two massless neutrinos and a massive neutrino with 0.06 eV mass \cite{Planck:2018vyg}.  
\subsection{Phenomenological non-cold dark matter}
\label{sec: nCDM theory}
 \begin{figure}[t]
     \centering
     \includegraphics[width=\linewidth]{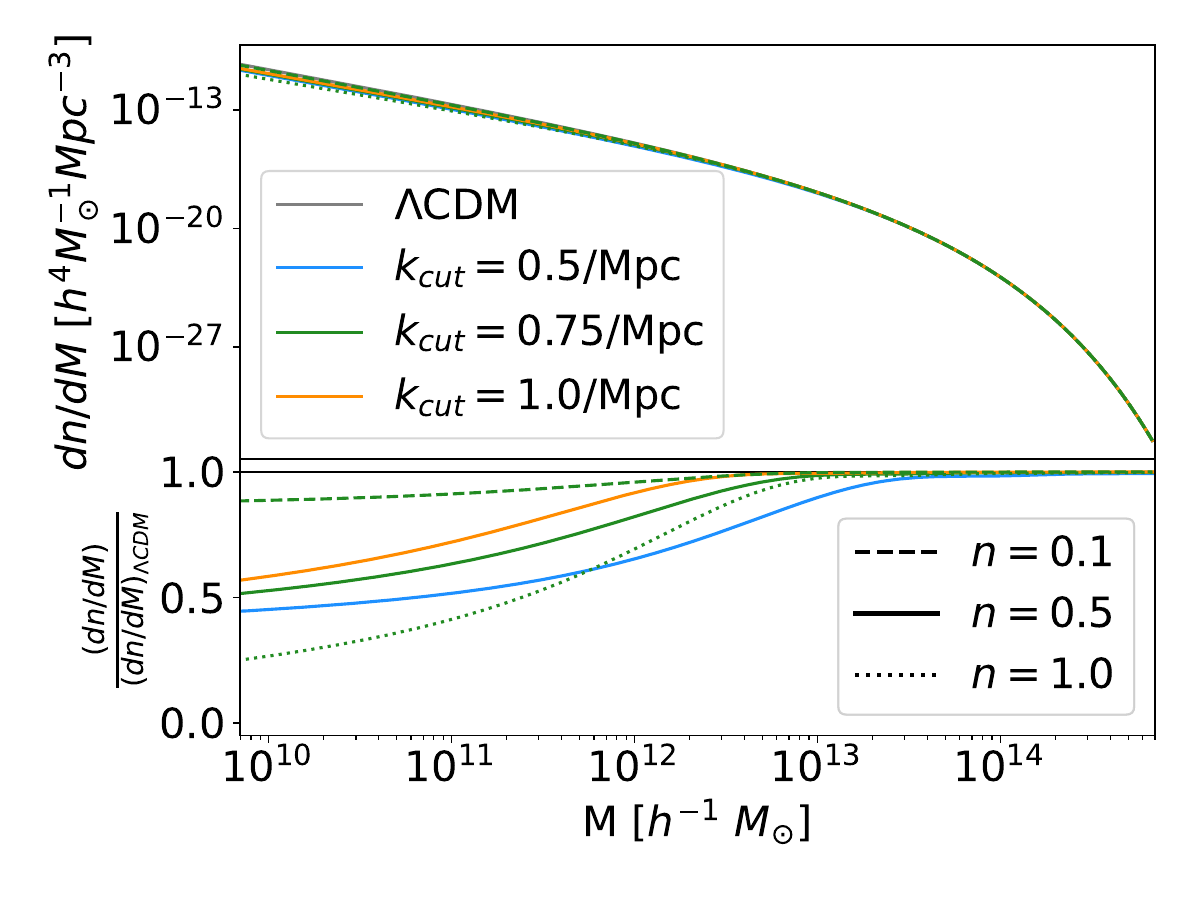}
     \caption{Halo-mass function (top) and ratio with respect to the $\Lambda$CDM prediction (bottom) for different choices of cut-off scale $k_\text{cut}$ and slope $n$ assuming a phenomenological non-cold dark matter cosmology. }
     \label{fig: ncdm HMF}
 \end{figure}
 \begin{figure*}[t]
     \centering
     \includegraphics[width=\textwidth]{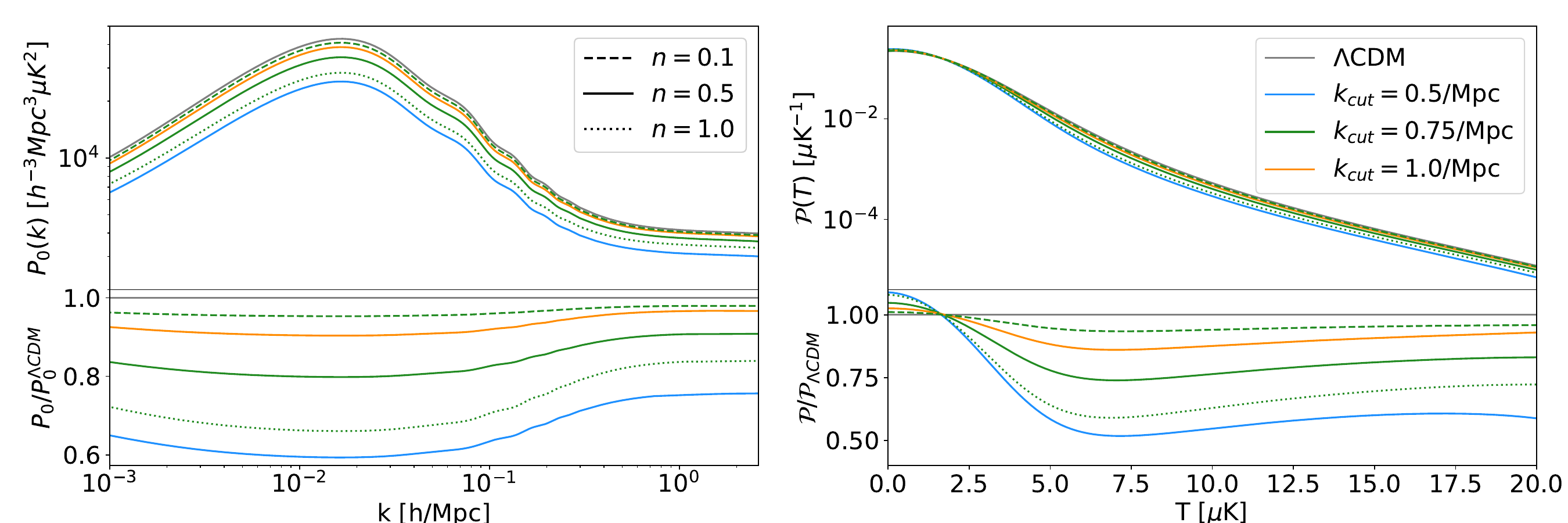}
     \caption{Observed power spectrum monopole (left), and temperature PDF (right) for the CO line observed at $z=2.9$ with a COMAP-Y5 type experiment assuming a phenomenological non-cold dark matter cosmology. Different colors represent different choices of the cut-off scale $k_{\rm cut}$, and different line styles show different choices of the slope $n$. Instrumental noise is not included in either observable, and the VID is not mean-subtracted. The $\Lambda$CDM model is shown in gray for comparison.}
     \label{fig: ncdm P0 and PT}
 \end{figure*}

We model the small-scale suppression of clustering characteristic of non-cold dark matter models by introducing a transfer function which cuts off the matter power spectrum at some specified scale $k_\text{cut}$:
\begin{equation}
\label{eq: ncdm transfer}
    \mathcal{T}^2(k) \equiv\frac{P_\text{nCDM}(k)}{P_\text{CDM}(k)} =
    \begin{cases}
    1 & \text{if } k\leq k_\text{cut},\\ 
    \left(\frac{k}{k_\text{cut}}\right)^{-n} & \text{if } k > k_\text{cut},
    \end{cases}
\end{equation}
where $n$ gives the slope or rate at which power is suppressed at small scales. 

We show the resulting halo-mass function as a function of the cut off scale and slope in Fig.~\ref{fig: ncdm HMF}. The suppression in matter power at small scales decreases the number density of low-mass halos, but high-mass halo densities are unaffected. 
As the matter power spectrum suppression is pushed to larger scales,
i.e. lower $k_{\rm cut}$, higher mass halos are similarly restricted. 
Decreasing the spectral index of the nCDM transfer function (i.e., increasing the sharpness of the suppression), intuitively decreases the overall number of halos formed, with a larger effect at lower mass. 

Figure \ref{fig: ncdm P0 and PT}, shows the CO power spectrum monopole and temperature PDF at $z=2.9$ for a COMAP-like experiment. The main effect of the nCDM model is an overall suppression in the CO power spectrum monopole relative to $\Lambda$CDM, due to the reduction in the number density of halos and the subsequent decrease in the mean temperature, which controls the amplitude of the clustering term. The suppression increases as $k_{\rm cut}$ decreases and $n$ increases (each producing a stronger cut off in the matter power spectrum). Secondly, we see that the monopole is more suppressed at scales larger than those dominated by shot noise. Non-cold dark matter affects the shot noise through the second moment of the luminosity function (see Eq.~\eqref{eq: Pshot}), the integral of which is dominated by the bright emitters and high-mass halos and therefore is not as sensitive to the cut off in the power spectrum. This is the reason for the apparent scale dependence in the suppression of the monopole.

The influence of the suppression on the VID signal is more complex to understand. The right panel of Fig.~\ref{fig: ncdm P0 and PT} shows the PDF of temperature for different choices of $k_{\rm cut}$ and $n$, with the $\Lambda$CDM model shown in gray. The main effect we see is an overall reduction in brightness temperature. As you decrease $k_{\rm cut}$ and increase $n$, producing a larger-scale or sharper cutoff in the matter power spectrum, the mean temperature is reduced. This follows from the overall lower density of emitters we see in Fig.~\ref{fig: ncdm HMF}. Since in our analysis we consider the PDF of the mean-subtracted temperature, this reduction in the mean temperature effectively shifts the zero-point of the PDF. 

\subsection{Axionic Dark Matter}
\label{sec: axion theory}
 \begin{figure*}[t]
     \centering
     \includegraphics[width=\textwidth]{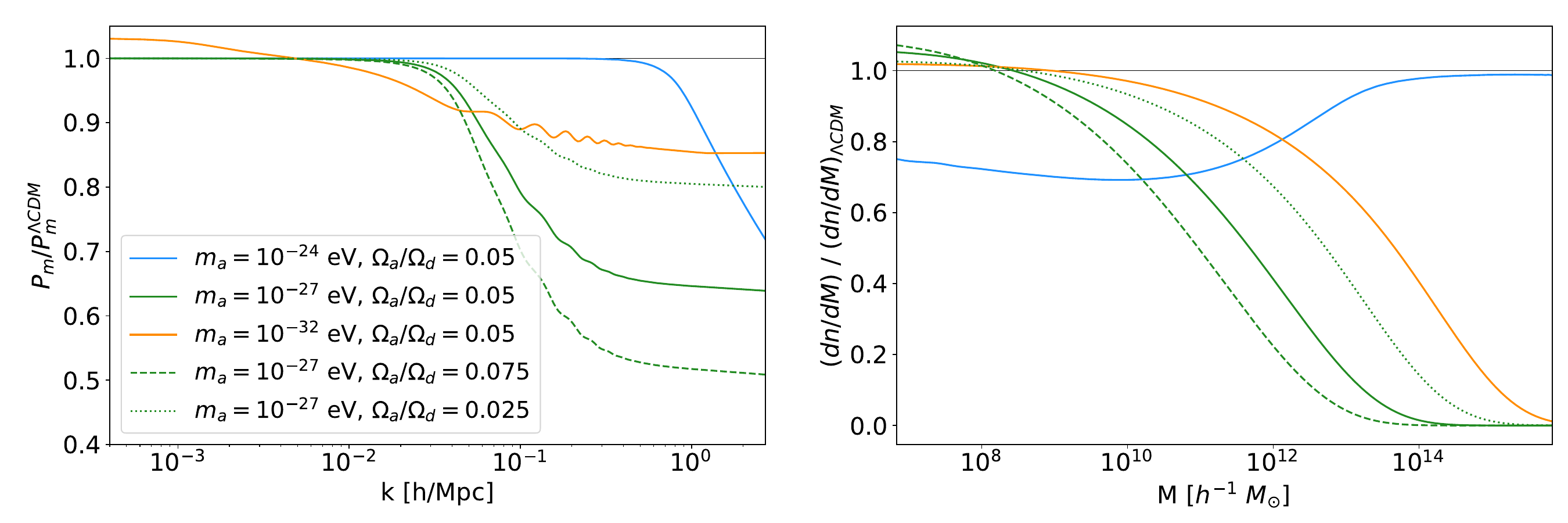}
     \caption{The ratio with respect to a $\Lambda$CDM model of the linear matter power spectrum at $z=2.9$ (left) and the halo mass function (right) in an axion DM cosmology assuming different choices of the axion mass $m_a$ and density fraction $\Omega_a/\Omega_d$ with a fixed total dark matter density fraction $\Omega_d$.}
     \label{fig: axion Pmk HMF}
 \end{figure*}
 \begin{figure*}[t]
     \centering
     \includegraphics[width=\textwidth]{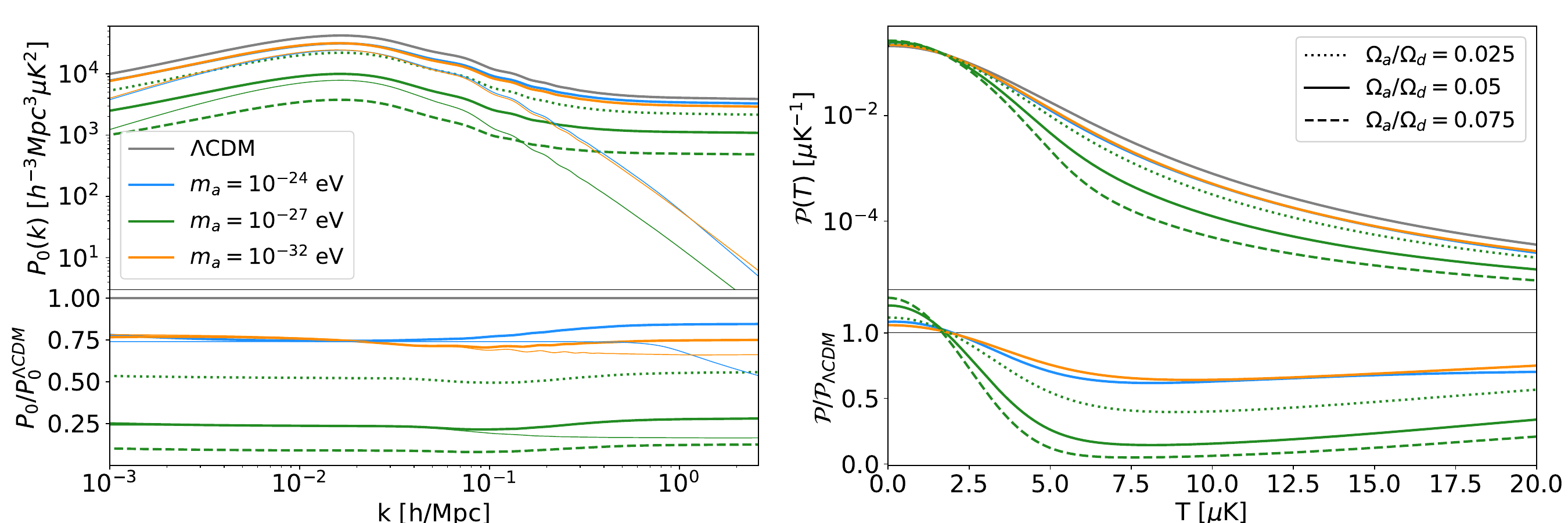}
     \caption{Power spectrum monopole (left), and temperature PDF (right) for the CO line observed at $z=2.9$ with a COMAP-Y5 type experiment assuming an axion DM cosmology. Different colors represent different choices of the axion mass $m_a$, and different line styles show different choices of the axion density fraction $\Omega_a/\Omega_d$. On the left plot, the thick solid lines show the total power spectrum monopole, whereas the thin solid lines give just the two-halo clustering term given by Eq.~(\ref{eqn: pred Pk}). Instrumental noise is not included in either observable, and the VID is not mean-subtracted. The $\Lambda$CDM model is shown in gray for comparison.}
     \label{fig: axion observables}
 \end{figure*}


Here we briefly review the relevant physics of axions included in \texttt{axionCAMB}\footnote{Publicly available at: \href{https://github.com/dgrin1/axionCAMB}{https://github.com/dgrin1/axionCAMB}}, which we use to generate the matter power spectra used for our calculations of the LIM observables. For a full description of the theory behind ultralight axions see Ref.~\cite{Hlozek:2014lca}. Ultralight axions  are described by a pseudo-scalar field $\phi$, which obeys the Klein-Gordon equation given in natural units by 
\begin{equation}
\phi''+2\mathcal{H}\phi'+m_a^2 a^2\phi=0,
\end{equation}
where $m_a$ is the axion mass in units of energy, $a$ is the cosmological scale factor, $\mathcal{H}=a'/a=aH$ is the conformal Hubble parameter, and primes denote derivatives with respect to conformal time.
At early times when $m_a\ll H$, the axion field is overdamped and evolves like a cosmological constant with equation-of-state parameter $w=-1$. As the universe cools, the axion field begins to oscillate about the minimum of its potential.  
For $a>a_{\rm osc}$ ---where $a_{\rm osc}$ fulfills $m_a\approx 3H(a_\text{osc})$--- the number of axions is roughly conserved and the axion energy density redshifts like matter, with $\rho_a\sim a^{-3}$. The relic axion density is thus $\Omega_a=\rho_a(a_\text{osc})a_\text{osc}^3/\rho_\text{crit}$ where $\rho_a(a_{\rm osc})$ is the background energy density of axions at $a_{\rm osc}$ and $\rho_\text{crit}$ is the present day critical density. We parameterize the axion abundance in relation to the total dark matter density with $\Omega_a/\Omega_d$ and $\Omega_d = \Omega_c+\Omega_a$.

When the axion field is in its oscillatory phase, the axion has a non-negligible sound speed arising from the large de Broglie wavelength of the axion:
\begin{equation}
c_s^2 = \frac{\frac{k^2}{4m_a^2 a^2}}{1+\frac{k^2}{4m_a^2 a^2}}. 
\end{equation}
From this equation we see that at large scales $c_s^2\rightarrow 0$, and the axions behave like pressureless CDM. However, at small scales there is an induced pressure leading to a suppression of clustering with respect to CDM, as shown in Fig.~\ref{fig: axion Pmk HMF}. The threshold scale at which the suppression is effective is referred to as the axion Jeans scale. 

Lighter axions ($m_a\lesssim 10^{-27}$ eV) thaw from the Hubble friction and begin oscillating during the matter or $\Lambda$-dominated eras at late times, and thus behave like dark energy at matter-radiation equality. We call these low mass axions ``DE-like''. Alternatively, heavier axions ($m_a \gtrsim 10^{-27}$ eV) begin their evolution during the radiation-domination epoch, behaving like DM much earlier on, and suppress clustering below their Jeans scale. We call these axions ``DM-like''. The effects of these axions are frozen into the matter-power spectrum at matter-radiation equality, leading to significantly different signatures in the LIM observables depending on the axion mass considered. 

For low-mass axions with $m_a\sim 10^{-32}$ eV, the matter power spectrum is enhanced at very large scales, due to their dark energy like behavior (orange curve). Alternatively, high-mass axions (blue curve) behave similarly to our nCDM model with high $k_{\rm cut}$, and only suppress power on very small scales. We show the resulting changes to the halo-mass function in the right plot of Fig.~\ref{fig: axion Pmk HMF}. Heavier axions with $m_a\sim 10^{-24}$ eV, suppress the formation of halos below their Jeans mass, thus only change the low-mass end of the HMF, very similarly to the phenomenological non-CDM model discussed in the previous Section. Conversely, the lighter axions have a similar effect as neutrinos, providing an added radiation pressure that shifts the halo-mass function towards lighter halo masses, introducing an enhancement at low-mass, and suppression at high-mass compared to $\Lambda$CDM. 
As can be expected, higher values of $\Omega_a/\Omega_d$ cause more severe suppression in the matter power spectrum and bigger impact in the halo mass function.

The total effect of these changes to the LIM observables can be seen in Fig.~\ref{fig: axion observables} where we plot the CO monopole and temperature PDF without the mean subtracted for the same variations in axion mass and density fraction.
As was the case with the phenomenological nCDM model, the most noticeable change in the monopole due to axion DM is an overall suppression of power, due to the net reduction in the number density of halos. However, for this model each axion mass exhibits distinct changes to both the clustering and shot noise terms. To understand these underlying changes we also show the two-halo clustering term given by Eq.~(\ref{eqn: pred Pk}) in the thin solid lines for each choice of $m_a$, see discussion in Appendix \ref{sec: noise/survey specs}. 
With qualitatively similar changes to the low mass end of the HMF, the heaviest axion with $m_a=10^{-24}$ eV has the same overall impact on the LIM observables as the non-CDM model. The shot noise given by Eq.~(\ref{eq: Pshot}) starts to dominate over the clustering term on larger scales than when the suppression of power due to the axions comes into effect, resulting in the difference in the ratio of the full monopole with respect to $\Lambda$CDM between large and small scales.
Interestingly, despite changes to the HMF at opposite ends of the mass spectrum, the lightest axion with $m_a=10^{-32}$ eV shows a net effect on the power spectrum monopole and temperature PDF that is very similar to the ``DM-like'' axion, however the cause for the changes are substantially different. In this case, the shot noise is even smaller than in $\Lambda$CDM due to the fewer number of high mass halos, which results in a lower value of the second moment of the luminosity function. Hence, the suppression in clustering which results from the damping of matter power is more prominently exhibited in the power spectrum monopole, than with the heavier axion. 
The intermediate mass axion intuitively follows the same qualitative behavior as the light axion, though the net effect on the monopole and temperature PDF is larger, following the changes to the HMF and matter power spectrum.

\subsection{Non-Gaussianity}
\label{sec: fnl theory}
 \begin{figure}[t]
     \centering
     \includegraphics[width=\linewidth]{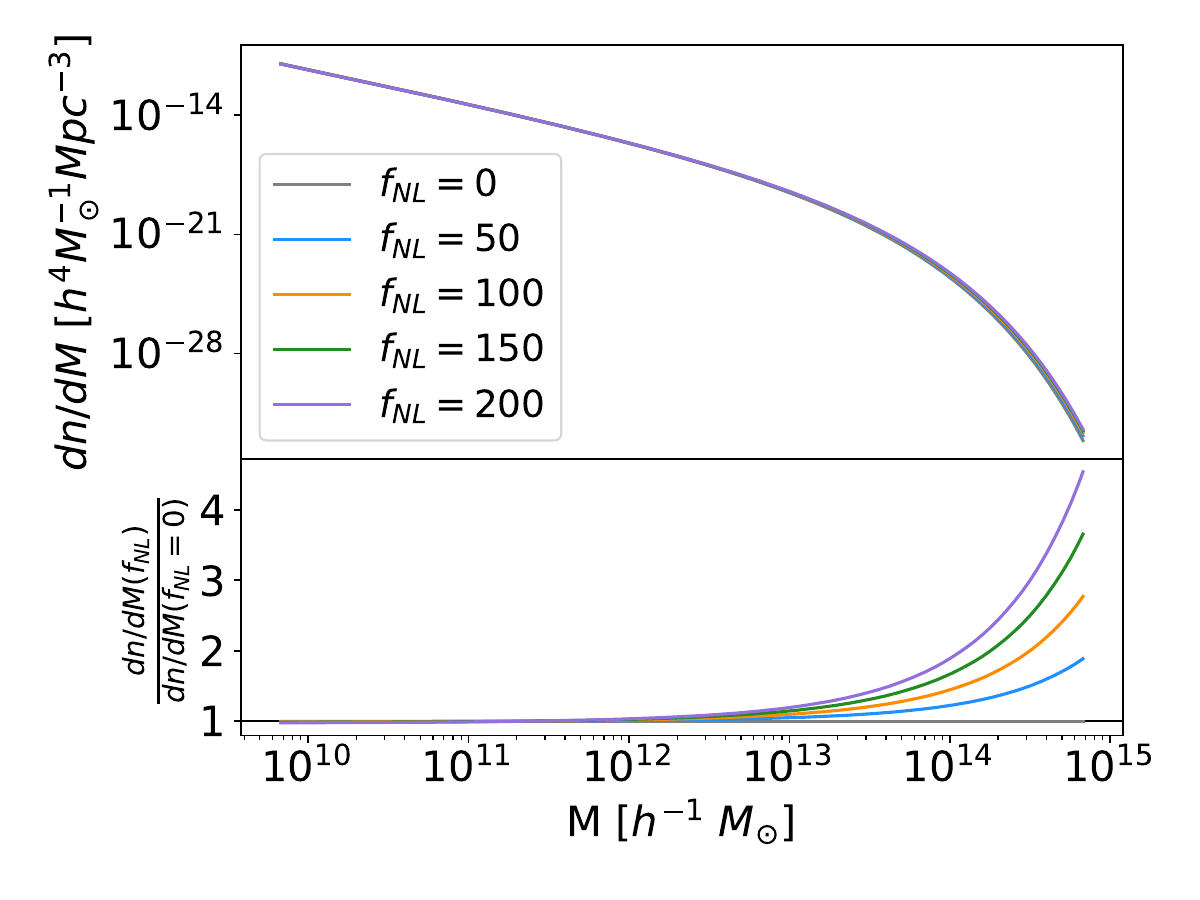}
     \caption{Halo mass function (top) for different choices of positive $f_{\rm NL}$ and the ratio with respect to the prediction for Gaussian initial conditions (bottom).  
     }
     \label{fig: fnl hmf}
 \end{figure}
 \begin{figure*}[t]
     \centering
     \includegraphics[width=\textwidth]{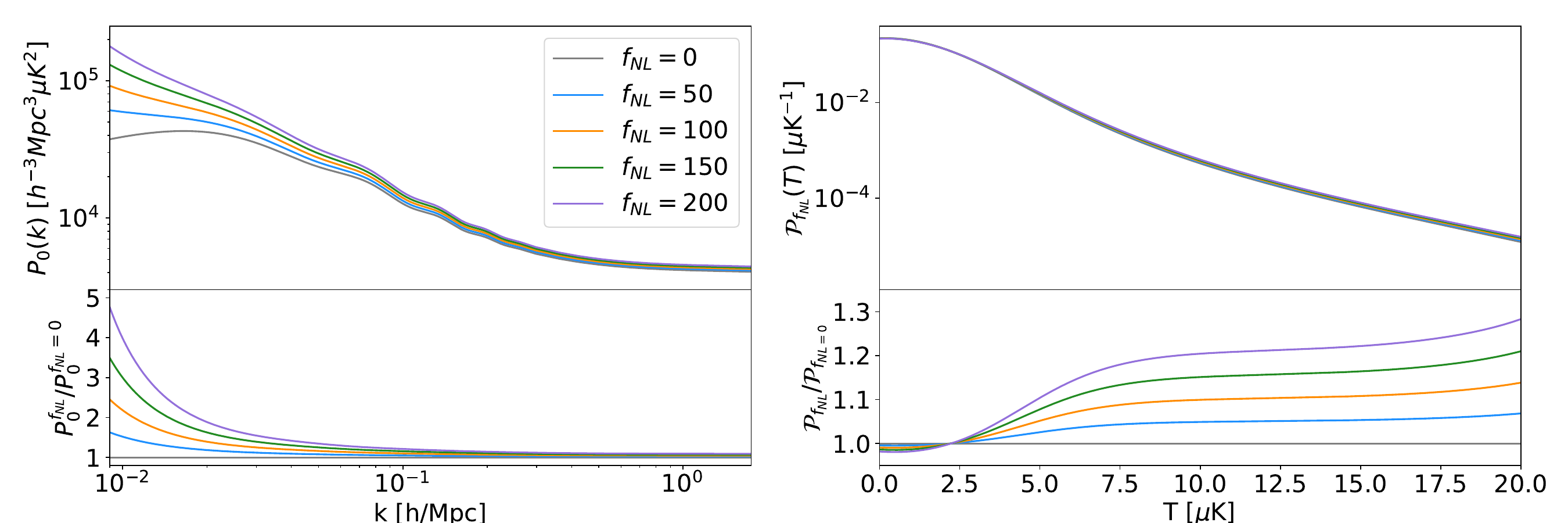}
     \caption{ The power spectrum monopole (left), and temperature PDF (right) for the CO line observed at $z=2.9$ with a COMAP-Y5 type experiment assuming different values of positive $f_{\rm NL}$. Instrumental noise is not included in either observable, and the VID is not mean-subtracted. }
     \label{fig: fnl P0 PT}
 \end{figure*}

In the local limit, primordial non-Gaussianity can be modeled in terms of the gauge-invariant Bardeen gravitational potential,
\begin{equation}
    \Phi = \phi + f_{\rm NL}(\phi^2 - \braket{\phi^2})\ ,
\end{equation}
where $\phi$ is a Gaussian random field, and $f_{\rm NL}$ quantifies the amount of non-Gaussianity. Note that in our notation $\Phi=-\Psi$ where $\Psi$ is the usual Newtonian potential. With this convention, positive $f_{\rm NL}$ corresponds to a positive skewness of the density probability distribution $S_3(M)\equiv \braket{\delta_M^3}$, and hence an increased number of overdense regions. As such, any deviations from Gaussianity will modify both the halo bias and the halo mass function. 

To model the effect of non-Gaussianity on the halo mass function, we follow Ref.~\cite{Sabti:2020ser} and use the Press-Schecter formalism \cite{LoVerde:2007ri}, which we outline here. For full details of this derivation refer to Ref.~\cite{Sabti:2020ser}.\footnote{We note that Ref.~\cite{Sabti:2020ser} introduces a cut-off scale in the primordial bispectrum, below which the bispectrum vanishes, in order to limit their analysis to the small-scale primordial non-Gaussianity. We do not introduce such a cut-off in our analysis.} In this framework, the primordial non-Gaussianity introduces a non-Gaussian correction into the standard halo mass function:
\begin{equation}
\label{eq: NG dndM}
    \left(\frac{dn}{dM}\right)_{NG} =  \left(\frac{dn}{dM}\right)_G(1+\Delta_{HMF})\,. 
\end{equation}
This non-Gaussian correction can be written as 
\begin{equation}
\label{eq: HMF NG correction}
    \Delta_{HMF} = \frac{\kappa_3 H_3(\nu_c)}{6} - \frac{H_2(\nu_c)}{6}\frac{\kappa_3'}{\nu_c'},
\end{equation}
where primes denote derivatives with respect to halo mass, $\nu_c=\sqrt{0.707}\delta_{ec}/\sigma_M=1.42/\sigma_M$ is a rescaling of the critical matter density for ellipsoidal collapse $\delta_{ec}=1.686$, and $H_n(\nu)$ are the Hermite polynomials given by
\begin{equation}
H_n(\nu) = (-1)^n \exp(\nu^2/2)\frac{d^n}{d\nu^n}\exp(-\nu^2/2), 
\end{equation}
with $\nu\equiv\delta_M/\sigma_M$. We define a normalized skewness $\kappa_3\equiv \braket{\delta_M^3}/\sigma_M^3$, where
\begin{equation}
    \sigma_M^2 = \int \frac{d^3k}{(2\pi^3)}W_M(k) \mathcal{T}_m^2(k,z) P_\Phi(k),
\end{equation}
is the variance of the linear density field with $P_\Phi(k)$ and $\mathcal{T}_m(k,z)$ being the primordial power spectrum of $\Phi(k)$ and the linear matter transfer function, respectively. The density perturbation $\delta_M$ smoothed over a mass scale $M$ is written as
\begin{equation}
    \delta_M(z) = \int \frac{d^3k}{(2\pi)^3}W_M(k) \mathcal{T}_m(k,z) \Phi(k),
\end{equation}
where 
\begin{equation}
    W_M(k) = \frac{3\sin(kR)}{(kR)^3} - \frac{3\cos(kR)}{(kR)^2}
\end{equation}
is a top-hat window function with comoving radius $R(M)=(3M/(4\pi\rho_m))^{1/3}$.  In the local non-Gaussian limit, the skewness of the density probability distribution becomes
\begin{multline}
    \braket{\delta_M^3} = 6f_{\rm NL}\int\frac{d^3k_1}{(2\pi)^3}\int\frac{d^3k_2}{(2\pi)^3}W_M(\vert k_1\vert)W_M(\vert k_2\vert)\\ 
    \times W_M(\vert k_1+k_2\vert)\mathcal{T}_m(\vert k_1\vert)\mathcal{T}_m(\vert k_2\vert)\mathcal{T}_m(\vert k_1+k_2\vert)\\
    \times P_\Phi(\vert k_1\vert)P_\Phi(\vert k_2\vert), 
\end{multline}
which can be calculated numerically. 
We have only kept terms to first order in $f_{\rm NL}$, making the correction to the halo mass function given in Eq.~(\ref{eq: HMF NG correction}) also linear in $f_{\rm NL}$. While higher-order terms exist, we neglect them following the arguments of Ref.~\cite{Sabti:2020ser}. 

The effects of this generalized non-Gaussian correction on the halo mass function can be seen in Fig.~\ref{fig: fnl hmf}. As $f_{\rm NL}$ increases, i.e. more positive skewness, the number of high-mass halos grows. Conversely, a negative value of $f_{\rm NL}$ would have the opposite effect, suppressing the formation of high-mass halos.

The total halo bias appearing in Eq.~(\ref{eqn: pred Pk}) can be written as $b_h = b_h^G +\Delta b_h$ where the effects from primordial non-Gaussianity are introduced as a correction on the Gaussian halo bias $b_h^G$ (see e.g., Refs.~\cite{{Matarrese:2000iz,Dalal:2007cu,Matarrese:2008nc,Desjacques:2010jw}}). In the local limit, the skewness introduced in the density probability distribution introduces a scale-dependent correction of the form,
\begin{equation}
\label{eq: halo bias correction}
    \Delta b_h = (b_h^G-1)f_{\rm NL} \delta_{ec} \frac{3\Omega_m H_0^2}{c^2 k^2 T(k,z)}\,.
\end{equation}
Here we assume universality of the halo mass function to derive the relation between the bias $b_\phi$ with respect to the primordial potential perturbations and the linear halo bias $b_h$. We acknowledge that universality does not hold, and that this relationship cannot be predicted a priori. This introduces a complete degeneracy between $f_{\rm NL}$ and $b_\phi$, forcing constraints to be framed in terms of $f_{\rm NL}b_\phi$~\cite{Barreira:2022sey}, also for the LIM power spectrum~\cite{Barreira:2021dpt}. 
Given the proof-of-concept nature of this work, we prefer to assume universality to avoid confusion and to ease the interpretability of our results. Nonetheless, we highlight that constraints on $f_{\rm NL}$ from the VID are not sensitive to $b_\phi$, providing a potential avenue to break the degeneracy affecting the power spectrum constraints.

We show the effect of $f_{\rm NL}\neq 0$ on the LIM measurements in Fig.~\ref{fig: fnl P0 PT}. Introducing the corrections to the halo mass function given in Eq.~(\ref{eq: HMF NG correction}), and the bias given in Eq.~(\ref{eq: halo bias correction}), we find that the non-Gaussian initial conditions lead to large scale enhancement in the CO power spectrum due to the $k^{-2}$ dependence of the correction to the halo bias. There is also a small overall increase of power due to the boost in the first and second moments of the luminosity functions (affecting the clustering and shot-noise contributions, respectively) due to the increased abundance of high-mass emitters; since in any case the abundance of these emitters is low, this effect is less significant than the effect caused by the correction in the halo bias. Similarly, we find high-intensity enhancement in the VID, due to the larger abundance of bright emitters. In turn, the amplitude of the VID decreases at low intensities to compensate. 

\section{LIM Forecasting}
\label{sec: forecasting}
In this Section, we present our forecasts on the models presented in the previous Section, and the improvement gained from a joint analysis of the LIM observables. 
To compute the LIM observables and all related quantities, we modify 
\texttt{lim}\footnote{Available 
at: \href{https://github.com/jl-bernal/lim}{https://github.com/jl-bernal/lim}} to account for the changes to the matter-power spectrum, halo bias, and halo mass function for each cosmology we consider. We assume efficient foreground subtraction and line-interloper cleaning, achieved following strategies developed in e.g., Refs.~\cite{Breysse:2015baa, 2018ApJ...856..107S, Cheng:2020asz, moriwaki2020deep, Cunnington:2023jpq, VanCuyck:2023uli, Bernal_deinterloping}, and neglect their effects on the forecast.

\subsection{Experimental setup and astrophysical model}
\label{sec: experiment}
We choose to center our analysis on the ground-based COMAP instrument \cite{Cleary:2021dsp}, which targets the CO(1-0) spectral line observed at a frequency of $\nu_{\rm obs}=29.6$ GHz ($z=2.9$). We follow the expected sensitivities after five years of observations \cite{COMAP:2021pxy,COMAP:2021lae}, which we delineate as `COMAP-Y5' throughout the text. We assume an effective system temperature of $T_{\rm sys}=45/\sqrt{69.4}$ K, where the factor of $\sqrt{69.4}$ accounts for the increase of sensitivity between the early science sensitivities of COMAP and the finished five-year survey. The remaining experimental parameters roughly correspond to the early-science specifications:
a single survey which covers 4 deg$^2$ of sky with 38 effective detectors (19 feeds with double polarization) observing for a cumulative time of $t_{\rm obs}=1000$ hours. We consider an angular and spectral resolutions of $\theta_{\rm FWHM}=4.5$ arcmin and $\delta\nu=31.25$ MHz, respectively, and a bandwidth of $\Delta\nu=7.7$ GHz corresponding to the redshift range $z \in 2.4-3.4$. 
To forecast the potential of a survey targeting a larger volume ---and compare a deep-narrow and a wide-shallow survey--- we also consider a hypothetical future CO survey which covers 200 deg$^2$ with only twice the total observing time, which we refer to as `COMAP-XL'. The corresponding instrumental noise per voxel of COMAP-Y5 and COMAP-XL is 1.45 $\mu$K and 5.09 $\mu$K, respectively. 


We model the relation between the total CO luminosity and the halo mass using the fiducial COMAP model \cite{COMAP:2021lae},
\begin{equation}
\frac{L_{\rm CO}}{L_\odot}(M) = 4.9\times 10^{-5} \frac{C}{(M/M_\ast)^A+(M/M_\ast)^B},
\end{equation}
with fiducial parameters $A=-2.85$, $B=-0.42$, $C=10^{10.63}$, and $M_\ast=10^{12.3}M_\odot$, obtained from a fit to results from Universe Machine \cite{Behroozi:2019kql}, COLDz \cite{Riechers:2018zjg}, and COPSS \cite{Keating:2016pka}. We include a mean-preserving logarithmic scatter $\sigma_L=0.42$, and assume a halo mass function and halo bias from the fits of Ref.~\cite{Tinker:2008ff, Tinker:2010my}.

\subsection{Fisher Formalism}
\label{sec: fisher formalism}
We forecast constraints on the models discussed above using a Fisher matrix formalism, which assumes a Gaussian distribution for the parameter likelihoods centered on some chosen fiducial values. To obtain the joint forecast from the power spectrum and VID, we construct a vector of our observables $\Theta=[\tilde P_0(k), \mathcal{B}_i]$. Then, the total Fisher matrix element corresponding to parameters $p_\alpha$ and $p_\beta$ is:
\begin{equation}
F_{\alpha\beta} = \frac{\partial \Theta^T}{\partial p_\alpha} \xi^{-1} \frac{\partial \Theta}{\partial p_\beta},
\end{equation}
where $\xi$ is a block matrix where the diagonal blocks correspond to the covariance of the power spectrum monopole and the VID, and off-diagonal blocks, to their covariance:
\begin{equation}
\xi = 
\begin{pmatrix}
\sigma_{P_0}^2 & \sigma_{P_0,B_i}\\
\sigma_{B_i,P_0} & \sigma_{B_i}^2
\end{pmatrix}
\end{equation}
where all elements have been defined in Sec.~\ref{sec: Pk+VID covariance theory}. 

In our Fisher analyses we vary the parameters of our line emission model, $p_\alpha=\{ A,B,\log C,\log(M_\ast/M_\odot)\}$, along with the different beyond-$\Lambda$CDM parameters for each case. We marginalize over the astrophysical parameters to produce our final constraints on the cosmological parameters of interest.

We compute the Fisher matrix for the power spectrum monopole, VID, and their combination for both the COMAP-Y5 and COMAP-XL experimental setups. For all forecasts we assume a $k$-range of $k_\text{min}=2\pi/L_\parallel$, representing the minimum $k$-accessible in the observed field, and $k_\text{max}=1.0$ Mpc$^{-1}$, and a temperature range 
$T\in [0\,\mu {\rm K},\, T_{\rm max}]$ with $T_\text{max}=20-50$ $\mu$K for COMAP-Y5, and $T_\text{max}=200$ $\mu$K for COMAP-XL. Note that resolution limits suppress the power spectrum at scales larger than $k_{\rm max}$, so that the specific maximum wave number used does not affect our results. We choose $T_{\rm max}$ such that the calculation of the VID remains stable and normalized with respect to the parameter changes needed for the numerical derivatives, and the signal-to-noise in each temperature bin remains above unity, meaning the exact value changes for each cosmological model we consider.\footnote{For $\Lambda$CDM, nCDM, and non-Gaussianity this informs $T_{\rm max}=20$ $\mu$K, for axion DM $T_{\rm max}=50$ $\mu$K.}

\subsection{Results}
\begin{table*}[t]
\begin{center}
\setstretch{1.2}
\begin{tabular}{| c | c | c | c | c | c | c | c |}
\hline \hline
  & & \multicolumn{3}{c|}{COMAP-Y5} & \multicolumn{3}{c|}{COMAP-XL}\\ 
  \hline
 Parameter & Fiducial & $P_0$ & $B_i$ & $P_0+B_i$ & $P_0$ & $B_i$ & $P_0+B_i$\\ 
 \hline \hline
 $k_{\rm cut}$ [Mpc$^{-1}$] & 0.5 & $\pm$8.74 & $\pm$41.93 & $\pm$2.42 & $\pm$11.56 & $\pm$1.48 & $\pm$1.24 \\ 
 $n$ & 0.1 & $\pm$15.95 & $\pm$12.16 & $\pm$0.24 & $\pm$25.75 & $\pm$0.15 & $\pm$0.11 \\ 
 \hline
 $\Omega_a/\Omega_d$ ($m_a=10^{-32}$ eV) & 0.04 & $\pm$0.76 & $\pm$0.52 & $\pm$0.04 & $\pm$0.17 & $\pm$0.31 & $\pm$0.02 \\
 $\Omega_a/\Omega_d$ ($m_a=10^{-27}$ eV) & 0.04 & $\pm$0.19 & $\pm$0.18 & $\pm$0.02 & $\pm$0.09 & $\pm$0.07 & $\pm$0.01 \\ 
 $\Omega_a/\Omega_d$ ($m_a=10^{-24}$ eV) & 0.04 & $\pm$78.2 & $\pm$0.14 & $\pm$0.06 & $\pm$20.4 & $\pm$0.06 & $\pm$0.02\\ 
 \hline
 $f_{\rm NL}$ & 0 & $\pm$3140 & $\pm$71 & $\pm$3.2 & $\pm$220 & $\pm$14.2 & $\pm$0.38 \\
 \hline \hline 
\end{tabular}
\end{center}
\caption{\label{tab: BSM constraints} Forecasted 68\% (1-$\sigma$) parameter constraints for all three BSM cosmologies considered using COMAP-Y5 and a hypothetical survey with a larger volume COMAP-XL. We compare constraints derived from just the power spectrum ($P_0$) and VID ($B_i$) alone, to the constraint from a combined analysis assuming the analytical covariance derived in Sec.~\ref{sec: Pk+VID covariance theory} ($P_0+B_i$).}
\end{table*}
 \begin{figure}[t]
     \centering
     \includegraphics[width=\linewidth]{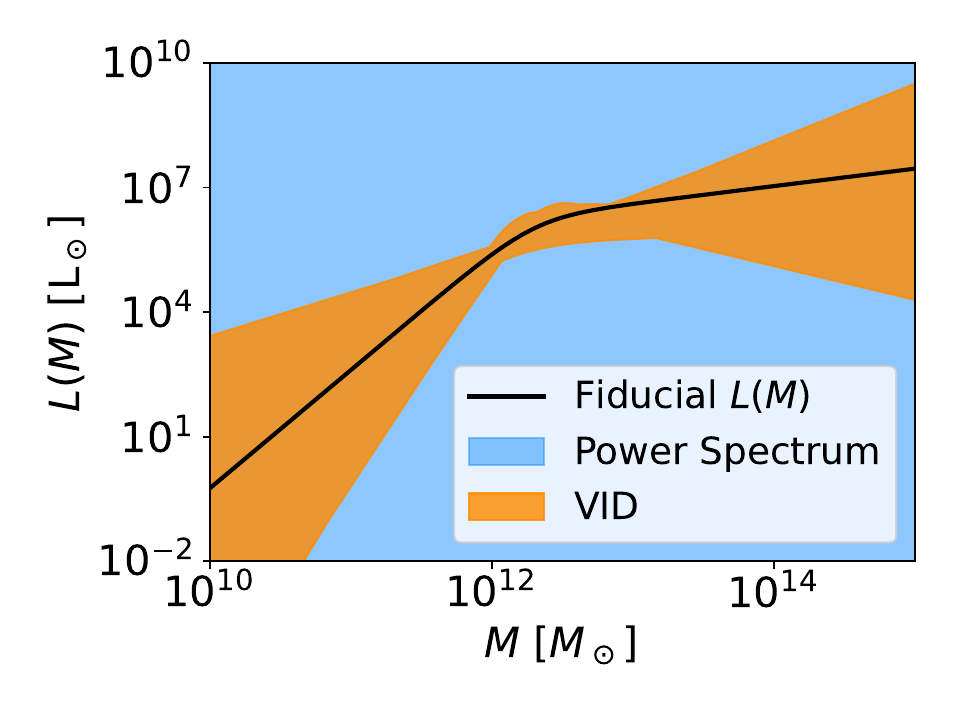}
     \caption{ 68\% confidence regions around our fiducial $L(M)$ obtained from the CO power spectrum monopole (blue) and the VID (orange) assuming a COMAP-Y5 experimental setup. The power spectrum alone offers virtually no constraining power on the mass and luminosity range considered in our analysis, hence the blue background. }
     \label{fig: L(M) constraints}
 \end{figure}
 \begin{figure}[t]
     \centering
     \includegraphics[width=\linewidth]{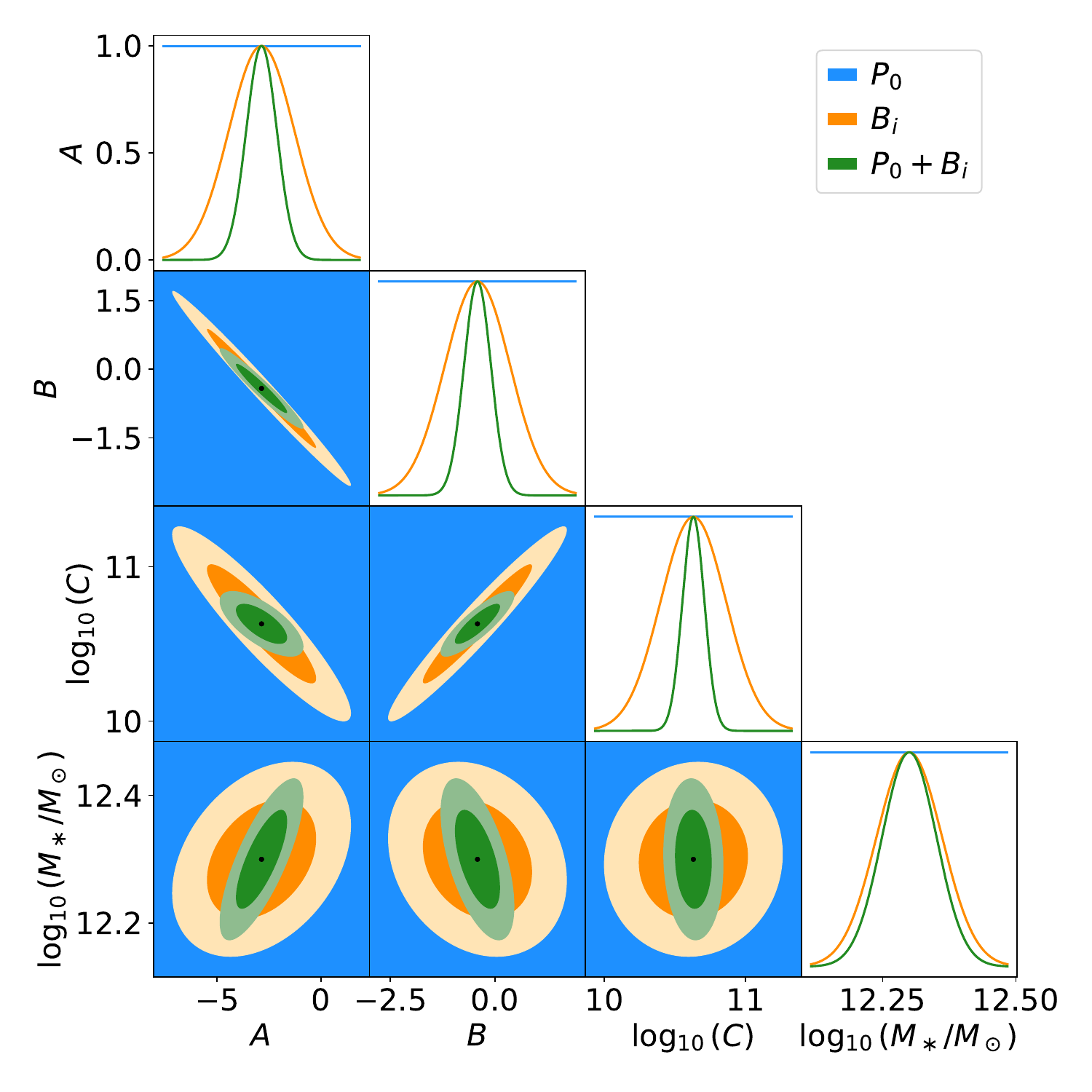}
     \caption{ 68\% and 95\% forecasted marginalized constraints on the $L(M)$ parameters assuming a COMAP-Y5 setup. We compare the results from the power spectrum monopole (blue), the VID (orange), and their combination including the analytical covariance (green).}
     \label{fig: L(M) contours}
 \end{figure}
 \begin{figure*}[t]
     \centering
      \includegraphics[width=\textwidth]{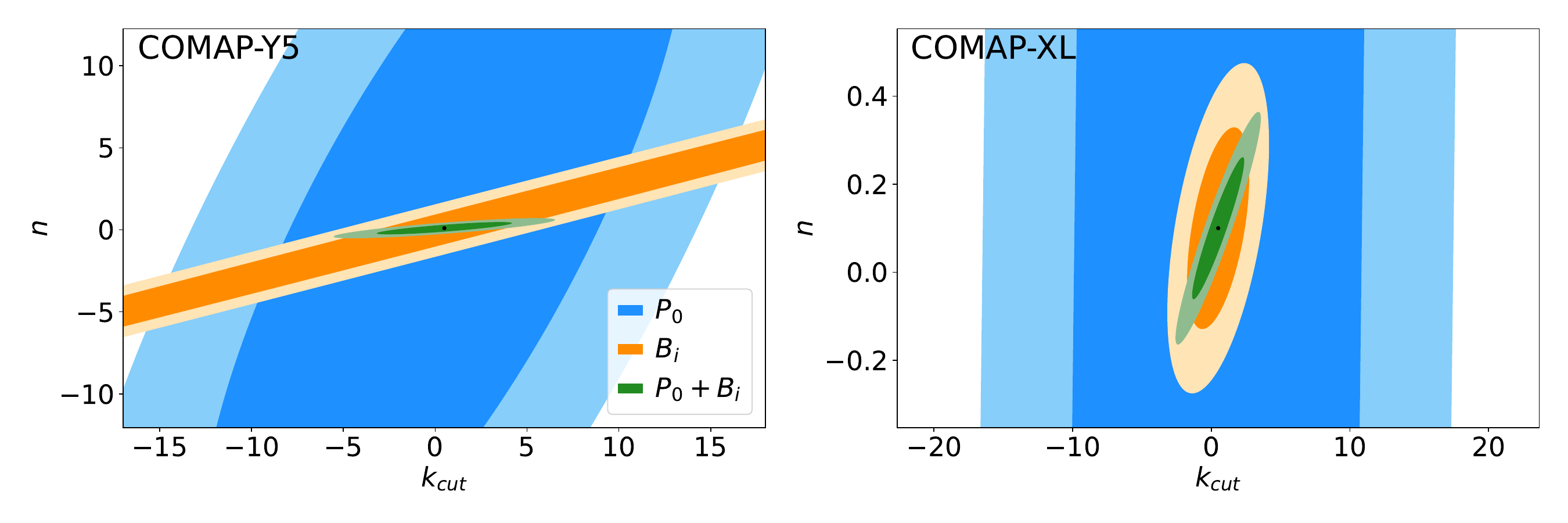}
     \caption{ 68\% and 95\% forecasted marginalized constraints on the nCDM parameters assuming fiducial values of $k_{cut}=0.5$ Mpc$^{-1}$ and $n=0.1$. We compare the results from the power spectrum monopole (blue), the VID (orange), and their combination including the analytical covariance (green). Constraints derived assuming a COMAP-Y5 setup are shown on the left, whereas constraints our hypothetical larger COMAP-XL setup are shown on the right.}
     \label{fig: ncdm contours}
 \end{figure*}
 \begin{figure}[t]
     \centering
     \includegraphics[width=\linewidth]{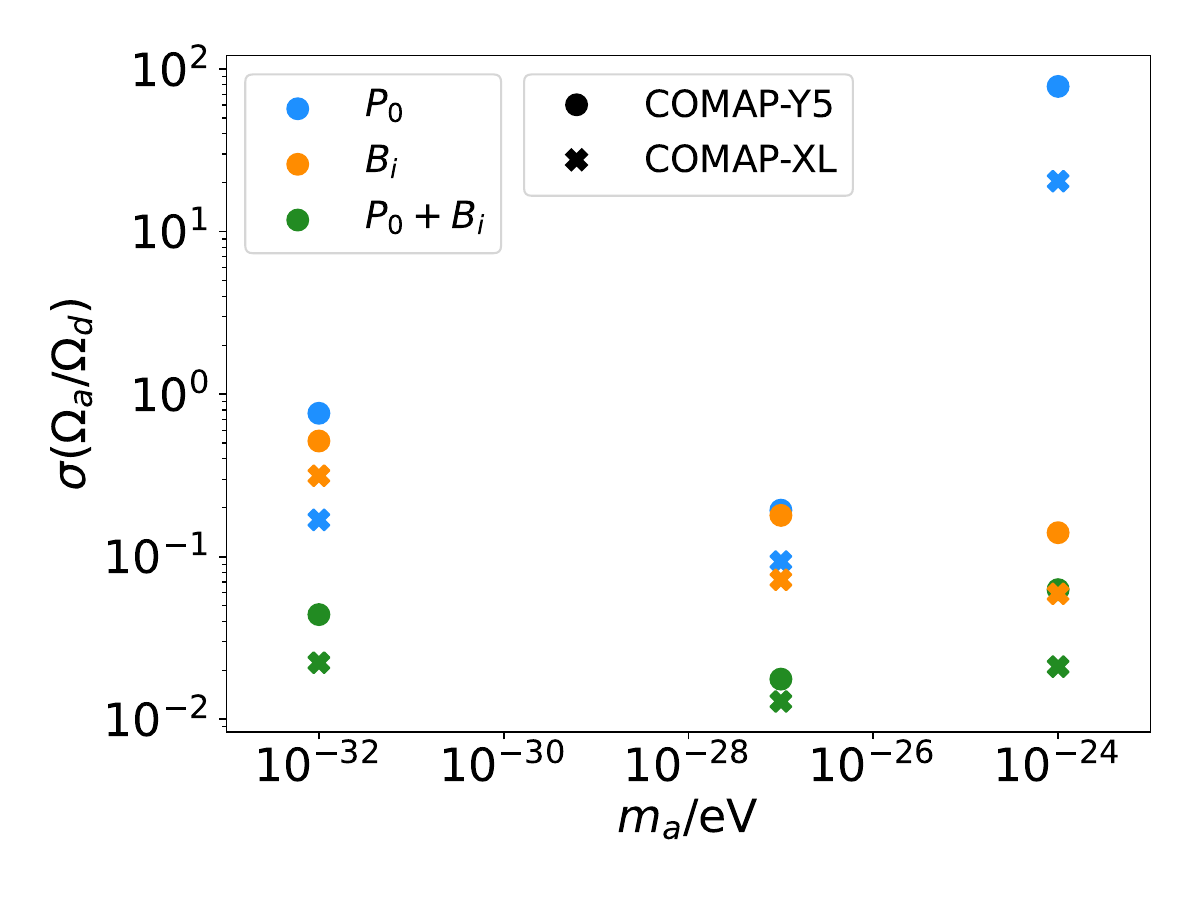}
     \caption{ 68\% marginalized errors on the axion density fraction assuming a fiducial value of $\Omega_a/\Omega_d=0.04$ for the three different axion masses considered. We compare the results of the power spectrum monopole (blue), the VID (orange), and their combination including the analytical covariance (green) for the COMAP-Y5 and COMAP-XL experimental configurations. }
     \label{fig: axion constraints}
 \end{figure}

A summary results of our Fisher forecasts on all 
cosmological parameters for both the COMAP-Y5 and COMAP-XL configurations are found in Table~\ref{tab: BSM constraints} where we give the 68\% confidence level forecast marginalized uncertainties using just the power spectrum, just the VID, and a combination of the two, labelled $P_0$, $B_i$, and $P_0+B_i$, respectively. As we can see, a joint analysis leads to significantly tighter constraints in all cases. The major factor in this is the ability of the VID to constrain the astrophysical parameters, allowing for access to cosmological information that would otherwise be inaccessible due to strong degeneracies between the astrophysics and cosmology being tested. The degeneracies can be seen in the figures of Sec.~\ref{sec: cosmologies}, where the main effects on the power spectrum at the scales probed are similar to those from changes in the first and second moment of the luminosity functions.  

In Fig.~\ref{fig: L(M) constraints}, we show the 68\% confidence regions around our fiducial $L(M)$ derived in $\Lambda$CDM assuming a COMAP-Y5 experiment, where we see quite how much better the VID statistic is at measuring the $L(M)$ relation. On the mass and luminosity ranges probed, the power spectrum provides virtually no constraining power relative to the VID. This can also be seen looking at the marginalized forecast constraints on the parameters controling the $L(M)$ relation, shown in Fig.~\ref{fig: L(M) contours}: the VID alone improves the marginalized constraints on these parameters by a factor ranging from 150 to over 1000.
Adding the power spectrum to the VID slightly improves the constraining power, especially reducing the degeneracies between the parameters. 
As such when beyond-$\Lambda$CDM cosmological parameters are included, their constraints are limited by the power spectrum's ability to distinguish changes to mass-luminosity relation from changes to the cosmology. Therefore, once the astrophysical parameters are constrained by the VID, the power spectrum adds significantly more constraining power than the naive combination of the independent constraints. We discuss these results in the context of each model in more detail in the subsequent paragraphs. 

For the phenomenological non-CDM model presented in Sec.~\ref{sec: cosmologies}, we choose fiducial values of $k_\text{cut}=0.5$ Mpc$^{-1}$ and $n=0.1$, giving similar small-scale suppression to an axion-DM model with $m_a\sim 10^{-24}$ eV, and $\Omega_a/\Omega_d\sim 0.025$. In Fig.~\ref{fig: ncdm contours}, we show the marginalized forecasted constraints on the $k_{\rm cut}$-$n$ plane for the power spectrum monopole (in blue), VID statistic $B_i$ (in orange), and their combination (in green), for the COMAP-Y5 and COMAP-XL experimental configurations. Focusing first on the Y5 scenario, we see that the VID is not very sensitive to the cut-off scale, but does provide a stronger constraint on the slope $n$ than the power spectrum alone. 
In turn, the joint analysis improves the Y5 constraint on $k_{\rm cut}$ by a factor of 3.6 over the power spectrum alone, and improves the constraint on the slope $n$ by a factor of more than 50 over the VID alone. When we consider a larger survey, the constraints on the nCDM parameters from the power spectrum alone actually worsen compared to the COMAP-Y5 configuration. This is because the influence of nCDM is limited to small scales, and there is a higher noise in the COMAP-XL case, which makes the measured power spectrum less sensitive to the small-scale changes.
The constraint on the VID however gets considerably better with a larger survey due to the increase in the total number of voxels, which scales linearly with $\Omega_{\rm field}$, and the added sensitivity coming from a larger $T_{\rm max}$. Overall, this leads to the combined COMAP-XL constraints on both nCDM parameters improving by $\sim 50\%$ when compared with the Y5 constraints.

For the axion DM model, we choose to include only the axion density fraction $\Omega_a/\Omega_d$ alongside the astrophysical parameters in our Fisher analysis. We follow Refs.~\cite{Hlozek:2017zzf,Bauer:2020zsj,Lague:2021frh,Rogers:2023ezo} by forecasting constraints on the axion fraction for fixed values of $m_a\in[10^{-32},10^{-24}]$ eV, since a highly non-trivial degeneracy exists in the $m_a$-$\Omega_a/\Omega_d$ plane. We choose three exemplary cases: first, we assume a very light axion with $m_a=10^{-32}$ eV as an example of ``DE-like'' axions; second, we choose a heavier axion with $m_a=10^{-24}$ eV as an example of ``DM-like'' axions; and finally, we choose an intermediate value of $m_a=10^{-27}$ eV, to show the behavior in between these two extremes. We do not consider cases outside of this range as lighter axions are indistinguishable from a cosmological constant, and heavier axions are indistinguishable from cold dark matter on the scales we probe. We consider a fiducial value of $\Omega_a/\Omega_d=0.04$, well within current bounds, and hold $\Omega_d$ fixed across all cases. 

We show the forecasted marginalized constraints on $\Omega_a/\Omega_d$ for the three different axion masses we consider in Fig.~\ref{fig: axion constraints} for COMAP-Y5 and COMAP-XL. Starting with COMAP-Y5 forecasts, we see that the axions with $m_a=10^{-32}$ eV and $m_a=10^{-27}$ eV are similarly constrained by the power spectrum and VID, with the VID being slightly more sensitive to the axion fraction in both cases. Since axions, especially towards the light end, change the shape of the power spectrum on all scales, 
the constraints from the power spectrum alone are comparable to those from the VID.
When the two statistics are combined, we see the sensitivity increase by a factor of 13 over the VID alone for the lighter axion, and by a factor of 9 for intermediate mass axion. 
The heavier ``DM-like'' axion, is a different story. In this case, much like with nCDM, this heavier axion influences the power spectrum mainly on small scales that are dominated by shot noise, giving stronger degeneracies with astrophysical parameters, hence the VID contributes way more constraining power than the power spectrum alone. Combining both statistics results in a 2.3 times improvement in sensitivity over the VID alone for the heaviest axion. When we size up to the COMAP-XL scenario, the power spectrum constraints do not get worse as they did for the nCDM case because the DM-like axions still introduce small changes in the shape of the power spectrum at all scales.
For each axion mass, the XL combined constraints on $\Omega_a/\Omega_d$ improve by $\sim 50\%$ when compared with the Y5 combined constraints.

Finally, for our forecast on $f_{\rm NL}$ we assume a fiducial value of $f_{\rm NL}=0$. We can see from Table \ref{tab: BSM constraints} that the power spectrum alone is not very sensitive to deviations from Gaussianity, especially using COMAP-Y5. This is due to the limited COMAP-Y5 survey volume cutting off the scales accessible by the power spectrum. As we see in Fig.~\ref{fig: fnl P0 PT}, non-zero $f_{\rm NL}$ mostly affects the largest scales (moreover, all-scale effects are degenerate with changes in the moments of the luminosity function), but the minimum $k-$value accessible by the power spectrum for COMAP-Y5 is $k_{\rm min}\sim0.006$/Mpc. As such, small deviations from Gaussianity are hard to distinguish with the scales available in this forecast. When we consider a larger volume with the COMAP-XL configuration, the minimum accessible scale is pushed to $k_{\rm min}\sim0.004$/Mpc and we see the constraint from the power spectrum alone improves by over an order of magnitude. 
The VID captures the effect of primordial non-Gaussianity in the abundance of massive emitters, hence providing some more constraining power. The power of a joint analysis becomes particularly clear in the case of non-Gaussianity where the combined constraint from the power spectrum and VID represents more than an order of magnitude of improvement over the individual constraints. 

In summary, our Fisher forecasts demonstrate that a joint analysis utilizing both the power spectrum and VID significantly enhances constraints on beyond-$\Lambda$CDM cosmologies compared to individual analyses. The combination of these statistics provides tighter constraints by breaking degeneracies between astrophysical parameters, and cosmological parameters, which we show for the cases of nCDM, axion dark matter, and non-Gaussianity.

\section{Discussion and Conclusion}
\label{sec: discussion}
Line-intensity mapping proposes a novel observational technique, capable of measuring large cosmological volumes at redshifts beyond the reach of conventional galaxy surveys. The fluctuations in an observed intensity map depend on the spatial distribution of galaxies and the luminosity function of the spectral line of interest. The LIM power spectrum carries the bulk of cosmological information available in intensity maps, but their sensitivity to the astrophysics is limited and degenerate with some of the cosmological parameters. On the other hand, the VID, which depends directly on the full luminosity function, carries non-Gaussian information, and is sensitive to the clustering at very small scales through its impact in the halo mass function. As such, analyses which combine these two summary statistics should significantly improve constraints breaking degeneracies between astrophysical uncertainties and cosmological features. 

The benefits of combining the power spectrum and the VID to constrain the line-luminosity function had been shown previously~\cite{COMAP:2018kem}. The potential of the VID to constrain changes in the matter power spectrum at small scales was also highlighted~\cite{Libanore:2022ntl,Adi:2023qdf}.
In this work, we take advantage of the analytical covariance between the power spectrum and VID (derived for the first time in Ref.~\cite{Sato-Polito:2022fkd} and updated in this work) to explore the improvement in sensitivity to beyond-$\Lambda$CDM physics due to their combination. We focus on cosmological models beyond $\Lambda$CDM which alter the halo-mass function at light and heavy masses, where the combination of the VID and the power spectrum shows the biggest improvement in comparison to an analysis using only the power spectrum. 
In all cases, the VID's ability to constrain the astrophysical parameters allows the power spectrum to greatly constrain the cosmological parameters, such that their combination reaches a constraining power beyond a naive addition of the individual marginalized constraints. 
We forecast constraints assuming a COMAP mission measuring the CO line at redshifts $2.4<z<3.4$, as an example of near-term observational capabilities (COMAP-Y5), as well as a shallower but wider hypothetical iteration for comparison (COMAP-XL). In all cases, the shallower but wider version obtains better constraints on the cosmological parameters once the VID and the power spectrum are combined. However, we note that for models which only change the shape of the power spectrum at small scales, a deeper and narrower survey, which has a lower instrumental noise, is preferred for analyses using only the power spectrum. 
 
Our phenomenological non-CDM model introduces two new cosmological parameters, the scale at which clustering begins to be suppressed $k_\text{cut}$, and the slope of the suppression $n$. We find that a joint Fisher analysis using the LIM power spectrum monopole and VID can lead to an increase in precision, with respect to the better single observable, on the estimation of $k_\text{cut}$ by a factor of 3.6(1.2), and an increase by a factor of 50(1.4) for $n$ using COMAP-Y5(COMAP-XL).  The increased sensitivity is a direct result of the different $k_\text{cut}$-$n$ degeneracy directions in the power spectrum monopole and VID, making their combination break the degeneracy and yield much tighter constraints. 

We consider three cases of axion dark matter with varying masses of the axion. First, we consider a ``DE-like'' axion with $m_a=10^{-32}$ eV which thaws from the Hubble friction after matter-radiation equality, making its effect on the matter power spectrum similar to that of dark energy or massive neutrinos. In this case we find that a joint analysis increases precision on the estimation of the axion density fraction $\Omega_a/\Omega_d$ by a factor of 13(8.5) with respect to an analysis using only one observable for COMAP-Y5(XL). Secondly, we consider a heavy ``DM-like'' axion with $m_a=10^{-24}$ eV, which becomes dynamical during the radiation era. As such, its effect on the matter power spectrum is similar to CDM, and we only see changes at very small scales which are nearly out of reach of the LIM power spectrum. For this reason, the inclusion of the VID, which is more sensitive to these small scale changes, increases precision on $\Omega_a/\Omega_d$ by two orders of magnitude when compared with constraints from the power spectrum alone. The final axion we consider exists somewhere in between DE and DM with $m_a=10^{-27}$ eV. This case is overall the most constrained of the three we consider as it has the greatest effect on both the power spectrum and VID affecting all accessible scales and intensities. Here the inclusion of the VID results in 9(7) times greater sensitivity to the the axion density for COMAP-Y5(XL). 

Finally, we consider deviations from Gaussianity in the primordial perturbations by introducing first-order local type non-Gaussian corrections to the halo-mass function and halo bias. The changes to the LIM observables only appear on large scales, leading to our derived constraints on $f_{\rm NL}$ being driven by VID's access to a wider range of scales when compared to the power spectrum. We find that our joint analysis increases sensitivity to $f_{\rm NL}$ by a factor of 22(37) for a COMAP-Y5(XL) experiment. Furthermore, information from the VID, which does not depend on the halo bias, may help to break the degeneracy between $b_\phi$ and $f_{\rm NL}$ that limits the constraining power of the power spectrum on primordial non Gaussianity. We note that although we consider scale-independent non-Gaussianity, the VID would also be sensitive to scale-dependent local primordial non-Gaussianity \cite{Stahl:2024stz}.

While we focus our forecast on a single experiment and spectral line, it can be generalized for others. With accurate models of the correlation between different emission lines, multiple lines could be added together to boost sensitivity to cosmological parameters even more. Furthermore, the joint analysis could be extended to other observables such as the velocity field reconstruction using the kinetic Sunyaev Zeldovic tomography with LIM~\cite{Sato-Polito:2020cil}, among many others.

This work aims to highlight the raw potential of the VID and its complementarity to the power spectrum to constrain fundamental physics with LIM. As such, it makes some simplifying assumptions that affect the exact value of the forecast sensitivities. First, we assume that foregrounds and line interlopers are under control and do not reduce the sensitivity of the LIM observables. Employing variations of the VID which are more robust against observational systematics, as the conditional VID~\cite{Breysse:2019cdw} or the deconvolved density estimator~\cite{Breysse:2022fdi,COMAP:2022sdg} are expected to return similar qualitative improvements, although likely resulting in a decline in the overall sensitivity which respect to the VID.
Second, while we introduce new cosmological parameters, our analysis assumes the standard model parameters are given by the \textit{Planck} best-fit $\Lambda$CDM cosmology. We hope that improvements in the modeling of the contaminants and the strategies to reduce their impact, as well as combinations of observations from different experiments, different spectral lines and different observables can return robust line-intensity maps without much information loss. The purpose of this work is to show the increased precision obtained in a joint analysis, not an accurate estimation of parameter values. Hence, we leave a full, more realistic forecast to future work.

In summary, we demonstrate how the combination of the LIM power spectrum monopole and voxel intensity distribution can boost the sensitivity in LIM surveys to beyond-standard-model physics by a factor $\sim 2-50$ in our examples. This gain is due to breaking the degeneracies between the astrophysical and cosmological parameters in the power spectrum, but also due to the different degeneracies between the actual cosmological parameters in the VID and the power spectrum. 

Many LIM pathfinder experiments are already observing and many others will come in the near future. With this experimental effort, a variety of emission lines sourced at redshifts reaching back to cosmic dawn will be targeted. We hope this work provides a useful framework to maximize the sensitivity of these experiments for probing beyond-$\Lambda$CDM cosmologies. 

\begin{acknowledgments}
VS acknowledges the support from the New College Oxford University/Johns Hopkins University Balzan Centre for Cosmological Studies Program. JLB acknowledges funding from the Ramón y Cajal Grant RYC2021-033191-I, financed by MCIN/AEI/10.13039/501100011033 and by
the European Union “NextGenerationEU”/PRTR, as well as the project UC-LIME (PID2022-140670NA-I00), financed by MCIN/AEI/ 10.13039/501100011033/FEDER, UE. GSP acknowledges support from the Friends of the Institute for Advanced Study Fund.  MK was supported by NSF Grant No.~2112699, the Simons Foundation, and the John Templeton Foundation.
\end{acknowledgments}

\appendix

\section{Noise \& Survey Specifications}
\label{sec: noise/survey specs}
Due to the limited resolution and finite observed volume of LIM experiments, the observed power spectrum will differ from the one predicted by Eqns.~\eqref{eqn: pred Pk} and~\eqref{eq: Pshot}, mainly by limiting the minimum and maximum accessible scales, respectively. Following Ref.~\cite{Sato-Polito:2022fkd}, we model these experimental limitations by applying window functions to our predicted power spectrum:
\begin{equation}
\label{eq: windows}
    \tilde P(k,\mu) = \int \frac{d^3\mathbf{q}}{(2\pi)^3} W^2_\text{vol}(\mathbf{k}) W^2_\text{res}(\mathbf{q}-\mathbf{k}) P(\mathbf{q}-\mathbf{k}),  
\end{equation}
where $W_\text{vol}$ and $W_\text{res}$ model the limited survey volume and voxel resolution, respectively, and the tilde denotes an observed quantity.
As $W_\text{res}$ captures the loss of information on scales smaller than the size of the voxel, it is applied as a convolution in real space, and a product in Fourier space. Conversely, $W_\text{vol}$ cuts off certain spatial positions, thus it is applied as a product in real space, and a convolution in Fourier space.

The spectral and angular resolutions of the spectrometer and the telescope define the resolution limits in the radial and transverse directions, respectively, and correspond to spatial scales
\begin{equation}
    \sigma_\parallel = \frac{c\delta\nu (1+z)}{H(z)\nu_\text{obs}}, \hspace{1cm} \sigma_\perp = D_M(z) \theta_\text{FWHM},
\end{equation}
where $D_M(z)$ is the comoving angular diameter distance. As in Ref.~\cite{Li:2015gqa}, we assume a Gaussian function in Fourier space to model the resolution window as
\begin{equation}
    W_\text{res}(k,\mu) = \exp \bigg \{ -k^2\left[ \sigma_\parallel^2\mu^2 + \sigma_\perp^2 (1-\mu^2)\right]\bigg \}. 
\end{equation}

We assume that the surveyed volume corresponds to a cylindrical volume aligned along the line of sight, with side length $L_\parallel=c\Delta \nu (1+z)/H(z)\nu_\text{obs}$ and radius $R_\perp=D_M(z)\sqrt{\Omega_\text{field}/\pi}$. Assuming that all spatial positions in the survey are observed with the same efficiency we model the volume window as a top hat in configuration space with values of 1 and 0 for points within and outside this volume respectively. In cylindrical coordinates this becomes,
\begin{equation}
    W_\text{vol}(k_\parallel,k_\perp) = \frac{1}{V_\text{field}}\frac{2\pi R_\perp L_\parallel}{k_\perp} J_1(k_\perp R_\perp) \text{sinc}\left(\frac{k_\parallel L_\parallel}{2} \right),
\end{equation}
where $V_\text{field}=L_\parallel \pi R_\perp^2$ is the comoving volume of the observed field, $k_\parallel = k\mu$, $k_\perp=k\sqrt{1-\mu^2}$, and $J_1$ is the Bessel function of the first kind.

The total observed LIM power spectrum will also include a component due to instrumental noise giving a final expression
\begin{equation}
\label{eq: observed power spectrum}
    \tilde P_\text{tot}(k,\mu) = \tilde P_\text{clust}(k,\mu,z)+\tilde P_\text{shot}(k,\mu,z) + P_N(z)\,.
\end{equation}
Assuming a spatially uniform, Gaussian-distributed instrumental noise, the noise power spectrum $P_N$ is given by
\begin{equation}
    P_N = V_\text{vox}\sigma_N^2,
\end{equation}
where $\sigma_N$ is the standard deviation of the instrumental noise per voxel, given by 
\begin{equation}
\sigma_N = \frac{T_{\rm sys}}{\sqrt{N_{\rm feeds} \delta\nu_{\rm FWHM} t_{\rm pix}}},
\end{equation}
where $N_{\rm feeds}$ is the total effective number of detectors, $t_{\rm pix}$ is the observing time per pixel, and $\delta\nu_{\rm FWHM} = \delta\nu \sqrt{8 \log 2}$. Values for $T_{\rm sys}$ are presented in Sec.~\ref{sec: experiment}. 

Lastly, in our analysis we consider only the monopole of the power spectrum, which can be simply computed from Eq.~(\ref{eq: observed power spectrum}) as
\begin{equation}
    \tilde P_0(k) = \frac{1}{2} \int d\mu \tilde P_\text{tot}(k,\mu). 
\end{equation}
The inclusion of higher-order multipoles would lead to a minimal increase in precision compared with the precision gained from the combination of the monopole and VID, hence for simplicity we consider only the monopole in this work.

\bibliography{main}

\begin{thebibliography}{90}%
\makeatletter
\providecommand \@ifxundefined [1]{%
 \@ifx{#1\undefined}
}%
\providecommand \@ifnum [1]{%
 \ifnum #1\expandafter \@firstoftwo
 \else \expandafter \@secondoftwo
 \fi
}%
\providecommand \@ifx [1]{%
 \ifx #1\expandafter \@firstoftwo
 \else \expandafter \@secondoftwo
 \fi
}%
\providecommand \natexlab [1]{#1}%
\providecommand \enquote  [1]{``#1''}%
\providecommand \bibnamefont  [1]{#1}%
\providecommand \bibfnamefont [1]{#1}%
\providecommand \citenamefont [1]{#1}%
\providecommand \href@noop [0]{\@secondoftwo}%
\providecommand \href [0]{\begingroup \@sanitize@url \@href}%
\providecommand \@href[1]{\@@startlink{#1}\@@href}%
\providecommand \@@href[1]{\endgroup#1\@@endlink}%
\providecommand \@sanitize@url [0]{\catcode `\\12\catcode `\$12\catcode `\&12\catcode `\#12\catcode `\^12\catcode `\_12\catcode `\%12\relax}%
\providecommand \@@startlink[1]{}%
\providecommand \@@endlink[0]{}%
\providecommand \url  [0]{\begingroup\@sanitize@url \@url }%
\providecommand \@url [1]{\endgroup\@href {#1}{\urlprefix }}%
\providecommand \urlprefix  [0]{URL }%
\providecommand \Eprint [0]{\href }%
\providecommand \doibase [0]{https://doi.org/}%
\providecommand \selectlanguage [0]{\@gobble}%
\providecommand \bibinfo  [0]{\@secondoftwo}%
\providecommand \bibfield  [0]{\@secondoftwo}%
\providecommand \translation [1]{[#1]}%
\providecommand \BibitemOpen [0]{}%
\providecommand \bibitemStop [0]{}%
\providecommand \bibitemNoStop [0]{.\EOS\space}%
\providecommand \EOS [0]{\spacefactor3000\relax}%
\providecommand \BibitemShut  [1]{\csname bibitem#1\endcsname}%
\let\auto@bib@innerbib\@empty
\bibitem [{\citenamefont {Kovetz}\ \emph {et~al.}(2017)\citenamefont {Kovetz} \emph {et~al.}}]{Kovetz:2017agg}%
  \BibitemOpen
  \bibfield  {author} {\bibinfo {author} {\bibfnamefont {E.~D.}\ \bibnamefont {Kovetz}} \emph {et~al.},\ }\bibfield  {title} {\bibinfo {title} {{Line-Intensity Mapping: 2017 Status Report}},\ }\href@noop {} {\  (\bibinfo {year} {2017})},\ \Eprint {https://arxiv.org/abs/1709.09066} {arXiv:1709.09066 [astro-ph.CO]} \BibitemShut {NoStop}%
\bibitem [{\citenamefont {Kovetz}\ \emph {et~al.}(2020)\citenamefont {Kovetz} \emph {et~al.}}]{Kovetz:2019uss}%
  \BibitemOpen
  \bibfield  {author} {\bibinfo {author} {\bibfnamefont {E.~D.}\ \bibnamefont {Kovetz}} \emph {et~al.},\ }\bibfield  {title} {\bibinfo {title} {{Astrophysics and Cosmology with Line-Intensity Mapping}},\ }\href@noop {} {\bibfield  {journal} {\bibinfo  {journal} {Bull. Am. Astron. Soc.}\ }\textbf {\bibinfo {volume} {51}},\ \bibinfo {pages} {101} (\bibinfo {year} {2020})},\ \Eprint {https://arxiv.org/abs/1903.04496} {arXiv:1903.04496 [astro-ph.CO]} \BibitemShut {NoStop}%
\bibitem [{\citenamefont {Bernal}\ and\ \citenamefont {Kovetz}(2022)}]{Bernal:2022jap}%
  \BibitemOpen
  \bibfield  {author} {\bibinfo {author} {\bibfnamefont {J.~L.}\ \bibnamefont {Bernal}}\ and\ \bibinfo {author} {\bibfnamefont {E.~D.}\ \bibnamefont {Kovetz}},\ }\bibfield  {title} {\bibinfo {title} {{Line-intensity mapping: theory review with a focus on star-formation lines}},\ }\href {https://doi.org/10.1007/s00159-022-00143-0} {\bibfield  {journal} {\bibinfo  {journal} {Astron. Astrophys. Rev.}\ }\textbf {\bibinfo {volume} {30}},\ \bibinfo {pages} {5} (\bibinfo {year} {2022})},\ \Eprint {https://arxiv.org/abs/2206.15377} {arXiv:2206.15377 [astro-ph.CO]} \BibitemShut {NoStop}%
\bibitem [{\citenamefont {Cheng}\ \emph {et~al.}(2019)\citenamefont {Cheng}, \citenamefont {de~Putter}, \citenamefont {Chang},\ and\ \citenamefont {Dore}}]{Cheng:2018hox}%
  \BibitemOpen
  \bibfield  {author} {\bibinfo {author} {\bibfnamefont {Y.-T.}\ \bibnamefont {Cheng}}, \bibinfo {author} {\bibfnamefont {R.}~\bibnamefont {de~Putter}}, \bibinfo {author} {\bibfnamefont {T.-C.}\ \bibnamefont {Chang}},\ and\ \bibinfo {author} {\bibfnamefont {O.}~\bibnamefont {Dore}},\ }\bibfield  {title} {\bibinfo {title} {{Optimally Mapping Large-Scale Structures with Luminous Sources}},\ }\href {https://doi.org/10.3847/1538-4357/ab1b2b} {\bibfield  {journal} {\bibinfo  {journal} {Astrophys. J.}\ }\textbf {\bibinfo {volume} {877}},\ \bibinfo {pages} {86} (\bibinfo {year} {2019})},\ \Eprint {https://arxiv.org/abs/1809.06384} {arXiv:1809.06384 [astro-ph.CO]} \BibitemShut {NoStop}%
\bibitem [{\citenamefont {Schaan}\ and\ \citenamefont {White}(2021)}]{Schaan:2021hhy}%
  \BibitemOpen
  \bibfield  {author} {\bibinfo {author} {\bibfnamefont {E.}~\bibnamefont {Schaan}}\ and\ \bibinfo {author} {\bibfnamefont {M.}~\bibnamefont {White}},\ }\bibfield  {title} {\bibinfo {title} {{Astrophysics \& Cosmology from Line Intensity Mapping vs Galaxy Surveys}},\ }\href {https://doi.org/10.1088/1475-7516/2021/05/067} {\bibfield  {journal} {\bibinfo  {journal} {JCAP}\ }\textbf {\bibinfo {volume} {05}},\ \bibinfo {pages} {067}},\ \Eprint {https://arxiv.org/abs/2103.01971} {arXiv:2103.01971 [astro-ph.CO]} \BibitemShut {NoStop}%
\bibitem [{\citenamefont {Chang}\ \emph {et~al.}(2008)\citenamefont {Chang}, \citenamefont {Pen}, \citenamefont {Peterson},\ and\ \citenamefont {McDonald}}]{Chang:2007xk}%
  \BibitemOpen
  \bibfield  {author} {\bibinfo {author} {\bibfnamefont {T.-C.}\ \bibnamefont {Chang}}, \bibinfo {author} {\bibfnamefont {U.-L.}\ \bibnamefont {Pen}}, \bibinfo {author} {\bibfnamefont {J.~B.}\ \bibnamefont {Peterson}},\ and\ \bibinfo {author} {\bibfnamefont {P.}~\bibnamefont {McDonald}},\ }\bibfield  {title} {\bibinfo {title} {{Baryon Acoustic Oscillation Intensity Mapping as a Test of Dark Energy}},\ }\href {https://doi.org/10.1103/PhysRevLett.100.091303} {\bibfield  {journal} {\bibinfo  {journal} {Phys. Rev. Lett.}\ }\textbf {\bibinfo {volume} {100}},\ \bibinfo {pages} {091303} (\bibinfo {year} {2008})},\ \Eprint {https://arxiv.org/abs/0709.3672} {arXiv:0709.3672 [astro-ph]} \BibitemShut {NoStop}%
\bibitem [{\citenamefont {Loeb}\ and\ \citenamefont {Wyithe}(2008)}]{Loeb:2008hg}%
  \BibitemOpen
  \bibfield  {author} {\bibinfo {author} {\bibfnamefont {A.}~\bibnamefont {Loeb}}\ and\ \bibinfo {author} {\bibfnamefont {S.}~\bibnamefont {Wyithe}},\ }\bibfield  {title} {\bibinfo {title} {{Precise Measurement of the Cosmological Power Spectrum With a Dedicated 21cm Survey After Reionization}},\ }\href {https://doi.org/10.1103/PhysRevLett.100.161301} {\bibfield  {journal} {\bibinfo  {journal} {Phys. Rev. Lett.}\ }\textbf {\bibinfo {volume} {100}},\ \bibinfo {pages} {161301} (\bibinfo {year} {2008})},\ \Eprint {https://arxiv.org/abs/0801.1677} {arXiv:0801.1677 [astro-ph]} \BibitemShut {NoStop}%
\bibitem [{\citenamefont {Visbal}\ \emph {et~al.}(2009)\citenamefont {Visbal}, \citenamefont {Loeb},\ and\ \citenamefont {Wyithe}}]{Visbal:2008rg}%
  \BibitemOpen
  \bibfield  {author} {\bibinfo {author} {\bibfnamefont {E.}~\bibnamefont {Visbal}}, \bibinfo {author} {\bibfnamefont {A.}~\bibnamefont {Loeb}},\ and\ \bibinfo {author} {\bibfnamefont {J.~S.~B.}\ \bibnamefont {Wyithe}},\ }\bibfield  {title} {\bibinfo {title} {{Cosmological Constraints from 21cm Surveys After Reionization}},\ }\href {https://doi.org/10.1088/1475-7516/2009/10/030} {\bibfield  {journal} {\bibinfo  {journal} {JCAP}\ }\textbf {\bibinfo {volume} {10}},\ \bibinfo {pages} {030}},\ \Eprint {https://arxiv.org/abs/0812.0419} {arXiv:0812.0419 [astro-ph]} \BibitemShut {NoStop}%
\bibitem [{\citenamefont {van Haarlem}\ \emph {et~al.}(2013)\citenamefont {van Haarlem} \emph {et~al.}}]{vanHaarlem:2013dsa}%
  \BibitemOpen
  \bibfield  {author} {\bibinfo {author} {\bibfnamefont {M.~P.}\ \bibnamefont {van Haarlem}} \emph {et~al.},\ }\bibfield  {title} {\bibinfo {title} {{LOFAR: The LOw-Frequency ARray}},\ }\href {https://doi.org/10.1051/0004-6361/201220873} {\bibfield  {journal} {\bibinfo  {journal} {Astron. Astrophys.}\ }\textbf {\bibinfo {volume} {556}},\ \bibinfo {pages} {A2} (\bibinfo {year} {2013})},\ \Eprint {https://arxiv.org/abs/1305.3550} {arXiv:1305.3550 [astro-ph.IM]} \BibitemShut {NoStop}%
\bibitem [{\citenamefont {Bandura}\ \emph {et~al.}(2014)\citenamefont {Bandura} \emph {et~al.}}]{Bandura:2014gwa}%
  \BibitemOpen
  \bibfield  {author} {\bibinfo {author} {\bibfnamefont {K.}~\bibnamefont {Bandura}} \emph {et~al.},\ }\bibfield  {title} {\bibinfo {title} {{Canadian Hydrogen Intensity Mapping Experiment (CHIME) Pathfinder}},\ }\href {https://doi.org/10.1117/12.2054950} {\bibfield  {journal} {\bibinfo  {journal} {Proc. SPIE Int. Soc. Opt. Eng.}\ }\textbf {\bibinfo {volume} {9145}},\ \bibinfo {pages} {22} (\bibinfo {year} {2014})},\ \Eprint {https://arxiv.org/abs/1406.2288} {arXiv:1406.2288 [astro-ph.IM]} \BibitemShut {NoStop}%
\bibitem [{\citenamefont {DeBoer}\ \emph {et~al.}(2017)\citenamefont {DeBoer} \emph {et~al.}}]{DeBoer:2016tnn}%
  \BibitemOpen
  \bibfield  {author} {\bibinfo {author} {\bibfnamefont {D.~R.}\ \bibnamefont {DeBoer}} \emph {et~al.},\ }\bibfield  {title} {\bibinfo {title} {{Hydrogen Epoch of Reionization Array (HERA)}},\ }\href {https://doi.org/10.1088/1538-3873/129/974/045001} {\bibfield  {journal} {\bibinfo  {journal} {Publ. Astron. Soc. Pac.}\ }\textbf {\bibinfo {volume} {129}},\ \bibinfo {pages} {045001} (\bibinfo {year} {2017})},\ \Eprint {https://arxiv.org/abs/1606.07473} {arXiv:1606.07473 [astro-ph.IM]} \BibitemShut {NoStop}%
\bibitem [{\citenamefont {Santos}\ \emph {et~al.}(2017)\citenamefont {Santos} \emph {et~al.}}]{MeerKLASS:2017vgf}%
  \BibitemOpen
  \bibfield  {author} {\bibinfo {author} {\bibfnamefont {M.~G.}\ \bibnamefont {Santos}} \emph {et~al.} (\bibinfo {collaboration} {MeerKLASS}),\ }\bibfield  {title} {\bibinfo {title} {{MeerKLASS: MeerKAT Large Area Synoptic Survey}},\ }in\ \href@noop {} {\emph {\bibinfo {booktitle} {{MeerKAT Science}: {On the Pathway to the SKA}}}}\ (\bibinfo {year} {2017})\ \Eprint {https://arxiv.org/abs/1709.06099} {arXiv:1709.06099 [astro-ph.CO]} \BibitemShut {NoStop}%
\bibitem [{\citenamefont {Keating}\ \emph {et~al.}(2020)\citenamefont {Keating}, \citenamefont {Marrone}, \citenamefont {Bower},\ and\ \citenamefont {Keenan}}]{Keating:2020wlx}%
  \BibitemOpen
  \bibfield  {author} {\bibinfo {author} {\bibfnamefont {G.~K.}\ \bibnamefont {Keating}}, \bibinfo {author} {\bibfnamefont {D.~P.}\ \bibnamefont {Marrone}}, \bibinfo {author} {\bibfnamefont {G.~C.}\ \bibnamefont {Bower}},\ and\ \bibinfo {author} {\bibfnamefont {R.~P.}\ \bibnamefont {Keenan}},\ }\bibfield  {title} {\bibinfo {title} {{An Intensity Mapping Detection of Aggregate CO Line Emission at 3 mm}},\ }\href {https://doi.org/10.3847/1538-4357/abb08e} {\bibfield  {journal} {\bibinfo  {journal} {Astrophys. J.}\ }\textbf {\bibinfo {volume} {901}},\ \bibinfo {pages} {141} (\bibinfo {year} {2020})},\ \Eprint {https://arxiv.org/abs/2008.08087} {arXiv:2008.08087 [astro-ph.GA]} \BibitemShut {NoStop}%
\bibitem [{\citenamefont {Keating}\ \emph {et~al.}(2016)\citenamefont {Keating}, \citenamefont {Marrone}, \citenamefont {Bower}, \citenamefont {Leitch}, \citenamefont {Carlstrom},\ and\ \citenamefont {DeBoer}}]{Keating:2016pka}%
  \BibitemOpen
  \bibfield  {author} {\bibinfo {author} {\bibfnamefont {G.~K.}\ \bibnamefont {Keating}}, \bibinfo {author} {\bibfnamefont {D.~P.}\ \bibnamefont {Marrone}}, \bibinfo {author} {\bibfnamefont {G.~C.}\ \bibnamefont {Bower}}, \bibinfo {author} {\bibfnamefont {E.}~\bibnamefont {Leitch}}, \bibinfo {author} {\bibfnamefont {J.~E.}\ \bibnamefont {Carlstrom}},\ and\ \bibinfo {author} {\bibfnamefont {D.~R.}\ \bibnamefont {DeBoer}},\ }\bibfield  {title} {\bibinfo {title} {{COPSS II: The molecular gas content of ten million cubic megaparsecs at redshift z \ensuremath{\sim} 3}},\ }\href {https://doi.org/10.3847/0004-637X/830/1/34} {\bibfield  {journal} {\bibinfo  {journal} {Astrophys. J.}\ }\textbf {\bibinfo {volume} {830}},\ \bibinfo {pages} {34} (\bibinfo {year} {2016})},\ \Eprint {https://arxiv.org/abs/1605.03971} {arXiv:1605.03971 [astro-ph.GA]} \BibitemShut {NoStop}%
\bibitem [{\citenamefont {Cleary}\ \emph {et~al.}(2021)\citenamefont {Cleary} \emph {et~al.}}]{Cleary:2021dsp}%
  \BibitemOpen
  \bibfield  {author} {\bibinfo {author} {\bibfnamefont {K.~A.}\ \bibnamefont {Cleary}} \emph {et~al.},\ }\bibfield  {title} {\bibinfo {title} {{COMAP Early Science: I. Overview}}\ }\href {https://doi.org/10.3847/1538-4357/ac63cc} {10.3847/1538-4357/ac63cc} (\bibinfo {year} {2021}),\ \Eprint {https://arxiv.org/abs/2111.05927} {arXiv:2111.05927 [astro-ph.CO]} \BibitemShut {NoStop}%
\bibitem [{\citenamefont {Ade}\ \emph {et~al.}(2020)\citenamefont {Ade} \emph {et~al.}}]{CONCERTO:2020ahk}%
  \BibitemOpen
  \bibfield  {author} {\bibinfo {author} {\bibfnamefont {P.}~\bibnamefont {Ade}} \emph {et~al.} (\bibinfo {collaboration} {CONCERTO}),\ }\bibfield  {title} {\bibinfo {title} {{A wide field-of-view low-resolution spectrometer at APEX: Instrument design and scientific forecast}},\ }\href {https://doi.org/10.1051/0004-6361/202038456} {\bibfield  {journal} {\bibinfo  {journal} {Astron. Astrophys.}\ }\textbf {\bibinfo {volume} {642}},\ \bibinfo {pages} {A60} (\bibinfo {year} {2020})},\ \Eprint {https://arxiv.org/abs/2007.14246} {arXiv:2007.14246 [astro-ph.IM]} \BibitemShut {NoStop}%
\bibitem [{\citenamefont {Gebhardt}\ \emph {et~al.}(2021)\citenamefont {Gebhardt} \emph {et~al.}}]{Gebhardt:2021vfo}%
  \BibitemOpen
  \bibfield  {author} {\bibinfo {author} {\bibfnamefont {K.}~\bibnamefont {Gebhardt}} \emph {et~al.},\ }\bibfield  {title} {\bibinfo {title} {{The Hobby\textendash{}Eberly Telescope Dark Energy Experiment (HETDEX) Survey Design, Reductions, and Detections*}},\ }\href {https://doi.org/10.3847/1538-4357/ac2e03} {\bibfield  {journal} {\bibinfo  {journal} {Astrophys. J.}\ }\textbf {\bibinfo {volume} {923}},\ \bibinfo {pages} {217} (\bibinfo {year} {2021})},\ \Eprint {https://arxiv.org/abs/2110.04298} {arXiv:2110.04298 [astro-ph.IM]} \BibitemShut {NoStop}%
\bibitem [{\citenamefont {Aravena}\ \emph {et~al.}(2023)\citenamefont {Aravena} \emph {et~al.}}]{CCAT-Prime:2021lly}%
  \BibitemOpen
  \bibfield  {author} {\bibinfo {author} {\bibfnamefont {M.}~\bibnamefont {Aravena}} \emph {et~al.} (\bibinfo {collaboration} {CCAT-Prime}),\ }\bibfield  {title} {\bibinfo {title} {{CCAT-prime Collaboration: Science Goals and Forecasts with Prime-Cam on the Fred Young Submillimeter Telescope}},\ }\href {https://doi.org/10.3847/1538-4365/ac9838} {\bibfield  {journal} {\bibinfo  {journal} {Astrophys. J. Suppl.}\ }\textbf {\bibinfo {volume} {264}},\ \bibinfo {pages} {7} (\bibinfo {year} {2023})},\ \Eprint {https://arxiv.org/abs/2107.10364} {arXiv:2107.10364 [astro-ph.CO]} \BibitemShut {NoStop}%
\bibitem [{\citenamefont {Sun}\ \emph {et~al.}(2021)\citenamefont {Sun} \emph {et~al.}}]{Sun:2020mco}%
  \BibitemOpen
  \bibfield  {author} {\bibinfo {author} {\bibfnamefont {G.}~\bibnamefont {Sun}} \emph {et~al.},\ }\bibfield  {title} {\bibinfo {title} {{Probing Cosmic Reionization and Molecular Gas Growth with TIME}},\ }\href {https://doi.org/10.3847/1538-4357/abfe62} {\bibfield  {journal} {\bibinfo  {journal} {Astrophys. J.}\ }\textbf {\bibinfo {volume} {915}},\ \bibinfo {pages} {33} (\bibinfo {year} {2021})},\ \Eprint {https://arxiv.org/abs/2012.09160} {arXiv:2012.09160 [astro-ph.GA]} \BibitemShut {NoStop}%
\bibitem [{\citenamefont {Switzer}\ \emph {et~al.}(2021)\citenamefont {Switzer} \emph {et~al.}}]{Switzer:2021jeg}%
  \BibitemOpen
  \bibfield  {author} {\bibinfo {author} {\bibfnamefont {E.~R.}\ \bibnamefont {Switzer}} \emph {et~al.},\ }\bibfield  {title} {\bibinfo {title} {{Experiment for cryogenic large-aperture intensity mapping: instrument design}},\ }\href {https://doi.org/10.1117/1.JATIS.7.4.044004} {\bibfield  {journal} {\bibinfo  {journal} {J. Astron. Telesc. Instrum. Syst.}\ }\textbf {\bibinfo {volume} {7}},\ \bibinfo {pages} {044004} (\bibinfo {year} {2021})}\BibitemShut {NoStop}%
\bibitem [{\citenamefont {{Vieira}}\ \emph {et~al.}(2020)\citenamefont {{Vieira}}, \citenamefont {{Aguirre}}, \citenamefont {{Bradford}}, \citenamefont {{Filippini}}, \citenamefont {{Groppi}}, \citenamefont {{Marrone}}, \citenamefont {{Bethermin}}, \citenamefont {{Chang}}, \citenamefont {{Devlin}}, \citenamefont {{Dore}}, \citenamefont {{Fu}}, \citenamefont {{Hailey Dunsheath}}, \citenamefont {{Holder}}, \citenamefont {{Keating}}, \citenamefont {{Keenan}}, \citenamefont {{Kovetz}}, \citenamefont {{Lagache}}, \citenamefont {{Mauskopf}}, \citenamefont {{Narayanan}}, \citenamefont {{Popping}}, \citenamefont {{Shirokoff}}, \citenamefont {{Somerville}}, \citenamefont {{Trumper}}, \citenamefont {{Uzgil}},\ and\ \citenamefont {{Zmuidzinas}}}]{Vieira:2020tim}%
  \BibitemOpen
  \bibfield  {author} {\bibinfo {author} {\bibfnamefont {J.}~\bibnamefont {{Vieira}}}, \bibinfo {author} {\bibfnamefont {J.}~\bibnamefont {{Aguirre}}}, \bibinfo {author} {\bibfnamefont {C.~M.}\ \bibnamefont {{Bradford}}}, \bibinfo {author} {\bibfnamefont {J.}~\bibnamefont {{Filippini}}}, \bibinfo {author} {\bibfnamefont {C.}~\bibnamefont {{Groppi}}}, \bibinfo {author} {\bibfnamefont {D.}~\bibnamefont {{Marrone}}}, \bibinfo {author} {\bibfnamefont {M.}~\bibnamefont {{Bethermin}}}, \bibinfo {author} {\bibfnamefont {T.-C.}\ \bibnamefont {{Chang}}}, \bibinfo {author} {\bibfnamefont {M.}~\bibnamefont {{Devlin}}}, \bibinfo {author} {\bibfnamefont {O.}~\bibnamefont {{Dore}}}, \bibinfo {author} {\bibfnamefont {J.~F.}\ \bibnamefont {{Fu}}}, \bibinfo {author} {\bibfnamefont {S.}~\bibnamefont {{Hailey Dunsheath}}}, \bibinfo {author} {\bibfnamefont {G.}~\bibnamefont {{Holder}}}, \bibinfo {author} {\bibfnamefont {G.}~\bibnamefont {{Keating}}}, \bibinfo {author} {\bibfnamefont {R.}~\bibnamefont {{Keenan}}}, \bibinfo
  {author} {\bibfnamefont {E.}~\bibnamefont {{Kovetz}}}, \bibinfo {author} {\bibfnamefont {G.}~\bibnamefont {{Lagache}}}, \bibinfo {author} {\bibfnamefont {P.}~\bibnamefont {{Mauskopf}}}, \bibinfo {author} {\bibfnamefont {D.}~\bibnamefont {{Narayanan}}}, \bibinfo {author} {\bibfnamefont {G.}~\bibnamefont {{Popping}}}, \bibinfo {author} {\bibfnamefont {E.}~\bibnamefont {{Shirokoff}}}, \bibinfo {author} {\bibfnamefont {R.}~\bibnamefont {{Somerville}}}, \bibinfo {author} {\bibfnamefont {I.}~\bibnamefont {{Trumper}}}, \bibinfo {author} {\bibfnamefont {B.}~\bibnamefont {{Uzgil}}},\ and\ \bibinfo {author} {\bibfnamefont {J.}~\bibnamefont {{Zmuidzinas}}},\ }\bibfield  {title} {\bibinfo {title} {{The Terahertz Intensity Mapper (TIM): a Next-Generation Experiment for Galaxy Evolution Studies}},\ }\href@noop {} {\bibfield  {journal} {\bibinfo  {journal} {arXiv e-prints}\ ,\ \bibinfo {eid} {arXiv:2009.14340}} (\bibinfo {year} {2020})},\ \Eprint {https://arxiv.org/abs/2009.14340} {arXiv:2009.14340 [astro-ph.IM]}
  \BibitemShut {NoStop}%
\bibitem [{\citenamefont {Dor\'e}\ \emph {et~al.}(2014)\citenamefont {Dor\'e} \emph {et~al.}}]{Dore:2014cca}%
  \BibitemOpen
  \bibfield  {author} {\bibinfo {author} {\bibfnamefont {O.}~\bibnamefont {Dor\'e}} \emph {et~al.} (\bibinfo {collaboration} {SPHEREx}),\ }\bibfield  {title} {\bibinfo {title} {{Cosmology with the SPHEREX All-Sky Spectral Survey}},\ }\href@noop {} {\  (\bibinfo {year} {2014})},\ \Eprint {https://arxiv.org/abs/1412.4872} {arXiv:1412.4872 [astro-ph.CO]} \BibitemShut {NoStop}%
\bibitem [{\citenamefont {Koopmans}\ \emph {et~al.}(2015)\citenamefont {Koopmans} \emph {et~al.}}]{Koopmans:2015sua}%
  \BibitemOpen
  \bibfield  {author} {\bibinfo {author} {\bibfnamefont {L.~V.~E.}\ \bibnamefont {Koopmans}} \emph {et~al.},\ }\bibfield  {title} {\bibinfo {title} {{The Cosmic Dawn and Epoch of Reionization with the Square Kilometre Array}},\ }\href {https://doi.org/10.22323/1.215.0001} {\bibfield  {journal} {\bibinfo  {journal} {PoS}\ }\textbf {\bibinfo {volume} {AASKA14}},\ \bibinfo {pages} {001} (\bibinfo {year} {2015})},\ \Eprint {https://arxiv.org/abs/1505.07568} {arXiv:1505.07568 [astro-ph.CO]} \BibitemShut {NoStop}%
\bibitem [{\citenamefont {Newburgh}\ \emph {et~al.}(2016)\citenamefont {Newburgh} \emph {et~al.}}]{Newburgh:2016mwi}%
  \BibitemOpen
  \bibfield  {author} {\bibinfo {author} {\bibfnamefont {L.~B.}\ \bibnamefont {Newburgh}} \emph {et~al.},\ }\bibfield  {title} {\bibinfo {title} {{HIRAX: A Probe of Dark Energy and Radio Transients}},\ }\href {https://doi.org/10.1117/12.2234286} {\bibfield  {journal} {\bibinfo  {journal} {Proc. SPIE Int. Soc. Opt. Eng.}\ }\textbf {\bibinfo {volume} {9906}},\ \bibinfo {pages} {99065X} (\bibinfo {year} {2016})},\ \Eprint {https://arxiv.org/abs/1607.02059} {arXiv:1607.02059 [astro-ph.IM]} \BibitemShut {NoStop}%
\bibitem [{\citenamefont {Padmanabhan}\ \emph {et~al.}(2019)\citenamefont {Padmanabhan}, \citenamefont {Refregier},\ and\ \citenamefont {Amara}}]{Padmanabhan:2018llf}%
  \BibitemOpen
  \bibfield  {author} {\bibinfo {author} {\bibfnamefont {H.}~\bibnamefont {Padmanabhan}}, \bibinfo {author} {\bibfnamefont {A.}~\bibnamefont {Refregier}},\ and\ \bibinfo {author} {\bibfnamefont {A.}~\bibnamefont {Amara}},\ }\bibfield  {title} {\bibinfo {title} {{Impact of astrophysics on cosmology forecasts for 21 cm surveys}},\ }\href {https://doi.org/10.1093/mnras/stz683} {\bibfield  {journal} {\bibinfo  {journal} {Mon. Not. Roy. Astron. Soc.}\ }\textbf {\bibinfo {volume} {485}},\ \bibinfo {pages} {4060} (\bibinfo {year} {2019})},\ \Eprint {https://arxiv.org/abs/1804.10627} {arXiv:1804.10627 [astro-ph.CO]} \BibitemShut {NoStop}%
\bibitem [{\citenamefont {Bernal}\ \emph {et~al.}(2019{\natexlab{a}})\citenamefont {Bernal}, \citenamefont {Breysse}, \citenamefont {Gil-Mar\'\i{}n},\ and\ \citenamefont {Kovetz}}]{Bernal:2019jdo}%
  \BibitemOpen
  \bibfield  {author} {\bibinfo {author} {\bibfnamefont {J.~L.}\ \bibnamefont {Bernal}}, \bibinfo {author} {\bibfnamefont {P.~C.}\ \bibnamefont {Breysse}}, \bibinfo {author} {\bibfnamefont {H.}~\bibnamefont {Gil-Mar\'\i{}n}},\ and\ \bibinfo {author} {\bibfnamefont {E.~D.}\ \bibnamefont {Kovetz}},\ }\bibfield  {title} {\bibinfo {title} {{User\textquoteright{}s guide to extracting cosmological information from line-intensity maps}},\ }\href {https://doi.org/10.1103/PhysRevD.100.123522} {\bibfield  {journal} {\bibinfo  {journal} {Phys. Rev. D}\ }\textbf {\bibinfo {volume} {100}},\ \bibinfo {pages} {123522} (\bibinfo {year} {2019}{\natexlab{a}})},\ \Eprint {https://arxiv.org/abs/1907.10067} {arXiv:1907.10067 [astro-ph.CO]} \BibitemShut {NoStop}%
\bibitem [{\citenamefont {Breysse}\ \emph {et~al.}(2017)\citenamefont {Breysse}, \citenamefont {Kovetz}, \citenamefont {Behroozi}, \citenamefont {Dai},\ and\ \citenamefont {Kamionkowski}}]{Breysse:2016szq}%
  \BibitemOpen
  \bibfield  {author} {\bibinfo {author} {\bibfnamefont {P.~C.}\ \bibnamefont {Breysse}}, \bibinfo {author} {\bibfnamefont {E.~D.}\ \bibnamefont {Kovetz}}, \bibinfo {author} {\bibfnamefont {P.~S.}\ \bibnamefont {Behroozi}}, \bibinfo {author} {\bibfnamefont {L.}~\bibnamefont {Dai}},\ and\ \bibinfo {author} {\bibfnamefont {M.}~\bibnamefont {Kamionkowski}},\ }\bibfield  {title} {\bibinfo {title} {{Insights from probability distribution functions of intensity maps}},\ }\href {https://doi.org/10.1093/mnras/stx203} {\bibfield  {journal} {\bibinfo  {journal} {Mon. Not. Roy. Astron. Soc.}\ }\textbf {\bibinfo {volume} {467}},\ \bibinfo {pages} {2996} (\bibinfo {year} {2017})},\ \Eprint {https://arxiv.org/abs/1609.01728} {arXiv:1609.01728 [astro-ph.CO]} \BibitemShut {NoStop}%
\bibitem [{\citenamefont {Ihle}\ \emph {et~al.}(2019)\citenamefont {Ihle} \emph {et~al.}}]{COMAP:2018kem}%
  \BibitemOpen
  \bibfield  {author} {\bibinfo {author} {\bibfnamefont {H.~T.}\ \bibnamefont {Ihle}} \emph {et~al.} (\bibinfo {collaboration} {COMAP}),\ }\bibfield  {title} {\bibinfo {title} {{Joint power spectrum and voxel intensity distribution forecast on the CO luminosity function with COMAP}},\ }\href {https://doi.org/10.3847/1538-4357/aaf4bc} {\bibfield  {journal} {\bibinfo  {journal} {Astrophys. J.}\ }\textbf {\bibinfo {volume} {871}},\ \bibinfo {pages} {75} (\bibinfo {year} {2019})},\ \Eprint {https://arxiv.org/abs/1808.07487} {arXiv:1808.07487 [astro-ph.CO]} \BibitemShut {NoStop}%
\bibitem [{\citenamefont {Chung}\ \emph {et~al.}(2022)\citenamefont {Chung} \emph {et~al.}}]{COMAP:2021lae}%
  \BibitemOpen
  \bibfield  {author} {\bibinfo {author} {\bibfnamefont {D.~T.}\ \bibnamefont {Chung}} \emph {et~al.} (\bibinfo {collaboration} {COMAP}),\ }\bibfield  {title} {\bibinfo {title} {{COMAP Early Science. V. Constraints and Forecasts at z \ensuremath{\sim} 3}},\ }\href {https://doi.org/10.3847/1538-4357/ac63c7} {\bibfield  {journal} {\bibinfo  {journal} {Astrophys. J.}\ }\textbf {\bibinfo {volume} {933}},\ \bibinfo {pages} {186} (\bibinfo {year} {2022})},\ \Eprint {https://arxiv.org/abs/2111.05931} {arXiv:2111.05931 [astro-ph.CO]} \BibitemShut {NoStop}%
\bibitem [{\citenamefont {Breysse}(2022)}]{Breysse:2022alx}%
  \BibitemOpen
  \bibfield  {author} {\bibinfo {author} {\bibfnamefont {P.~C.}\ \bibnamefont {Breysse}},\ }\bibfield  {title} {\bibinfo {title} {{Breaking the intensity-bias degeneracy in line intensity mapping}},\ }\href@noop {} {\  (\bibinfo {year} {2022})},\ \Eprint {https://arxiv.org/abs/2209.01223} {arXiv:2209.01223 [astro-ph.CO]} \BibitemShut {NoStop}%
\bibitem [{\citenamefont {Sato-Polito}\ and\ \citenamefont {Bernal}(2022)}]{Sato-Polito:2022fkd}%
  \BibitemOpen
  \bibfield  {author} {\bibinfo {author} {\bibfnamefont {G.}~\bibnamefont {Sato-Polito}}\ and\ \bibinfo {author} {\bibfnamefont {J.~L.}\ \bibnamefont {Bernal}},\ }\bibfield  {title} {\bibinfo {title} {{Analytical covariance between the voxel intensity distribution and the power spectrum of line-intensity maps}},\ }\href {https://doi.org/10.1103/PhysRevD.106.103534} {\bibfield  {journal} {\bibinfo  {journal} {Phys. Rev. D}\ }\textbf {\bibinfo {volume} {106}},\ \bibinfo {pages} {103534} (\bibinfo {year} {2022})},\ \Eprint {https://arxiv.org/abs/2202.02330} {arXiv:2202.02330 [astro-ph.CO]} \BibitemShut {NoStop}%
\bibitem [{\citenamefont {Sato-Polito}\ \emph {et~al.}(2023)\citenamefont {Sato-Polito}, \citenamefont {Kokron},\ and\ \citenamefont {Bernal}}]{Sato-Polito:2022wiq}%
  \BibitemOpen
  \bibfield  {author} {\bibinfo {author} {\bibfnamefont {G.}~\bibnamefont {Sato-Polito}}, \bibinfo {author} {\bibfnamefont {N.}~\bibnamefont {Kokron}},\ and\ \bibinfo {author} {\bibfnamefont {J.~L.}\ \bibnamefont {Bernal}},\ }\bibfield  {title} {\bibinfo {title} {{A multitracer empirically driven approach to line-intensity mapping light cones}},\ }\href {https://doi.org/10.1093/mnras/stad2498} {\bibfield  {journal} {\bibinfo  {journal} {Mon. Not. Roy. Astron. Soc.}\ }\textbf {\bibinfo {volume} {526}},\ \bibinfo {pages} {5883} (\bibinfo {year} {2023})},\ \Eprint {https://arxiv.org/abs/2212.08056} {arXiv:2212.08056 [astro-ph.CO]} \BibitemShut {NoStop}%
\bibitem [{\citenamefont {Libanore}\ \emph {et~al.}(2022)\citenamefont {Libanore}, \citenamefont {Unal}, \citenamefont {Sarkar},\ and\ \citenamefont {Kovetz}}]{Libanore:2022ntl}%
  \BibitemOpen
  \bibfield  {author} {\bibinfo {author} {\bibfnamefont {S.}~\bibnamefont {Libanore}}, \bibinfo {author} {\bibfnamefont {C.}~\bibnamefont {Unal}}, \bibinfo {author} {\bibfnamefont {D.}~\bibnamefont {Sarkar}},\ and\ \bibinfo {author} {\bibfnamefont {E.~D.}\ \bibnamefont {Kovetz}},\ }\bibfield  {title} {\bibinfo {title} {{Unveiling cosmological information on small scales with line intensity mapping}},\ }\href {https://doi.org/10.1103/PhysRevD.106.123512} {\bibfield  {journal} {\bibinfo  {journal} {Phys. Rev. D}\ }\textbf {\bibinfo {volume} {106}},\ \bibinfo {pages} {123512} (\bibinfo {year} {2022})},\ \Eprint {https://arxiv.org/abs/2208.01658} {arXiv:2208.01658 [astro-ph.CO]} \BibitemShut {NoStop}%
\bibitem [{\citenamefont {Adi}\ \emph {et~al.}(2023)\citenamefont {Adi}, \citenamefont {Libanore}, \citenamefont {Cruz},\ and\ \citenamefont {Kovetz}}]{Adi:2023qdf}%
  \BibitemOpen
  \bibfield  {author} {\bibinfo {author} {\bibfnamefont {T.}~\bibnamefont {Adi}}, \bibinfo {author} {\bibfnamefont {S.}~\bibnamefont {Libanore}}, \bibinfo {author} {\bibfnamefont {H.~A.~G.}\ \bibnamefont {Cruz}},\ and\ \bibinfo {author} {\bibfnamefont {E.~D.}\ \bibnamefont {Kovetz}},\ }\bibfield  {title} {\bibinfo {title} {{Constraining primordial magnetic fields with line-intensity mapping}},\ }\href {https://doi.org/10.1088/1475-7516/2023/09/035} {\bibfield  {journal} {\bibinfo  {journal} {JCAP}\ }\textbf {\bibinfo {volume} {09}},\ \bibinfo {pages} {035}},\ \Eprint {https://arxiv.org/abs/2305.06440} {arXiv:2305.06440 [astro-ph.CO]} \BibitemShut {NoStop}%
\bibitem [{\citenamefont {Bernal}\ \emph {et~al.}(2021{\natexlab{a}})\citenamefont {Bernal}, \citenamefont {Caputo},\ and\ \citenamefont {Kamionkowski}}]{Bernal:2020lkd}%
  \BibitemOpen
  \bibfield  {author} {\bibinfo {author} {\bibfnamefont {J.~L.}\ \bibnamefont {Bernal}}, \bibinfo {author} {\bibfnamefont {A.}~\bibnamefont {Caputo}},\ and\ \bibinfo {author} {\bibfnamefont {M.}~\bibnamefont {Kamionkowski}},\ }\bibfield  {title} {\bibinfo {title} {{Strategies to Detect Dark-Matter Decays with Line-Intensity Mapping}},\ }\href {https://doi.org/10.1103/PhysRevD.103.063523} {\bibfield  {journal} {\bibinfo  {journal} {Phys. Rev. D}\ }\textbf {\bibinfo {volume} {103}},\ \bibinfo {pages} {063523} (\bibinfo {year} {2021}{\natexlab{a}})},\ \bibinfo {note} {[Erratum: Phys.Rev.D 105, 089901 (2022)]},\ \Eprint {https://arxiv.org/abs/2012.00771} {arXiv:2012.00771 [astro-ph.CO]} \BibitemShut {NoStop}%
\bibitem [{\citenamefont {Bernal}\ \emph {et~al.}(2021{\natexlab{b}})\citenamefont {Bernal}, \citenamefont {Caputo}, \citenamefont {Villaescusa-Navarro},\ and\ \citenamefont {Kamionkowski}}]{Bernal:2021ylz}%
  \BibitemOpen
  \bibfield  {author} {\bibinfo {author} {\bibfnamefont {J.~L.}\ \bibnamefont {Bernal}}, \bibinfo {author} {\bibfnamefont {A.}~\bibnamefont {Caputo}}, \bibinfo {author} {\bibfnamefont {F.}~\bibnamefont {Villaescusa-Navarro}},\ and\ \bibinfo {author} {\bibfnamefont {M.}~\bibnamefont {Kamionkowski}},\ }\bibfield  {title} {\bibinfo {title} {{Searching for the Radiative Decay of the Cosmic Neutrino Background with Line-Intensity Mapping}},\ }\href {https://doi.org/10.1103/PhysRevLett.127.131102} {\bibfield  {journal} {\bibinfo  {journal} {Phys. Rev. Lett.}\ }\textbf {\bibinfo {volume} {127}},\ \bibinfo {pages} {131102} (\bibinfo {year} {2021}{\natexlab{b}})},\ \Eprint {https://arxiv.org/abs/2103.12099} {arXiv:2103.12099 [hep-ph]} \BibitemShut {NoStop}%
\bibitem [{\citenamefont {Bernal}\ \emph {et~al.}(2019{\natexlab{b}})\citenamefont {Bernal}, \citenamefont {Breysse},\ and\ \citenamefont {Kovetz}}]{Bernal:2019gfq}%
  \BibitemOpen
  \bibfield  {author} {\bibinfo {author} {\bibfnamefont {J.~L.}\ \bibnamefont {Bernal}}, \bibinfo {author} {\bibfnamefont {P.~C.}\ \bibnamefont {Breysse}},\ and\ \bibinfo {author} {\bibfnamefont {E.~D.}\ \bibnamefont {Kovetz}},\ }\bibfield  {title} {\bibinfo {title} {{Cosmic Expansion History from Line-Intensity Mapping}},\ }\href {https://doi.org/10.1103/PhysRevLett.123.251301} {\bibfield  {journal} {\bibinfo  {journal} {Phys. Rev. Lett.}\ }\textbf {\bibinfo {volume} {123}},\ \bibinfo {pages} {251301} (\bibinfo {year} {2019}{\natexlab{b}})},\ \Eprint {https://arxiv.org/abs/1907.10065} {arXiv:1907.10065 [astro-ph.CO]} \BibitemShut {NoStop}%
\bibitem [{\citenamefont {Del~Popolo}\ and\ \citenamefont {Le~Delliou}(2021)}]{DelPopolo:2021bom}%
  \BibitemOpen
  \bibfield  {author} {\bibinfo {author} {\bibfnamefont {A.}~\bibnamefont {Del~Popolo}}\ and\ \bibinfo {author} {\bibfnamefont {M.}~\bibnamefont {Le~Delliou}},\ }\bibfield  {title} {\bibinfo {title} {{Review of Solutions to the Cusp-Core Problem of the \({\Lambda}\)CDM Model}},\ }\href {https://doi.org/10.3390/galaxies9040123} {\bibfield  {journal} {\bibinfo  {journal} {Galaxies}\ }\textbf {\bibinfo {volume} {9}},\ \bibinfo {pages} {123} (\bibinfo {year} {2021})},\ \Eprint {https://arxiv.org/abs/2209.14151} {arXiv:2209.14151 [astro-ph.CO]} \BibitemShut {NoStop}%
\bibitem [{\citenamefont {Murgia}\ \emph {et~al.}(2017)\citenamefont {Murgia}, \citenamefont {Merle}, \citenamefont {Viel}, \citenamefont {Totzauer},\ and\ \citenamefont {Schneider}}]{Murgia:2017lwo}%
  \BibitemOpen
  \bibfield  {author} {\bibinfo {author} {\bibfnamefont {R.}~\bibnamefont {Murgia}}, \bibinfo {author} {\bibfnamefont {A.}~\bibnamefont {Merle}}, \bibinfo {author} {\bibfnamefont {M.}~\bibnamefont {Viel}}, \bibinfo {author} {\bibfnamefont {M.}~\bibnamefont {Totzauer}},\ and\ \bibinfo {author} {\bibfnamefont {A.}~\bibnamefont {Schneider}},\ }\bibfield  {title} {\bibinfo {title} {{''Non-cold'' dark matter at small scales: a general approach}},\ }\href {https://doi.org/10.1088/1475-7516/2017/11/046} {\bibfield  {journal} {\bibinfo  {journal} {JCAP}\ }\textbf {\bibinfo {volume} {11}},\ \bibinfo {pages} {046}},\ \Eprint {https://arxiv.org/abs/1704.07838} {arXiv:1704.07838 [astro-ph.CO]} \BibitemShut {NoStop}%
\bibitem [{\citenamefont {Marsh}(2016)}]{Marsh:2015xka}%
  \BibitemOpen
  \bibfield  {author} {\bibinfo {author} {\bibfnamefont {D.~J.~E.}\ \bibnamefont {Marsh}},\ }\bibfield  {title} {\bibinfo {title} {{Axion Cosmology}},\ }\href {https://doi.org/10.1016/j.physrep.2016.06.005} {\bibfield  {journal} {\bibinfo  {journal} {Phys. Rept.}\ }\textbf {\bibinfo {volume} {643}},\ \bibinfo {pages} {1} (\bibinfo {year} {2016})},\ \Eprint {https://arxiv.org/abs/1510.07633} {arXiv:1510.07633 [astro-ph.CO]} \BibitemShut {NoStop}%
\bibitem [{\citenamefont {Kim}\ and\ \citenamefont {Marsh}(2016)}]{Kim:2015yna}%
  \BibitemOpen
  \bibfield  {author} {\bibinfo {author} {\bibfnamefont {J.~E.}\ \bibnamefont {Kim}}\ and\ \bibinfo {author} {\bibfnamefont {D.~J.~E.}\ \bibnamefont {Marsh}},\ }\bibfield  {title} {\bibinfo {title} {{An ultralight pseudoscalar boson}},\ }\href {https://doi.org/10.1103/PhysRevD.93.025027} {\bibfield  {journal} {\bibinfo  {journal} {Phys. Rev. D}\ }\textbf {\bibinfo {volume} {93}},\ \bibinfo {pages} {025027} (\bibinfo {year} {2016})},\ \Eprint {https://arxiv.org/abs/1510.01701} {arXiv:1510.01701 [hep-ph]} \BibitemShut {NoStop}%
\bibitem [{\citenamefont {Chun}(2014)}]{Chun:2014xva}%
  \BibitemOpen
  \bibfield  {author} {\bibinfo {author} {\bibfnamefont {E.~J.}\ \bibnamefont {Chun}},\ }\bibfield  {title} {\bibinfo {title} {{Axion Dark Matter with High-Scale Inflation}},\ }\href {https://doi.org/10.1016/j.physletb.2014.06.017} {\bibfield  {journal} {\bibinfo  {journal} {Phys. Lett. B}\ }\textbf {\bibinfo {volume} {735}},\ \bibinfo {pages} {164} (\bibinfo {year} {2014})},\ \Eprint {https://arxiv.org/abs/1404.4284} {arXiv:1404.4284 [hep-ph]} \BibitemShut {NoStop}%
\bibitem [{\citenamefont {Kawasaki}\ \emph {et~al.}(2015)\citenamefont {Kawasaki}, \citenamefont {Saikawa},\ and\ \citenamefont {Sekiguchi}}]{Kawasaki:2014sqa}%
  \BibitemOpen
  \bibfield  {author} {\bibinfo {author} {\bibfnamefont {M.}~\bibnamefont {Kawasaki}}, \bibinfo {author} {\bibfnamefont {K.}~\bibnamefont {Saikawa}},\ and\ \bibinfo {author} {\bibfnamefont {T.}~\bibnamefont {Sekiguchi}},\ }\bibfield  {title} {\bibinfo {title} {{Axion dark matter from topological defects}},\ }\href {https://doi.org/10.1103/PhysRevD.91.065014} {\bibfield  {journal} {\bibinfo  {journal} {Phys. Rev. D}\ }\textbf {\bibinfo {volume} {91}},\ \bibinfo {pages} {065014} (\bibinfo {year} {2015})},\ \Eprint {https://arxiv.org/abs/1412.0789} {arXiv:1412.0789 [hep-ph]} \BibitemShut {NoStop}%
\bibitem [{\citenamefont {Hwang}\ and\ \citenamefont {Noh}(2009)}]{Hwang:2009js}%
  \BibitemOpen
  \bibfield  {author} {\bibinfo {author} {\bibfnamefont {J.-c.}\ \bibnamefont {Hwang}}\ and\ \bibinfo {author} {\bibfnamefont {H.}~\bibnamefont {Noh}},\ }\bibfield  {title} {\bibinfo {title} {{Axion as a Cold Dark Matter candidate}},\ }\href {https://doi.org/10.1016/j.physletb.2009.08.031} {\bibfield  {journal} {\bibinfo  {journal} {Phys. Lett. B}\ }\textbf {\bibinfo {volume} {680}},\ \bibinfo {pages} {1} (\bibinfo {year} {2009})},\ \Eprint {https://arxiv.org/abs/0902.4738} {arXiv:0902.4738 [astro-ph.CO]} \BibitemShut {NoStop}%
\bibitem [{\citenamefont {Visinelli}\ and\ \citenamefont {Gondolo}(2010)}]{Visinelli:2009kt}%
  \BibitemOpen
  \bibfield  {author} {\bibinfo {author} {\bibfnamefont {L.}~\bibnamefont {Visinelli}}\ and\ \bibinfo {author} {\bibfnamefont {P.}~\bibnamefont {Gondolo}},\ }\bibfield  {title} {\bibinfo {title} {{Axion cold dark matter in non-standard cosmologies}},\ }\href {https://doi.org/10.1103/PhysRevD.81.063508} {\bibfield  {journal} {\bibinfo  {journal} {Phys. Rev. D}\ }\textbf {\bibinfo {volume} {81}},\ \bibinfo {pages} {063508} (\bibinfo {year} {2010})},\ \Eprint {https://arxiv.org/abs/0912.0015} {arXiv:0912.0015 [astro-ph.CO]} \BibitemShut {NoStop}%
\bibitem [{\citenamefont {Kamionkowski}(1997)}]{Kamionkowski:1997zb}%
  \BibitemOpen
  \bibfield  {author} {\bibinfo {author} {\bibfnamefont {M.}~\bibnamefont {Kamionkowski}},\ }\bibfield  {title} {\bibinfo {title} {{WIMP and axion dark matter}},\ }in\ \href@noop {} {\emph {\bibinfo {booktitle} {{ICTP Summer School in High-Energy Physics and Cosmology}}}}\ (\bibinfo {year} {1997})\ pp.\ \bibinfo {pages} {394--411},\ \Eprint {https://arxiv.org/abs/hep-ph/9710467} {arXiv:hep-ph/9710467} \BibitemShut {NoStop}%
\bibitem [{\citenamefont {Preskill}\ \emph {et~al.}(1983)\citenamefont {Preskill}, \citenamefont {Wise},\ and\ \citenamefont {Wilczek}}]{Preskill:1982cy}%
  \BibitemOpen
  \bibfield  {author} {\bibinfo {author} {\bibfnamefont {J.}~\bibnamefont {Preskill}}, \bibinfo {author} {\bibfnamefont {M.~B.}\ \bibnamefont {Wise}},\ and\ \bibinfo {author} {\bibfnamefont {F.}~\bibnamefont {Wilczek}},\ }\bibfield  {title} {\bibinfo {title} {{Cosmology of the Invisible Axion}},\ }\href {https://doi.org/10.1016/0370-2693(83)90637-8} {\bibfield  {journal} {\bibinfo  {journal} {Phys. Lett. B}\ }\textbf {\bibinfo {volume} {120}},\ \bibinfo {pages} {127} (\bibinfo {year} {1983})}\BibitemShut {NoStop}%
\bibitem [{\citenamefont {Bauer}\ \emph {et~al.}(2020)\citenamefont {Bauer}, \citenamefont {Marsh}, \citenamefont {Hlo\v{z}ek}, \citenamefont {Padmanabhan},\ and\ \citenamefont {Lagu\"e}}]{Bauer:2020zsj}%
  \BibitemOpen
  \bibfield  {author} {\bibinfo {author} {\bibfnamefont {J.~B.}\ \bibnamefont {Bauer}}, \bibinfo {author} {\bibfnamefont {D.~J.~E.}\ \bibnamefont {Marsh}}, \bibinfo {author} {\bibfnamefont {R.}~\bibnamefont {Hlo\v{z}ek}}, \bibinfo {author} {\bibfnamefont {H.}~\bibnamefont {Padmanabhan}},\ and\ \bibinfo {author} {\bibfnamefont {A.}~\bibnamefont {Lagu\"e}},\ }\bibfield  {title} {\bibinfo {title} {{Intensity Mapping as a Probe of Axion Dark Matter}},\ }\href {https://doi.org/10.1093/mnras/staa3300} {\bibfield  {journal} {\bibinfo  {journal} {Mon. Not. Roy. Astron. Soc.}\ }\textbf {\bibinfo {volume} {500}},\ \bibinfo {pages} {3162} (\bibinfo {year} {2020})},\ \Eprint {https://arxiv.org/abs/2003.09655} {arXiv:2003.09655 [astro-ph.CO]} \BibitemShut {NoStop}%
\bibitem [{\citenamefont {Carucci}\ \emph {et~al.}(2015)\citenamefont {Carucci}, \citenamefont {Villaescusa-Navarro}, \citenamefont {Viel},\ and\ \citenamefont {Lapi}}]{Carucci:2015bra}%
  \BibitemOpen
  \bibfield  {author} {\bibinfo {author} {\bibfnamefont {I.~P.}\ \bibnamefont {Carucci}}, \bibinfo {author} {\bibfnamefont {F.}~\bibnamefont {Villaescusa-Navarro}}, \bibinfo {author} {\bibfnamefont {M.}~\bibnamefont {Viel}},\ and\ \bibinfo {author} {\bibfnamefont {A.}~\bibnamefont {Lapi}},\ }\bibfield  {title} {\bibinfo {title} {{Warm dark matter signatures on the 21cm power spectrum: Intensity mapping forecasts for SKA}},\ }\href {https://doi.org/10.1088/1475-7516/2015/07/047} {\bibfield  {journal} {\bibinfo  {journal} {JCAP}\ }\textbf {\bibinfo {volume} {07}},\ \bibinfo {pages} {047}},\ \Eprint {https://arxiv.org/abs/1502.06961} {arXiv:1502.06961 [astro-ph.CO]} \BibitemShut {NoStop}%
\bibitem [{\citenamefont {Creque-Sarbinowski}\ and\ \citenamefont {Kamionkowski}(2018)}]{Creque-Sarbinowski:2018ebl}%
  \BibitemOpen
  \bibfield  {author} {\bibinfo {author} {\bibfnamefont {C.}~\bibnamefont {Creque-Sarbinowski}}\ and\ \bibinfo {author} {\bibfnamefont {M.}~\bibnamefont {Kamionkowski}},\ }\bibfield  {title} {\bibinfo {title} {{Searching for Decaying and Annihilating Dark Matter with Line Intensity Mapping}},\ }\href {https://doi.org/10.1103/PhysRevD.98.063524} {\bibfield  {journal} {\bibinfo  {journal} {Phys. Rev. D}\ }\textbf {\bibinfo {volume} {98}},\ \bibinfo {pages} {063524} (\bibinfo {year} {2018})},\ \Eprint {https://arxiv.org/abs/1806.11119} {arXiv:1806.11119 [astro-ph.CO]} \BibitemShut {NoStop}%
\bibitem [{\citenamefont {Murakami}\ \emph {et~al.}(2024)\citenamefont {Murakami}, \citenamefont {Kadota}, \citenamefont {Nishizawa}, \citenamefont {Nagamine},\ and\ \citenamefont {Shimizu}}]{Murakami:2024jyi}%
  \BibitemOpen
  \bibfield  {author} {\bibinfo {author} {\bibfnamefont {K.}~\bibnamefont {Murakami}}, \bibinfo {author} {\bibfnamefont {K.}~\bibnamefont {Kadota}}, \bibinfo {author} {\bibfnamefont {A.~J.}\ \bibnamefont {Nishizawa}}, \bibinfo {author} {\bibfnamefont {K.}~\bibnamefont {Nagamine}},\ and\ \bibinfo {author} {\bibfnamefont {I.}~\bibnamefont {Shimizu}},\ }\bibfield  {title} {\bibinfo {title} {{Constraining dark matter model using 21cm line intensity mapping}},\ }\href@noop {} {\  (\bibinfo {year} {2024})},\ \Eprint {https://arxiv.org/abs/2403.06203} {arXiv:2403.06203 [astro-ph.CO]} \BibitemShut {NoStop}%
\bibitem [{\citenamefont {Byrnes}\ and\ \citenamefont {Choi}(2010)}]{Byrnes:2010em}%
  \BibitemOpen
  \bibfield  {author} {\bibinfo {author} {\bibfnamefont {C.~T.}\ \bibnamefont {Byrnes}}\ and\ \bibinfo {author} {\bibfnamefont {K.-Y.}\ \bibnamefont {Choi}},\ }\bibfield  {title} {\bibinfo {title} {{Review of local non-Gaussianity from multi-field inflation}},\ }\href {https://doi.org/10.1155/2010/724525} {\bibfield  {journal} {\bibinfo  {journal} {Adv. Astron.}\ }\textbf {\bibinfo {volume} {2010}},\ \bibinfo {pages} {724525} (\bibinfo {year} {2010})},\ \Eprint {https://arxiv.org/abs/1002.3110} {arXiv:1002.3110 [astro-ph.CO]} \BibitemShut {NoStop}%
\bibitem [{\citenamefont {Li}\ \emph {et~al.}(2016)\citenamefont {Li}, \citenamefont {Wechsler}, \citenamefont {Devaraj},\ and\ \citenamefont {Church}}]{Li:2015gqa}%
  \BibitemOpen
  \bibfield  {author} {\bibinfo {author} {\bibfnamefont {T.~Y.}\ \bibnamefont {Li}}, \bibinfo {author} {\bibfnamefont {R.~H.}\ \bibnamefont {Wechsler}}, \bibinfo {author} {\bibfnamefont {K.}~\bibnamefont {Devaraj}},\ and\ \bibinfo {author} {\bibfnamefont {S.~E.}\ \bibnamefont {Church}},\ }\bibfield  {title} {\bibinfo {title} {{Connecting CO Intensity Mapping to Molecular Gas and Star Formation in the Epoch of Galaxy Assembly}},\ }\href {https://doi.org/10.3847/0004-637X/817/2/169} {\bibfield  {journal} {\bibinfo  {journal} {Astrophys. J.}\ }\textbf {\bibinfo {volume} {817}},\ \bibinfo {pages} {169} (\bibinfo {year} {2016})},\ \Eprint {https://arxiv.org/abs/1503.08833} {arXiv:1503.08833 [astro-ph.CO]} \BibitemShut {NoStop}%
\bibitem [{\citenamefont {Cooray}\ and\ \citenamefont {Sheth}(2002)}]{Cooray:2002dia}%
  \BibitemOpen
  \bibfield  {author} {\bibinfo {author} {\bibfnamefont {A.}~\bibnamefont {Cooray}}\ and\ \bibinfo {author} {\bibfnamefont {R.~K.}\ \bibnamefont {Sheth}},\ }\bibfield  {title} {\bibinfo {title} {{Halo Models of Large Scale Structure}},\ }\href {https://doi.org/10.1016/S0370-1573(02)00276-4} {\bibfield  {journal} {\bibinfo  {journal} {Phys. Rept.}\ }\textbf {\bibinfo {volume} {372}},\ \bibinfo {pages} {1} (\bibinfo {year} {2002})},\ \Eprint {https://arxiv.org/abs/astro-ph/0206508} {arXiv:astro-ph/0206508} \BibitemShut {NoStop}%
\bibitem [{\citenamefont {Asgari}\ \emph {et~al.}(2023)\citenamefont {Asgari}, \citenamefont {Mead},\ and\ \citenamefont {Heymans}}]{Asgari:2023mej}%
  \BibitemOpen
  \bibfield  {author} {\bibinfo {author} {\bibfnamefont {M.}~\bibnamefont {Asgari}}, \bibinfo {author} {\bibfnamefont {A.~J.}\ \bibnamefont {Mead}},\ and\ \bibinfo {author} {\bibfnamefont {C.}~\bibnamefont {Heymans}},\ }\bibfield  {title} {\bibinfo {title} {{The halo model for cosmology: a pedagogical review}}\ }\href {https://doi.org/10.21105/astro.2303.08752} {10.21105/astro.2303.08752} (\bibinfo {year} {2023}),\ \Eprint {https://arxiv.org/abs/2303.08752} {arXiv:2303.08752 [astro-ph.CO]} \BibitemShut {NoStop}%
\bibitem [{\citenamefont {Diemer}\ and\ \citenamefont {Joyce}(2019)}]{Diemer:2018vmz}%
  \BibitemOpen
  \bibfield  {author} {\bibinfo {author} {\bibfnamefont {B.}~\bibnamefont {Diemer}}\ and\ \bibinfo {author} {\bibfnamefont {M.}~\bibnamefont {Joyce}},\ }\bibfield  {title} {\bibinfo {title} {{An accurate physical model for halo concentrations}},\ }\href {https://doi.org/10.3847/1538-4357/aafad6} {\bibfield  {journal} {\bibinfo  {journal} {Astrophys. J.}\ }\textbf {\bibinfo {volume} {871}},\ \bibinfo {pages} {168} (\bibinfo {year} {2019})},\ \Eprint {https://arxiv.org/abs/1809.07326} {arXiv:1809.07326 [astro-ph.CO]} \BibitemShut {NoStop}%
\bibitem [{\citenamefont {Moradinezhad~Dizgah}\ \emph {et~al.}(2022)\citenamefont {Moradinezhad~Dizgah}, \citenamefont {Nikakhtar}, \citenamefont {Keating},\ and\ \citenamefont {Castorina}}]{MoradinezhadDizgah:2021dei}%
  \BibitemOpen
  \bibfield  {author} {\bibinfo {author} {\bibfnamefont {A.}~\bibnamefont {Moradinezhad~Dizgah}}, \bibinfo {author} {\bibfnamefont {F.}~\bibnamefont {Nikakhtar}}, \bibinfo {author} {\bibfnamefont {G.~K.}\ \bibnamefont {Keating}},\ and\ \bibinfo {author} {\bibfnamefont {E.}~\bibnamefont {Castorina}},\ }\bibfield  {title} {\bibinfo {title} {{Precision tests of CO and CII power spectra models against simulated intensity maps}},\ }\href {https://doi.org/10.1088/1475-7516/2022/02/026} {\bibfield  {journal} {\bibinfo  {journal} {JCAP}\ }\textbf {\bibinfo {volume} {02}}\bibfield  {number} {\bibinfo  {number} { (02)},\ \bibinfo {pages} {026}},\ }\Eprint {https://arxiv.org/abs/2111.03717} {arXiv:2111.03717 [astro-ph.CO]} \BibitemShut {NoStop}%
\bibitem [{\citenamefont {Obuljen}\ \emph {et~al.}(2023)\citenamefont {Obuljen}, \citenamefont {Simonovi\'c}, \citenamefont {Schneider},\ and\ \citenamefont {Feldmann}}]{Obuljen:2022cjo}%
  \BibitemOpen
  \bibfield  {author} {\bibinfo {author} {\bibfnamefont {A.}~\bibnamefont {Obuljen}}, \bibinfo {author} {\bibfnamefont {M.}~\bibnamefont {Simonovi\'c}}, \bibinfo {author} {\bibfnamefont {A.}~\bibnamefont {Schneider}},\ and\ \bibinfo {author} {\bibfnamefont {R.}~\bibnamefont {Feldmann}},\ }\bibfield  {title} {\bibinfo {title} {{Modeling HI at the field level}},\ }\href {https://doi.org/10.1103/PhysRevD.108.083528} {\bibfield  {journal} {\bibinfo  {journal} {Phys. Rev. D}\ }\textbf {\bibinfo {volume} {108}},\ \bibinfo {pages} {083528} (\bibinfo {year} {2023})},\ \Eprint {https://arxiv.org/abs/2207.12398} {arXiv:2207.12398 [astro-ph.CO]} \BibitemShut {NoStop}%
\bibitem [{\citenamefont {Bernal}(2024)}]{Bernal:2023ovz}%
  \BibitemOpen
  \bibfield  {author} {\bibinfo {author} {\bibfnamefont {J.~L.}\ \bibnamefont {Bernal}},\ }\bibfield  {title} {\bibinfo {title} {{Toward accurate modeling of line-intensity mapping one-point statistics: Including extended profiles}},\ }\href {https://doi.org/10.1103/PhysRevD.109.043517} {\bibfield  {journal} {\bibinfo  {journal} {Phys. Rev. D}\ }\textbf {\bibinfo {volume} {109}},\ \bibinfo {pages} {043517} (\bibinfo {year} {2024})},\ \Eprint {https://arxiv.org/abs/2309.06481} {arXiv:2309.06481 [astro-ph.CO]} \BibitemShut {NoStop}%
\bibitem [{\citenamefont {Vernstrom}\ \emph {et~al.}(2014)\citenamefont {Vernstrom}, \citenamefont {Scott}, \citenamefont {Wall}, \citenamefont {Condon}, \citenamefont {Cotton}, \citenamefont {Fomalont}, \citenamefont {Kellermann}, \citenamefont {Miller},\ and\ \citenamefont {Perley}}]{Vernstrom:2013vva}%
  \BibitemOpen
  \bibfield  {author} {\bibinfo {author} {\bibfnamefont {T.}~\bibnamefont {Vernstrom}}, \bibinfo {author} {\bibfnamefont {D.}~\bibnamefont {Scott}}, \bibinfo {author} {\bibfnamefont {J.~V.}\ \bibnamefont {Wall}}, \bibinfo {author} {\bibfnamefont {J.~J.}\ \bibnamefont {Condon}}, \bibinfo {author} {\bibfnamefont {W.~D.}\ \bibnamefont {Cotton}}, \bibinfo {author} {\bibfnamefont {E.~B.}\ \bibnamefont {Fomalont}}, \bibinfo {author} {\bibfnamefont {K.~I.}\ \bibnamefont {Kellermann}}, \bibinfo {author} {\bibfnamefont {N.}~\bibnamefont {Miller}},\ and\ \bibinfo {author} {\bibfnamefont {R.~A.}\ \bibnamefont {Perley}},\ }\bibfield  {title} {\bibinfo {title} {{Deep 3 GHz number counts from a P(D) fluctuation analysis}},\ }\href {https://doi.org/10.1093/mnras/stu470} {\bibfield  {journal} {\bibinfo  {journal} {Mon. Not. Roy. Astron. Soc.}\ }\textbf {\bibinfo {volume} {440}},\ \bibinfo {pages} {2791} (\bibinfo {year} {2014})},\ \Eprint {https://arxiv.org/abs/1311.7451} {arXiv:1311.7451 [astro-ph.CO]} \BibitemShut
  {NoStop}%
\bibitem [{\citenamefont {Aghanim}\ \emph {et~al.}(2020)\citenamefont {Aghanim} \emph {et~al.}}]{Planck:2018vyg}%
  \BibitemOpen
  \bibfield  {author} {\bibinfo {author} {\bibfnamefont {N.}~\bibnamefont {Aghanim}} \emph {et~al.} (\bibinfo {collaboration} {Planck}),\ }\bibfield  {title} {\bibinfo {title} {{Planck 2018 results. VI. Cosmological parameters}},\ }\href {https://doi.org/10.1051/0004-6361/201833910} {\bibfield  {journal} {\bibinfo  {journal} {Astron. Astrophys.}\ }\textbf {\bibinfo {volume} {641}},\ \bibinfo {pages} {A6} (\bibinfo {year} {2020})},\ \bibinfo {note} {[Erratum: Astron.Astrophys. 652, C4 (2021)]},\ \Eprint {https://arxiv.org/abs/1807.06209} {arXiv:1807.06209 [astro-ph.CO]} \BibitemShut {NoStop}%
\bibitem [{\citenamefont {Hlozek}\ \emph {et~al.}(2015)\citenamefont {Hlozek}, \citenamefont {Grin}, \citenamefont {Marsh},\ and\ \citenamefont {Ferreira}}]{Hlozek:2014lca}%
  \BibitemOpen
  \bibfield  {author} {\bibinfo {author} {\bibfnamefont {R.}~\bibnamefont {Hlozek}}, \bibinfo {author} {\bibfnamefont {D.}~\bibnamefont {Grin}}, \bibinfo {author} {\bibfnamefont {D.~J.~E.}\ \bibnamefont {Marsh}},\ and\ \bibinfo {author} {\bibfnamefont {P.~G.}\ \bibnamefont {Ferreira}},\ }\bibfield  {title} {\bibinfo {title} {{A search for ultralight axions using precision cosmological data}},\ }\href {https://doi.org/10.1103/PhysRevD.91.103512} {\bibfield  {journal} {\bibinfo  {journal} {Phys. Rev. D}\ }\textbf {\bibinfo {volume} {91}},\ \bibinfo {pages} {103512} (\bibinfo {year} {2015})},\ \Eprint {https://arxiv.org/abs/1410.2896} {arXiv:1410.2896 [astro-ph.CO]} \BibitemShut {NoStop}%
\bibitem [{\citenamefont {Sabti}\ \emph {et~al.}(2021)\citenamefont {Sabti}, \citenamefont {Mu\~noz},\ and\ \citenamefont {Blas}}]{Sabti:2020ser}%
  \BibitemOpen
  \bibfield  {author} {\bibinfo {author} {\bibfnamefont {N.}~\bibnamefont {Sabti}}, \bibinfo {author} {\bibfnamefont {J.~B.}\ \bibnamefont {Mu\~noz}},\ and\ \bibinfo {author} {\bibfnamefont {D.}~\bibnamefont {Blas}},\ }\bibfield  {title} {\bibinfo {title} {{First Constraints on Small-Scale Non-Gaussianity from UV Galaxy Luminosity Functions}},\ }\href {https://doi.org/10.1088/1475-7516/2021/01/010} {\bibfield  {journal} {\bibinfo  {journal} {JCAP}\ }\textbf {\bibinfo {volume} {01}},\ \bibinfo {pages} {010}},\ \Eprint {https://arxiv.org/abs/2009.01245} {arXiv:2009.01245 [astro-ph.CO]} \BibitemShut {NoStop}%
\bibitem [{\citenamefont {LoVerde}\ \emph {et~al.}(2008)\citenamefont {LoVerde}, \citenamefont {Miller}, \citenamefont {Shandera},\ and\ \citenamefont {Verde}}]{LoVerde:2007ri}%
  \BibitemOpen
  \bibfield  {author} {\bibinfo {author} {\bibfnamefont {M.}~\bibnamefont {LoVerde}}, \bibinfo {author} {\bibfnamefont {A.}~\bibnamefont {Miller}}, \bibinfo {author} {\bibfnamefont {S.}~\bibnamefont {Shandera}},\ and\ \bibinfo {author} {\bibfnamefont {L.}~\bibnamefont {Verde}},\ }\bibfield  {title} {\bibinfo {title} {{Effects of Scale-Dependent Non-Gaussianity on Cosmological Structures}},\ }\href {https://doi.org/10.1088/1475-7516/2008/04/014} {\bibfield  {journal} {\bibinfo  {journal} {JCAP}\ }\textbf {\bibinfo {volume} {04}},\ \bibinfo {pages} {014}},\ \Eprint {https://arxiv.org/abs/0711.4126} {arXiv:0711.4126 [astro-ph]} \BibitemShut {NoStop}%
\bibitem [{\citenamefont {Matarrese}\ \emph {et~al.}(2000)\citenamefont {Matarrese}, \citenamefont {Verde},\ and\ \citenamefont {Jimenez}}]{Matarrese:2000iz}%
  \BibitemOpen
  \bibfield  {author} {\bibinfo {author} {\bibfnamefont {S.}~\bibnamefont {Matarrese}}, \bibinfo {author} {\bibfnamefont {L.}~\bibnamefont {Verde}},\ and\ \bibinfo {author} {\bibfnamefont {R.}~\bibnamefont {Jimenez}},\ }\bibfield  {title} {\bibinfo {title} {{The Abundance of high-redshift objects as a probe of non-Gaussian initial conditions}},\ }\href {https://doi.org/10.1086/309412} {\bibfield  {journal} {\bibinfo  {journal} {Astrophys. J.}\ }\textbf {\bibinfo {volume} {541}},\ \bibinfo {pages} {10} (\bibinfo {year} {2000})},\ \Eprint {https://arxiv.org/abs/astro-ph/0001366} {arXiv:astro-ph/0001366} \BibitemShut {NoStop}%
\bibitem [{\citenamefont {Dalal}\ \emph {et~al.}(2008)\citenamefont {Dalal}, \citenamefont {Dore}, \citenamefont {Huterer},\ and\ \citenamefont {Shirokov}}]{Dalal:2007cu}%
  \BibitemOpen
  \bibfield  {author} {\bibinfo {author} {\bibfnamefont {N.}~\bibnamefont {Dalal}}, \bibinfo {author} {\bibfnamefont {O.}~\bibnamefont {Dore}}, \bibinfo {author} {\bibfnamefont {D.}~\bibnamefont {Huterer}},\ and\ \bibinfo {author} {\bibfnamefont {A.}~\bibnamefont {Shirokov}},\ }\bibfield  {title} {\bibinfo {title} {{The imprints of primordial non-gaussianities on large-scale structure: scale dependent bias and abundance of virialized objects}},\ }\href {https://doi.org/10.1103/PhysRevD.77.123514} {\bibfield  {journal} {\bibinfo  {journal} {Phys. Rev. D}\ }\textbf {\bibinfo {volume} {77}},\ \bibinfo {pages} {123514} (\bibinfo {year} {2008})},\ \Eprint {https://arxiv.org/abs/0710.4560} {arXiv:0710.4560 [astro-ph]} \BibitemShut {NoStop}%
\bibitem [{\citenamefont {Matarrese}\ and\ \citenamefont {Verde}(2008)}]{Matarrese:2008nc}%
  \BibitemOpen
  \bibfield  {author} {\bibinfo {author} {\bibfnamefont {S.}~\bibnamefont {Matarrese}}\ and\ \bibinfo {author} {\bibfnamefont {L.}~\bibnamefont {Verde}},\ }\bibfield  {title} {\bibinfo {title} {{The effect of primordial non-Gaussianity on halo bias}},\ }\href {https://doi.org/10.1086/587840} {\bibfield  {journal} {\bibinfo  {journal} {Astrophys. J. Lett.}\ }\textbf {\bibinfo {volume} {677}},\ \bibinfo {pages} {L77} (\bibinfo {year} {2008})},\ \Eprint {https://arxiv.org/abs/0801.4826} {arXiv:0801.4826 [astro-ph]} \BibitemShut {NoStop}%
\bibitem [{\citenamefont {Desjacques}\ and\ \citenamefont {Seljak}(2010)}]{Desjacques:2010jw}%
  \BibitemOpen
  \bibfield  {author} {\bibinfo {author} {\bibfnamefont {V.}~\bibnamefont {Desjacques}}\ and\ \bibinfo {author} {\bibfnamefont {U.}~\bibnamefont {Seljak}},\ }\bibfield  {title} {\bibinfo {title} {{Primordial non-Gaussianity from the large scale structure}},\ }\href {https://doi.org/10.1088/0264-9381/27/12/124011} {\bibfield  {journal} {\bibinfo  {journal} {Class. Quant. Grav.}\ }\textbf {\bibinfo {volume} {27}},\ \bibinfo {pages} {124011} (\bibinfo {year} {2010})},\ \Eprint {https://arxiv.org/abs/1003.5020} {arXiv:1003.5020 [astro-ph.CO]} \BibitemShut {NoStop}%
\bibitem [{\citenamefont {Barreira}(2022{\natexlab{a}})}]{Barreira:2022sey}%
  \BibitemOpen
  \bibfield  {author} {\bibinfo {author} {\bibfnamefont {A.}~\bibnamefont {Barreira}},\ }\bibfield  {title} {\bibinfo {title} {{Can we actually constrain f$_{NL}$ using the scale-dependent bias effect? An illustration of the impact of galaxy bias uncertainties using the BOSS DR12 galaxy power spectrum}},\ }\href {https://doi.org/10.1088/1475-7516/2022/11/013} {\bibfield  {journal} {\bibinfo  {journal} {JCAP}\ }\textbf {\bibinfo {volume} {11}},\ \bibinfo {pages} {013}},\ \Eprint {https://arxiv.org/abs/2205.05673} {arXiv:2205.05673 [astro-ph.CO]} \BibitemShut {NoStop}%
\bibitem [{\citenamefont {Barreira}(2022{\natexlab{b}})}]{Barreira:2021dpt}%
  \BibitemOpen
  \bibfield  {author} {\bibinfo {author} {\bibfnamefont {A.}~\bibnamefont {Barreira}},\ }\bibfield  {title} {\bibinfo {title} {{The local PNG bias of neutral Hydrogen, H$_{I}$}},\ }\href {https://doi.org/10.1088/1475-7516/2022/04/057} {\bibfield  {journal} {\bibinfo  {journal} {JCAP}\ }\textbf {\bibinfo {volume} {04}}\bibfield  {number} {\bibinfo  {number} { (04)},\ \bibinfo {pages} {057}},\ }\Eprint {https://arxiv.org/abs/2112.03253} {arXiv:2112.03253 [astro-ph.CO]} \BibitemShut {NoStop}%
\bibitem [{\citenamefont {Breysse}\ \emph {et~al.}(2015)\citenamefont {Breysse}, \citenamefont {Kovetz},\ and\ \citenamefont {Kamionkowski}}]{Breysse:2015baa}%
  \BibitemOpen
  \bibfield  {author} {\bibinfo {author} {\bibfnamefont {P.~C.}\ \bibnamefont {Breysse}}, \bibinfo {author} {\bibfnamefont {E.~D.}\ \bibnamefont {Kovetz}},\ and\ \bibinfo {author} {\bibfnamefont {M.}~\bibnamefont {Kamionkowski}},\ }\bibfield  {title} {\bibinfo {title} {{Masking line foregrounds in intensity mapping surveys}},\ }\href {https://doi.org/10.1093/mnras/stv1476} {\bibfield  {journal} {\bibinfo  {journal} {Mon. Not. Roy. Astron. Soc.}\ }\textbf {\bibinfo {volume} {452}},\ \bibinfo {pages} {3408} (\bibinfo {year} {2015})},\ \Eprint {https://arxiv.org/abs/1503.05202} {arXiv:1503.05202 [astro-ph.CO]} \BibitemShut {NoStop}%
\bibitem [{\citenamefont {{Sun}}\ \emph {et~al.}(2018)\citenamefont {{Sun}}, \citenamefont {{Moncelsi}}, \citenamefont {{Viero}}, \citenamefont {{Silva}}, \citenamefont {{Bock}}, \citenamefont {{Bradford}}, \citenamefont {{Chang}}, \citenamefont {{Cheng}}, \citenamefont {{Cooray}}, \citenamefont {{Crites}}, \citenamefont {{Hailey-Dunsheath}}, \citenamefont {{Uzgil}}, \citenamefont {{Hunacek}},\ and\ \citenamefont {{Zemcov}}}]{2018ApJ...856..107S}%
  \BibitemOpen
  \bibfield  {author} {\bibinfo {author} {\bibfnamefont {G.}~\bibnamefont {{Sun}}}, \bibinfo {author} {\bibfnamefont {L.}~\bibnamefont {{Moncelsi}}}, \bibinfo {author} {\bibfnamefont {M.~P.}\ \bibnamefont {{Viero}}}, \bibinfo {author} {\bibfnamefont {M.~B.}\ \bibnamefont {{Silva}}}, \bibinfo {author} {\bibfnamefont {J.}~\bibnamefont {{Bock}}}, \bibinfo {author} {\bibfnamefont {C.~M.}\ \bibnamefont {{Bradford}}}, \bibinfo {author} {\bibfnamefont {T.~C.}\ \bibnamefont {{Chang}}}, \bibinfo {author} {\bibfnamefont {Y.~T.}\ \bibnamefont {{Cheng}}}, \bibinfo {author} {\bibfnamefont {A.~R.}\ \bibnamefont {{Cooray}}}, \bibinfo {author} {\bibfnamefont {A.}~\bibnamefont {{Crites}}}, \bibinfo {author} {\bibfnamefont {S.}~\bibnamefont {{Hailey-Dunsheath}}}, \bibinfo {author} {\bibfnamefont {B.}~\bibnamefont {{Uzgil}}}, \bibinfo {author} {\bibfnamefont {J.~R.}\ \bibnamefont {{Hunacek}}},\ and\ \bibinfo {author} {\bibfnamefont {M.}~\bibnamefont {{Zemcov}}},\ }\bibfield  {title} {\bibinfo {title} {{A Foreground Masking
  Strategy for [C II] Intensity Mapping Experiments Using Galaxies Selected by Stellar Mass and Redshift}},\ }\href {https://doi.org/10.3847/1538-4357/aab3e3} {\bibfield  {journal} {\bibinfo  {journal} {\apj}\ }\textbf {\bibinfo {volume} {856}},\ \bibinfo {eid} {107} (\bibinfo {year} {2018})},\ \Eprint {https://arxiv.org/abs/1610.10095} {arXiv:1610.10095 [astro-ph.GA]} \BibitemShut {NoStop}%
\bibitem [{\citenamefont {Cheng}\ \emph {et~al.}(2020)\citenamefont {Cheng}, \citenamefont {Chang},\ and\ \citenamefont {Bock}}]{Cheng:2020asz}%
  \BibitemOpen
  \bibfield  {author} {\bibinfo {author} {\bibfnamefont {Y.-T.}\ \bibnamefont {Cheng}}, \bibinfo {author} {\bibfnamefont {T.-C.}\ \bibnamefont {Chang}},\ and\ \bibinfo {author} {\bibfnamefont {J.~J.}\ \bibnamefont {Bock}},\ }\bibfield  {title} {\bibinfo {title} {{Phase-space Spectral Line Deconfusion in Intensity Mapping}},\ }\href {https://doi.org/10.3847/1538-4357/abb023} {\bibfield  {journal} {\bibinfo  {journal} {Astrophys. J.}\ }\textbf {\bibinfo {volume} {901}},\ \bibinfo {pages} {142} (\bibinfo {year} {2020})},\ \Eprint {https://arxiv.org/abs/2005.05341} {arXiv:2005.05341 [astro-ph.CO]} \BibitemShut {NoStop}%
\bibitem [{\citenamefont {Moriwaki}\ \emph {et~al.}(2020)\citenamefont {Moriwaki}, \citenamefont {Filippova}, \citenamefont {Shirasaki},\ and\ \citenamefont {Yoshida}}]{moriwaki2020deep}%
  \BibitemOpen
  \bibfield  {author} {\bibinfo {author} {\bibfnamefont {K.}~\bibnamefont {Moriwaki}}, \bibinfo {author} {\bibfnamefont {N.}~\bibnamefont {Filippova}}, \bibinfo {author} {\bibfnamefont {M.}~\bibnamefont {Shirasaki}},\ and\ \bibinfo {author} {\bibfnamefont {N.}~\bibnamefont {Yoshida}},\ }\bibfield  {title} {\bibinfo {title} {Deep learning for intensity mapping observations: component extraction},\ }\href@noop {} {\bibfield  {journal} {\bibinfo  {journal} {Monthly Notices of the Royal Astronomical Society: Letters}\ }\textbf {\bibinfo {volume} {496}},\ \bibinfo {pages} {L54} (\bibinfo {year} {2020})}\BibitemShut {NoStop}%
\bibitem [{\citenamefont {Cunnington}\ \emph {et~al.}(2023)\citenamefont {Cunnington} \emph {et~al.}}]{Cunnington:2023jpq}%
  \BibitemOpen
  \bibfield  {author} {\bibinfo {author} {\bibfnamefont {S.}~\bibnamefont {Cunnington}} \emph {et~al.},\ }\bibfield  {title} {\bibinfo {title} {{The foreground transfer function for H\,i intensity mapping signal reconstruction: MeerKLASS and precision cosmology applications}},\ }\href {https://doi.org/10.1093/mnras/stad1567} {\bibfield  {journal} {\bibinfo  {journal} {Mon. Not. Roy. Astron. Soc.}\ }\textbf {\bibinfo {volume} {523}},\ \bibinfo {pages} {2453} (\bibinfo {year} {2023})},\ \Eprint {https://arxiv.org/abs/2302.07034} {arXiv:2302.07034 [astro-ph.CO]} \BibitemShut {NoStop}%
\bibitem [{\citenamefont {Van~Cuyck}\ \emph {et~al.}(2023)\citenamefont {Van~Cuyck} \emph {et~al.}}]{VanCuyck:2023uli}%
  \BibitemOpen
  \bibfield  {author} {\bibinfo {author} {\bibfnamefont {M.}~\bibnamefont {Van~Cuyck}} \emph {et~al.},\ }\bibfield  {title} {\bibinfo {title} {{CONCERTO: Extracting the power spectrum of the [C II ] emission line}},\ }\href {https://doi.org/10.1051/0004-6361/202346270} {\bibfield  {journal} {\bibinfo  {journal} {Astron. Astrophys.}\ }\textbf {\bibinfo {volume} {676}},\ \bibinfo {pages} {A62} (\bibinfo {year} {2023})},\ \Eprint {https://arxiv.org/abs/2306.01568} {arXiv:2306.01568 [astro-ph.CO]} \BibitemShut {NoStop}%
\bibitem [{\citenamefont {Bernal}\ and\ \citenamefont {Baleato~Lizancos}()}]{Bernal_deinterloping}%
  \BibitemOpen
  \bibfield  {author} {\bibinfo {author} {\bibfnamefont {J.~L.}\ \bibnamefont {Bernal}}\ and\ \bibinfo {author} {\bibfnamefont {A.}~\bibnamefont {Baleato~Lizancos}},\ }\bibfield  {title} {\bibinfo {title} {{}},\ }\href@noop {} {\bibinfo  {journal} {In preparation}\ }\BibitemShut {NoStop}%
\bibitem [{\citenamefont {Foss}\ \emph {et~al.}(2022)\citenamefont {Foss} \emph {et~al.}}]{COMAP:2021pxy}%
  \BibitemOpen
\bibfield  {journal} {  }\bibfield  {author} {\bibinfo {author} {\bibfnamefont {M.~K.}\ \bibnamefont {Foss}} \emph {et~al.} (\bibinfo {collaboration} {COMAP}),\ }\bibfield  {title} {\bibinfo {title} {{COMAP Early Science. III. CO Data Processing}},\ }\href {https://doi.org/10.3847/1538-4357/ac63ca} {\bibfield  {journal} {\bibinfo  {journal} {Astrophys. J.}\ }\textbf {\bibinfo {volume} {933}},\ \bibinfo {pages} {184} (\bibinfo {year} {2022})},\ \Eprint {https://arxiv.org/abs/2111.05929} {arXiv:2111.05929 [astro-ph.IM]} \BibitemShut {NoStop}%
\bibitem [{\citenamefont {Behroozi}\ \emph {et~al.}(2019)\citenamefont {Behroozi}, \citenamefont {Wechsler}, \citenamefont {Hearin},\ and\ \citenamefont {Conroy}}]{Behroozi:2019kql}%
  \BibitemOpen
  \bibfield  {author} {\bibinfo {author} {\bibfnamefont {P.}~\bibnamefont {Behroozi}}, \bibinfo {author} {\bibfnamefont {R.~H.}\ \bibnamefont {Wechsler}}, \bibinfo {author} {\bibfnamefont {A.~P.}\ \bibnamefont {Hearin}},\ and\ \bibinfo {author} {\bibfnamefont {C.}~\bibnamefont {Conroy}},\ }\bibfield  {title} {\bibinfo {title} {{UniverseMachine: The correlation between galaxy growth and dark matter halo assembly from z = 0\ensuremath{-}10}},\ }\href {https://doi.org/10.1093/mnras/stz1182} {\bibfield  {journal} {\bibinfo  {journal} {Mon. Not. Roy. Astron. Soc.}\ }\textbf {\bibinfo {volume} {488}},\ \bibinfo {pages} {3143} (\bibinfo {year} {2019})},\ \Eprint {https://arxiv.org/abs/1806.07893} {arXiv:1806.07893} \BibitemShut {NoStop}%
\bibitem [{\citenamefont {Riechers}\ \emph {et~al.}(2019)\citenamefont {Riechers} \emph {et~al.}}]{Riechers:2018zjg}%
  \BibitemOpen
  \bibfield  {author} {\bibinfo {author} {\bibfnamefont {D.~A.}\ \bibnamefont {Riechers}} \emph {et~al.},\ }\bibfield  {title} {\bibinfo {title} {{COLDz: Shape of the CO Luminosity Function at High Redshift and the Cold Gas History of the Universe}},\ }\href {https://doi.org/10.3847/1538-4357/aafc27} {\bibfield  {journal} {\bibinfo  {journal} {Astrophys. J.}\ }\textbf {\bibinfo {volume} {872}},\ \bibinfo {pages} {7} (\bibinfo {year} {2019})},\ \Eprint {https://arxiv.org/abs/1808.04371} {arXiv:1808.04371 [astro-ph.GA]} \BibitemShut {NoStop}%
\bibitem [{\citenamefont {Tinker}\ \emph {et~al.}(2008)\citenamefont {Tinker}, \citenamefont {Kravtsov}, \citenamefont {Klypin}, \citenamefont {Abazajian}, \citenamefont {Warren}, \citenamefont {Yepes}, \citenamefont {Gottlober},\ and\ \citenamefont {Holz}}]{Tinker:2008ff}%
  \BibitemOpen
  \bibfield  {author} {\bibinfo {author} {\bibfnamefont {J.~L.}\ \bibnamefont {Tinker}}, \bibinfo {author} {\bibfnamefont {A.~V.}\ \bibnamefont {Kravtsov}}, \bibinfo {author} {\bibfnamefont {A.}~\bibnamefont {Klypin}}, \bibinfo {author} {\bibfnamefont {K.}~\bibnamefont {Abazajian}}, \bibinfo {author} {\bibfnamefont {M.~S.}\ \bibnamefont {Warren}}, \bibinfo {author} {\bibfnamefont {G.}~\bibnamefont {Yepes}}, \bibinfo {author} {\bibfnamefont {S.}~\bibnamefont {Gottlober}},\ and\ \bibinfo {author} {\bibfnamefont {D.~E.}\ \bibnamefont {Holz}},\ }\bibfield  {title} {\bibinfo {title} {{Toward a halo mass function for precision cosmology: The Limits of universality}},\ }\href {https://doi.org/10.1086/591439} {\bibfield  {journal} {\bibinfo  {journal} {Astrophys. J.}\ }\textbf {\bibinfo {volume} {688}},\ \bibinfo {pages} {709} (\bibinfo {year} {2008})},\ \Eprint {https://arxiv.org/abs/0803.2706} {arXiv:0803.2706 [astro-ph]} \BibitemShut {NoStop}%
\bibitem [{\citenamefont {Tinker}\ \emph {et~al.}(2010)\citenamefont {Tinker}, \citenamefont {Robertson}, \citenamefont {Kravtsov}, \citenamefont {Klypin}, \citenamefont {Warren}, \citenamefont {Yepes},\ and\ \citenamefont {Gottlober}}]{Tinker:2010my}%
  \BibitemOpen
  \bibfield  {author} {\bibinfo {author} {\bibfnamefont {J.~L.}\ \bibnamefont {Tinker}}, \bibinfo {author} {\bibfnamefont {B.~E.}\ \bibnamefont {Robertson}}, \bibinfo {author} {\bibfnamefont {A.~V.}\ \bibnamefont {Kravtsov}}, \bibinfo {author} {\bibfnamefont {A.}~\bibnamefont {Klypin}}, \bibinfo {author} {\bibfnamefont {M.~S.}\ \bibnamefont {Warren}}, \bibinfo {author} {\bibfnamefont {G.}~\bibnamefont {Yepes}},\ and\ \bibinfo {author} {\bibfnamefont {S.}~\bibnamefont {Gottlober}},\ }\bibfield  {title} {\bibinfo {title} {{The Large Scale Bias of Dark Matter Halos: Numerical Calibration and Model Tests}},\ }\href {https://doi.org/10.1088/0004-637X/724/2/878} {\bibfield  {journal} {\bibinfo  {journal} {Astrophys. J.}\ }\textbf {\bibinfo {volume} {724}},\ \bibinfo {pages} {878} (\bibinfo {year} {2010})},\ \Eprint {https://arxiv.org/abs/1001.3162} {arXiv:1001.3162 [astro-ph.CO]} \BibitemShut {NoStop}%
\bibitem [{\citenamefont {Hlozek}\ \emph {et~al.}(2018)\citenamefont {Hlozek}, \citenamefont {Marsh},\ and\ \citenamefont {Grin}}]{Hlozek:2017zzf}%
  \BibitemOpen
  \bibfield  {author} {\bibinfo {author} {\bibfnamefont {R.}~\bibnamefont {Hlozek}}, \bibinfo {author} {\bibfnamefont {D.~J.~E.}\ \bibnamefont {Marsh}},\ and\ \bibinfo {author} {\bibfnamefont {D.}~\bibnamefont {Grin}},\ }\bibfield  {title} {\bibinfo {title} {{Using the Full Power of the Cosmic Microwave Background to Probe Axion Dark Matter}},\ }\href {https://doi.org/10.1093/mnras/sty271} {\bibfield  {journal} {\bibinfo  {journal} {Mon. Not. Roy. Astron. Soc.}\ }\textbf {\bibinfo {volume} {476}},\ \bibinfo {pages} {3063} (\bibinfo {year} {2018})},\ \Eprint {https://arxiv.org/abs/1708.05681} {arXiv:1708.05681 [astro-ph.CO]} \BibitemShut {NoStop}%
\bibitem [{\citenamefont {Lagu\"e}\ \emph {et~al.}(2022)\citenamefont {Lagu\"e}, \citenamefont {Bond}, \citenamefont {Hlo\v{z}ek}, \citenamefont {Rogers}, \citenamefont {Marsh},\ and\ \citenamefont {Grin}}]{Lague:2021frh}%
  \BibitemOpen
  \bibfield  {author} {\bibinfo {author} {\bibfnamefont {A.}~\bibnamefont {Lagu\"e}}, \bibinfo {author} {\bibfnamefont {J.~R.}\ \bibnamefont {Bond}}, \bibinfo {author} {\bibfnamefont {R.}~\bibnamefont {Hlo\v{z}ek}}, \bibinfo {author} {\bibfnamefont {K.~K.}\ \bibnamefont {Rogers}}, \bibinfo {author} {\bibfnamefont {D.~J.~E.}\ \bibnamefont {Marsh}},\ and\ \bibinfo {author} {\bibfnamefont {D.}~\bibnamefont {Grin}},\ }\bibfield  {title} {\bibinfo {title} {{Constraining ultralight axions with galaxy surveys}},\ }\href {https://doi.org/10.1088/1475-7516/2022/01/049} {\bibfield  {journal} {\bibinfo  {journal} {JCAP}\ }\textbf {\bibinfo {volume} {01}}\bibfield  {number} {\bibinfo  {number} { (01)},\ \bibinfo {pages} {049}},\ }\Eprint {https://arxiv.org/abs/2104.07802} {arXiv:2104.07802 [astro-ph.CO]} \BibitemShut {NoStop}%
\bibitem [{\citenamefont {Rogers}\ \emph {et~al.}(2023)\citenamefont {Rogers}, \citenamefont {Hlo\v{z}ek}, \citenamefont {Lagu\"e}, \citenamefont {Ivanov}, \citenamefont {Philcox}, \citenamefont {Cabass}, \citenamefont {Akitsu},\ and\ \citenamefont {Marsh}}]{Rogers:2023ezo}%
  \BibitemOpen
  \bibfield  {author} {\bibinfo {author} {\bibfnamefont {K.~K.}\ \bibnamefont {Rogers}}, \bibinfo {author} {\bibfnamefont {R.}~\bibnamefont {Hlo\v{z}ek}}, \bibinfo {author} {\bibfnamefont {A.}~\bibnamefont {Lagu\"e}}, \bibinfo {author} {\bibfnamefont {M.~M.}\ \bibnamefont {Ivanov}}, \bibinfo {author} {\bibfnamefont {O.~H.~E.}\ \bibnamefont {Philcox}}, \bibinfo {author} {\bibfnamefont {G.}~\bibnamefont {Cabass}}, \bibinfo {author} {\bibfnamefont {K.}~\bibnamefont {Akitsu}},\ and\ \bibinfo {author} {\bibfnamefont {D.~J.~E.}\ \bibnamefont {Marsh}},\ }\bibfield  {title} {\bibinfo {title} {{Ultra-light axions and the S $_{8}$ tension: joint constraints from the cosmic microwave background and galaxy clustering}},\ }\href {https://doi.org/10.1088/1475-7516/2023/06/023} {\bibfield  {journal} {\bibinfo  {journal} {JCAP}\ }\textbf {\bibinfo {volume} {06}},\ \bibinfo {pages} {023}},\ \Eprint {https://arxiv.org/abs/2301.08361} {arXiv:2301.08361 [astro-ph.CO]} \BibitemShut {NoStop}%
\bibitem [{\citenamefont {Stahl}\ \emph {et~al.}(2024)\citenamefont {Stahl}, \citenamefont {Famaey}, \citenamefont {Ibata}, \citenamefont {Hahn}, \citenamefont {Martinet},\ and\ \citenamefont {Montandon}}]{Stahl:2024stz}%
  \BibitemOpen
  \bibfield  {author} {\bibinfo {author} {\bibfnamefont {C.}~\bibnamefont {Stahl}}, \bibinfo {author} {\bibfnamefont {B.}~\bibnamefont {Famaey}}, \bibinfo {author} {\bibfnamefont {R.}~\bibnamefont {Ibata}}, \bibinfo {author} {\bibfnamefont {O.}~\bibnamefont {Hahn}}, \bibinfo {author} {\bibfnamefont {N.}~\bibnamefont {Martinet}},\ and\ \bibinfo {author} {\bibfnamefont {T.}~\bibnamefont {Montandon}},\ }\bibfield  {title} {\bibinfo {title} {{Scale-dependent local primordial non-Gaussianity as a solution to the $S_8$ tension}},\ }\href@noop {} {\  (\bibinfo {year} {2024})},\ \Eprint {https://arxiv.org/abs/2404.03244} {arXiv:2404.03244 [astro-ph.CO]} \BibitemShut {NoStop}%
\bibitem [{\citenamefont {Sato-Polito}\ \emph {et~al.}(2021)\citenamefont {Sato-Polito}, \citenamefont {Bernal}, \citenamefont {Boddy},\ and\ \citenamefont {Kamionkowski}}]{Sato-Polito:2020cil}%
  \BibitemOpen
  \bibfield  {author} {\bibinfo {author} {\bibfnamefont {G.}~\bibnamefont {Sato-Polito}}, \bibinfo {author} {\bibfnamefont {J.~L.}\ \bibnamefont {Bernal}}, \bibinfo {author} {\bibfnamefont {K.~K.}\ \bibnamefont {Boddy}},\ and\ \bibinfo {author} {\bibfnamefont {M.}~\bibnamefont {Kamionkowski}},\ }\bibfield  {title} {\bibinfo {title} {{Kinetic Sunyaev-Zel\textquoteright{}dovich tomography with line-intensity mapping}},\ }\href {https://doi.org/10.1103/PhysRevD.103.083519} {\bibfield  {journal} {\bibinfo  {journal} {Phys. Rev. D}\ }\textbf {\bibinfo {volume} {103}},\ \bibinfo {pages} {083519} (\bibinfo {year} {2021})},\ \Eprint {https://arxiv.org/abs/2011.08193} {arXiv:2011.08193 [astro-ph.CO]} \BibitemShut {NoStop}%
\bibitem [{\citenamefont {Breysse}\ \emph {et~al.}(2019)\citenamefont {Breysse}, \citenamefont {Anderson},\ and\ \citenamefont {Berger}}]{Breysse:2019cdw}%
  \BibitemOpen
  \bibfield  {author} {\bibinfo {author} {\bibfnamefont {P.~C.}\ \bibnamefont {Breysse}}, \bibinfo {author} {\bibfnamefont {C.~J.}\ \bibnamefont {Anderson}},\ and\ \bibinfo {author} {\bibfnamefont {P.}~\bibnamefont {Berger}},\ }\bibfield  {title} {\bibinfo {title} {{Canceling out intensity mapping foregrounds}},\ }\href {https://doi.org/10.1103/PhysRevLett.123.231105} {\bibfield  {journal} {\bibinfo  {journal} {Phys. Rev. Lett.}\ }\textbf {\bibinfo {volume} {123}},\ \bibinfo {pages} {231105} (\bibinfo {year} {2019})},\ \Eprint {https://arxiv.org/abs/1907.04369} {arXiv:1907.04369 [astro-ph.CO]} \BibitemShut {NoStop}%
\bibitem [{\citenamefont {Breysse}\ \emph {et~al.}(2023)\citenamefont {Breysse}, \citenamefont {Chung},\ and\ \citenamefont {Ihle}}]{Breysse:2022fdi}%
  \BibitemOpen
  \bibfield  {author} {\bibinfo {author} {\bibfnamefont {P.~C.}\ \bibnamefont {Breysse}}, \bibinfo {author} {\bibfnamefont {D.~T.}\ \bibnamefont {Chung}},\ and\ \bibinfo {author} {\bibfnamefont {H.~T.}\ \bibnamefont {Ihle}},\ }\bibfield  {title} {\bibinfo {title} {{Characteristic functions for cosmological cross-correlations}},\ }\href {https://doi.org/10.1093/mnras/stad2350} {\bibfield  {journal} {\bibinfo  {journal} {Mon. Not. Roy. Astron. Soc.}\ }\textbf {\bibinfo {volume} {525}},\ \bibinfo {pages} {1824} (\bibinfo {year} {2023})},\ \Eprint {https://arxiv.org/abs/2210.14902} {arXiv:2210.14902 [astro-ph.CO]} \BibitemShut {NoStop}%
\bibitem [{\citenamefont {Chung}\ \emph {et~al.}(2023)\citenamefont {Chung}, \citenamefont {Bangari}, \citenamefont {Breysse}, \citenamefont {Ihle}, \citenamefont {Bond}, \citenamefont {Dunne}, \citenamefont {Padmanabhan}, \citenamefont {Philip}, \citenamefont {Rennie},\ and\ \citenamefont {Viero}}]{COMAP:2022sdg}%
  \BibitemOpen
  \bibfield  {author} {\bibinfo {author} {\bibfnamefont {D.~T.}\ \bibnamefont {Chung}}, \bibinfo {author} {\bibfnamefont {I.}~\bibnamefont {Bangari}}, \bibinfo {author} {\bibfnamefont {P.~C.}\ \bibnamefont {Breysse}}, \bibinfo {author} {\bibfnamefont {H.~T.}\ \bibnamefont {Ihle}}, \bibinfo {author} {\bibfnamefont {J.~R.}\ \bibnamefont {Bond}}, \bibinfo {author} {\bibfnamefont {D.~A.}\ \bibnamefont {Dunne}}, \bibinfo {author} {\bibfnamefont {H.}~\bibnamefont {Padmanabhan}}, \bibinfo {author} {\bibfnamefont {L.}~\bibnamefont {Philip}}, \bibinfo {author} {\bibfnamefont {T.~J.}\ \bibnamefont {Rennie}},\ and\ \bibinfo {author} {\bibfnamefont {M.~P.}\ \bibnamefont {Viero}} (\bibinfo {collaboration} {COMAP}),\ }\bibfield  {title} {\bibinfo {title} {{The deconvolved distribution estimator: enhancing reionization-era CO line-intensity mapping analyses with a cross-correlation analogue for one-point statistics}},\ }\href {https://doi.org/10.1093/mnras/stad359} {\bibfield  {journal} {\bibinfo  {journal} {Mon. Not. Roy.
  Astron. Soc.}\ }\textbf {\bibinfo {volume} {520}},\ \bibinfo {pages} {5305} (\bibinfo {year} {2023})},\ \Eprint {https://arxiv.org/abs/2210.14890} {arXiv:2210.14890 [astro-ph.CO]} \BibitemShut {NoStop}%
\end{thebibliography}%

\end{document}